\documentclass[preprint,11pt,a4wide,nopreprintline]{elsarticle}

\usepackage[latin1]{inputenc} 
\usepackage{leqno}
\usepackage{amssymb}
\usepackage{amsmath}
\usepackage{xspace}
\usepackage{marvosym}
\usepackage{subfig} 
\usepackage{graphicx}
\usepackage{hyperref} 
\usepackage{txfonts}
\usepackage{multirow}
\usepackage{rotating} 

\journal{Astroparticle Physics}

\def\ruo{\rule{0mm}{6mm}}

\def\eqb{\begin{equation}}
\def\eqe{\end{equation}}
\def\itb{\begin{itemize}}
\def\ite{\end{itemize}}
\def\enb{\begin{enumerate}}
\def\ene{\end{enumerate}}

\def\lbra{\left (}
\def\rbra{\right )}

\newcommand{\unit}[1]{\,\mathrm{#1}}
\newcommand{\cher}{\xspace{Cherenkov}\xspace}
\newcommand{\cherl}{\xspace{Cherenkov light}\xspace}
\newcommand{\chera}{\xspace{Cherenkov angle}\xspace}
\newcommand{\cherthr}{\xspace{Cherenkov threshold}\xspace}
\newcommand{\cherpho}{\xspace{Cherenkov photon}\xspace}
\newcommand{\cherphos}{\xspace{Cherenkov photons}\xspace}
\newcommand{\cherc}{\xspace{Cherenkov cone}\xspace}
\newcommand{\tamm}{\xspace{Frank-Tamm}\xspace}
\newcommand{\tammf}{\xspace{\emph{Frank-Tamm factor}}\xspace}

\newcommand{\geantt}{\xspace{Geant toolkit}\xspace}

\newcommand{\geantdrei}{\xspace{Geant3.16}\xspace}
\newcommand{\geantfour}{\xspace{Geant4}\xspace}
\newcommand{\geantfourt}{\xspace{Geant4 toolkit}\xspace}
\newcommand{\geantfours}{\xspace{Geant4 simulation}\xspace}
\newcommand{\RM}[1]{\MakeUppercase{\romannumeral #1}}

\graphicspath{{./Submit/}}

\begin{document}

\begin{frontmatter}
\title{{\bf Calculation of the \cherl yield from low energetic secondary particles accompanying high-energy muons in ice and water with \geantfour simulations}}

\author[rwth]{Leif~R\"adel}
\ead{Leif.Raedel@physik.rwth-aachen.de}

\author[rwth]{Christopher~Wiebusch\corref{cor1}}
\ead{Christopher.Wiebusch@physik.rwth-aachen.de}

\cortext[cor1]{Corresponding author}


\address[rwth]{III. Physikalisches Institut, RWTH Aachen University, Otto Blumenthalstrasse, 52074 Aachen, Germany}

\date{Version of \today}


\begin{abstract}

In this work we investigate and parameterize the amount and angular
distribution of Cherenkov photons, which are generated  
by low-energy secondary particles (typically $\lesssim 500 $\,MeV), 
which accompany a muon 
track in water or ice. These secondary particles originate from small energy
loss processes. We simulte these processes with Geant-4 and investigate the contributions of the different 
energy loss processes as a function of the muon energy and the  maximum 
transferred energy.
For the calculation of the angular distribution we have developed 
 a  generic 
transformation method, which allows us to derive the angular distribution of \cherphos
for an arbitrary  distribution of track directions and their velocities.

\end{abstract}

\end{frontmatter}

\section{Introduction}

Muons, which originate from charged current interactions of muon neutrinos 
are a primary 
detection channel for high-energy neutrino telescopes such as IceCube, 
Baikal or Antares \cite{ICECUBE,BAIKAL,ANTARES}.
These muons propagate with a speed $v$ close to the vacuum speed 
of light $c$
 through the detection medium, which is water or ice.
In these media the  refraction index is typically 
$n \approx 1.33$. The phase velocity of 
light is given
 by $ c_{med} = c/n $ \cite{PRICEGROUP}.  
Particles with a  speed $ v> c_{med}$ 
will emit optical \cherl,
which is detected by photo-detectors.

 The \cherl is emitted into a cone, which half 
opening angle $\theta_c$ is
given by 
\eqb \label{eq:cerangle}
\cos{(\theta_c)} = \frac{1}{n\beta}
\eqe
Here, $\beta = v/c$ is the Lorentz factor of  the particle \cite{PDG}. 
The number of emitted photons per unit track $x$ and wavelength interval
 is given by the \tamm formula \cite{PDG,TAMM}
\eqb \label{eq:tamm}
{d^2 N \over dx d\lambda } = {2 \pi \alpha z^2 \over \lambda^2 }\cdot 
sin^2 (\theta_c )
\eqe
For optical wavelengths, these photons can propagate through  
an optically transparent medium and can be detected by photo-detectors.
Note, that the  following discussion focuses on the injected light yield 
and propagation effects such as
absorption, scattering \cite{ICEPAPER} 
or chromatic dispersion \cite{KUZMICHEV,PRICEGROUP} are not considered.

A 
relativistic track ($\beta = 1$) in water or ice ($n \approx 1.33$)  produces
typically  $N_0 \approx 250 \, \mathrm{cm}^{-1} $ optical photons
in a wavelength interval between $300$\,nm and
 $500$\,nm, which is a typical sensitive region of photo-detectors used in 
 neutrino telescopes like IceCube \cite{ICPMT}.

The \chera for a relativistic track ($\beta =1 $) is  
$\theta_{c,0} = \arccos (1/n)  \approx 41^\circ $.
For smaller particle velocities the opening angle of the cone shrinks
according to equation \ref{eq:cerangle} and 
the number of photons decreases according to the factor $sin^2 (\theta_c )$
in equation \ref{eq:tamm}.

The \cherthr is given by $v \le c_{med} $, which is equivalent to
$ \beta \le \frac{1}{n} $ and $\theta_c =0$.
In terms of kinetic energy  
the \cherthr is
\eqb
E_{c} \ge m \cdot \lbra \frac{1}{\sqrt{1-{1\over n^2}}} -1\rbra
\eqe
In water and ice, the threshold is $E_{c,\mu } \approx 55 $\,MeV
for muons and  $E_{c,e^- } \approx 0.26$\,MeV for electrons.

\begin{figure}[htp]
\centering
\includegraphics*[width=.99\textwidth, page=1]{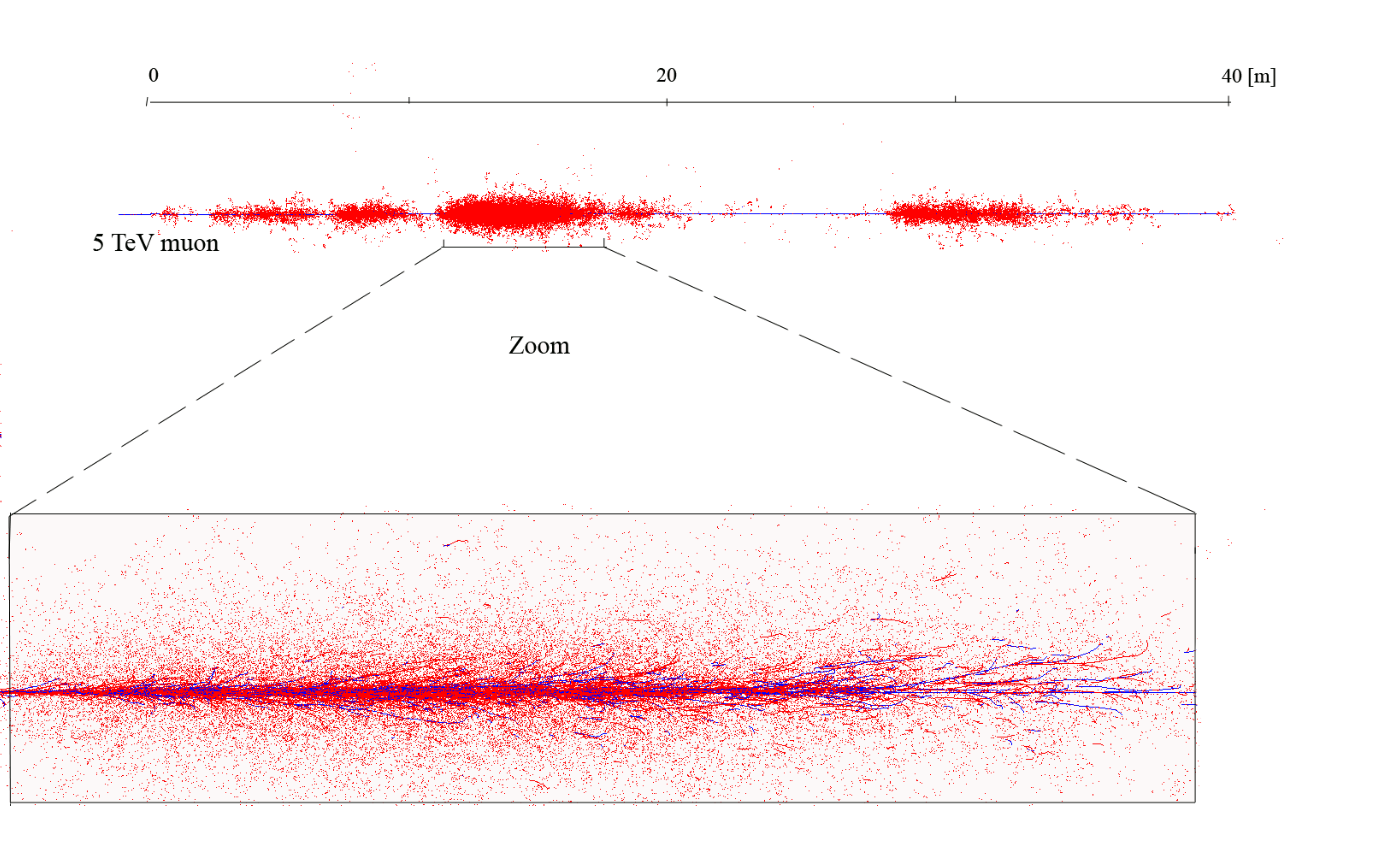}
\caption{Visualization of a 5 TeV
muon, $\mu^+$, which propagates a distance of $40\unit{m}$ 
through water (figure from \cite{LAIHEM}). 
Shown are all generated charged secondary particles (red for 
negative and blue for positive charge) as the result of a \geantfour 
simulation.
Neutral particles, like photons are not shown. 
No lower energy threshold has been applied to the \geantfours.
 \label{fig:intro:tpm}}
\end{figure}

During propagation through the medium muons suffer energy loss by
ionization of atoms and stochastic 
radiative processes such as
bremsstrahlung, pair production and photonuclear interaction \cite{PDG,MMC}. 
Charged secondary particles with  an energy above the 
 \cherthr will also emit \cherl and 
contribute to the total light yield. Secondary 
electrons and positrons,
for which the \cherthr is  relatively low 
 will considerably enhance the brightness of the muon track.
In this picture, a  muon, which propagates through water or ice can be 
considered to be accompanied by a glowing 
halo of secondary particles.
This is  illustrated in figure \ref{fig:intro:tpm}. 
Shown are all charged secondary particles.
Many of these particles, in particular those close to the 
muon track, have an energy sufficiently above the
\cherthr.

For high energies of TeV and above, 
radiative processes dominate the energy loss. 
These occur stochastically and 
can lead to a large,  \emph{catastrophic} energy loss. Therefore,
large local fluctuations 
of the density of secondary tracks occur along the path of the muon.
 This can also be seen in
 figure \ref{fig:intro:tpm}.

For neutrino telescopes such stochastic energy losses have to be considered
and can be  accurately simulated by muon propagation 
programs like \texttt{MMC} or MUM \cite{MMC,MUM}. 
Usually, because of computing limitations,  only  
catastrophic energy losses bigger than
 a predefined energy threshold $E_{max}$ are simulated stochastically 
whereas the large number of 
small  energy losses are approximated as continuous losses.
A typical value  of  $E_{max} \simeq 0.5$\,GeV has been established
in \cite{CHWPHD} and is used to date in IceCube 
as  a reasonable compromise between accuracy and
computing resources.

The amount of additional \cherl due to energy losses below 
$E_{max} $ has been parameterized in \cite{CHWPHD}.
The  relative 
amount of  Cherenkov radiating track length $l_{sec} $ from secondary processes 
 per unit muon track $l_{\mu} $ was found as
\eqb \label{eq:cwresult}
\frac{l_{sec}}{l_{\mu}} = 0.172 + 0.032 \cdot \ln \lbra E/\mathrm{GeV} \rbra
\eqe
for $E_{max} =0.5$\,GeV for $e^+$, $e^-$ and  $E_{max} =1$\,GeV for $\gamma$. 
This  calculation has been done with an older version
of  the \geantt (3.16,  \cite{GEANT316}) and the reduction in light yield due to particles
with $\beta < 1$ was neglected.

In this paper these calculations are repeated and extended\footnote{An initial attempt of this work has been  started in \cite{HAKAN}.}
  using the \geantfourt \cite{GEANT4}.
With today's increased computing power, the accuracy of the simulations
can be improved substantially.
Muons are simulated with higher statistics and the 
\tammf $\sin^2(\theta_c) $ in equation \ref{eq:tamm} is  
used to calculate the correct light yield. The light yield is investigated 
as a function of the muon energy $E_\mu $ and the chosen $E_{max} $.
Furthermore, the angular distribution of the additional \cherphos is
calculated and parameterized.

For the calculation of the angular distribution 
an improved transformation method is used, which was originally 
developed in \cite{CHWPHD}. It is described in  \ref{sec:trafo}.
The advantage of this transformation method is that \cherphos do not need to be generated and propagated
 during the \geantfours. 


\section{The Monte Carlo method \label{sec:mcframe}}

The \geantfour (GEometry ANd Tracking) toolkit is designed for the Monte Carlo simulation of particle-interactions
with matter in high-energy physics \cite{GEANT4}. 
It is object-oriented and programmed in the C++
language in contrast to its predecessors \cite{GEANT316}, which were 
programmed in FORTRAN-77.

\begin{figure}[htp]
    \centering 
	\includegraphics*[width=.79\textwidth]{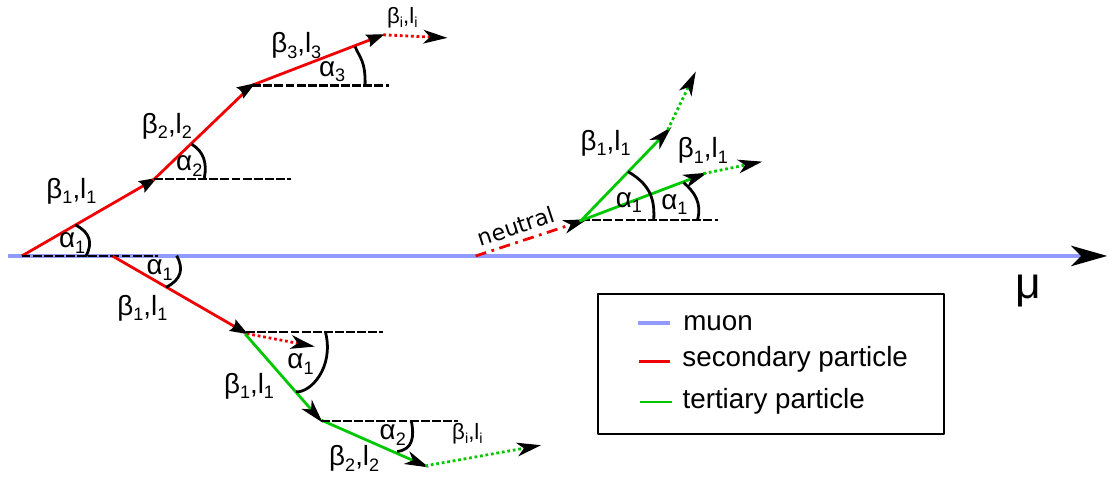}
	\caption{Basic principle of the simulation
 \label{fig:intro:principle}}
\end{figure}

The  simulation principle of this work is illustrated in figure \ref{fig:intro:principle}. For the calculation of the yield of photons, the muon is propagated through the simulation medium and secondary particles are created, which again produce further particles.
For each secondary track segment $i$ we store the length $l_i$, the Lorentz factor $\beta_i $ and the inclination angle $\alpha_i$ of the direction with respect to the muon direction.
In the following $\alpha_i$ is called zenith angle.
Summing over all track segments allows to calculate the Cherenkov photon yield and the corresponding angular distribution.

\begin{figure}[htp]
    \centering
	\includegraphics*[width=.49\textwidth]{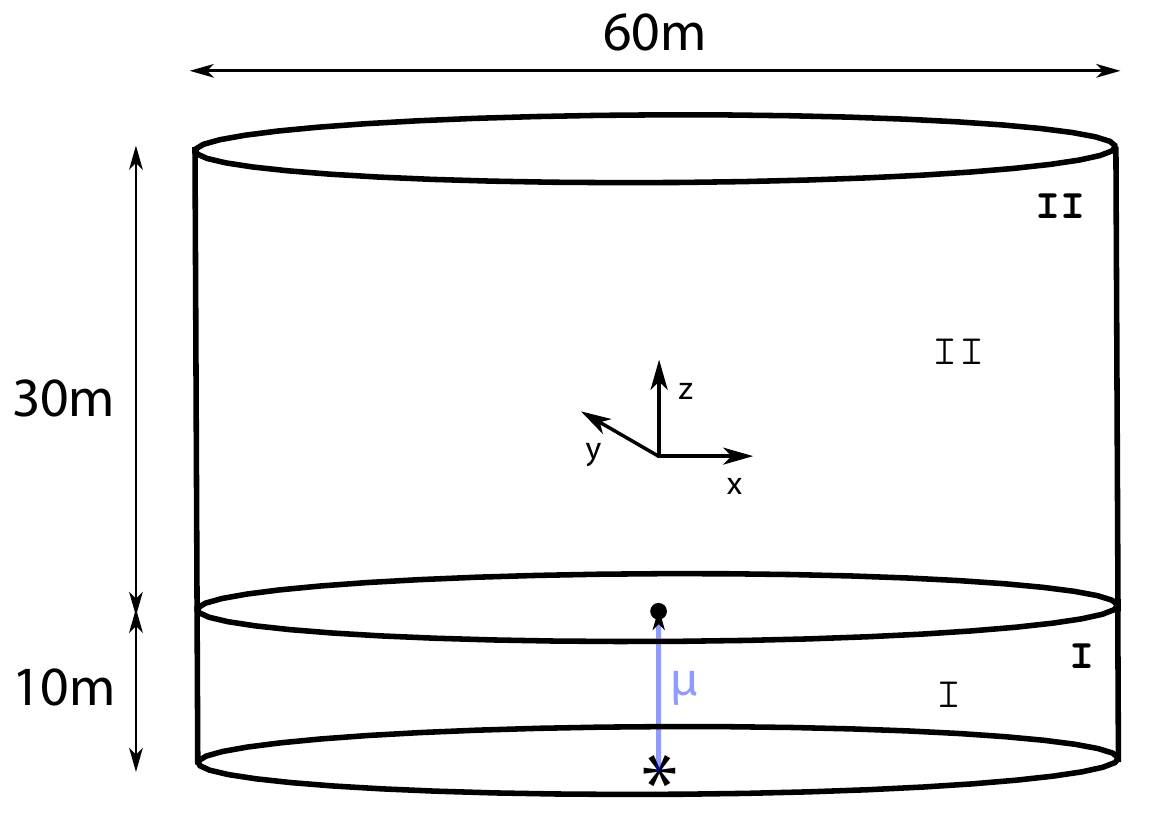}	
\caption{Geometrical layout of the simulation
 \label{fig:intro:geometry}}
\end{figure}

For the simulations we have defined a volume,
which is filled with water or ice.  The geometry is  a cylinder with a total length
of $40\unit{m}$ and a radius of $30\unit{m}$, as shown in figure \ref{fig:intro:geometry}. The dimensions
are chosen such that all secondary particles are well confined  within the 
geometry and  fully tracked. The cylinder is logically
divided into two cylindrical sub-volumes \texttt{\RM{1}} and \texttt{\RM{2}}. The first sub-volume has a length of $10\unit{m}$.
 The used media properties are given in  \ref{app:geant:conf}.

The simulated muon is injected from the bottom center into this volume
with its initial momentum pointing into positive z-direction. It is killed once it reaches the end of this first sub-volume, after having propagated  $10\unit{m}$.
A propagation length of $10\unit{m}$ is chosen as a compromise to optimize
simulation speed between  an approximately constant 
muon energy along the muon's track and to allow for a higher probability of rare processes per event.
The average energy loss of the muon
is small compared to the here considered muon energies, substantially 
above $100$\,GeV and
therefore the muon energy can be considered approximately constant.

However, if only 
the first sub-volume would be considered the total additional track length from secondary particles would be underestimated, because the injected
the muon is not accompanied by secondary particles from upstream processes. 
In order to compensate for  this effect,  secondary 
particles are, unlike the muon, allowed to enter the second 
sub-volume and are fully considered in the simulation. 
By this method the missing ``\emph{spill-in}'' 
secondaries are compensated by additional ``\emph{spill-out}'' 
secondaries.

For the here described simulation it is important to simulate all 
particles with energies above the 
\cherthr. Details on the simulated  physics processes are given in  \ref{app:geant:conf}.
 In \geantfour some electromagnetic processes require production thresholds to avoid 
infrared divergences \cite{GEANT4UserGuideApplication}. 
These production 
thresholds are specified as a cut-in-range threshold, using the 
\texttt{SetCuts()} method of \texttt{G4VUserPhysicsList}.  
Here,  particles are
 tracked if their mean expected range is larger than this cut-in-range threshold.
For each material and particle type, this  cut-in-range is 
transformed into a corresponding energy threshold.
Here, a cut-in-range of $100\,\mu\mathrm{m}$ is chosen. 
This corresponds to a kinetic energy threshold
of $E_{\mathrm{cut},e^{\pm}} \approx 80\unit{keV}$ for electrons, 
which is well below the \cherthr 
$E_{\mathrm{c},e^{\pm}}\approx 264\unit{keV}$. 
Once produced, all  secondary particles are tracked until 
they stop. 

In order not to falsely include catastrophic energy losses an upper energy threshold $E_{max} $ is implemented. Technically in the implementation,   secondary particles are only created, if they correspond to a process by which the 
muon loses less than $E_{max} $ in energy.
In this work we use the default value  $ E_{max} = 0.5$\,GeV but also investigate
the effect of  different values.

Electro-magnetic   cross-sections in Geant4 are certified up to a maximum muon energy of
$10$\,TeV. In order to extend the  simulations
beyond this limit, the cross-sections are extrapolated by \geantfour up to $100$\,TeV
(see  \ref{app:geant:conf}).

Special care was taken about multiple scattering. In order to avoid 
unphysical biases in the resulting  angular
distributions, we simulated each single scattering process individually
(see  \ref{app:geant:conf}).

For $\beta=1 $, the number of emitted \cherphos is proportional to the length
of the track and can be calculated using
 equation \ref{eq:tamm}. For  $\beta<1$ the yield is smaller and proportional
 to the factor
\eqb
\sin^2 (\theta_c ) = 1- \cos^2 (\theta_c ) = 1-\frac{1}{\beta^2 \cdot n^2} 
\eqe
In order to properly account for the smaller yield, the length of each track segment $l$ can be scaled with the \tammf
\eqb \label{eq:tamm:factor}
\hat{l} = \frac{\sin^2 (\theta_c )}{sin^2(\theta_{c,0}) } \cdot l
\quad \mbox{with} \quad
\sin^2 ( \theta_{c,0} ) = 1-\frac{1}{n^2} 
\eqe
The value $\hat{l} $ thus corresponds to the equivalent length of a 
relativistic track with the same photon yield as the track length $l$.

The use of the equivalent length $\hat{l} $ instead of an explicit 
calculation of photons has the advantage that the here presented
results can be rescaled 
 to slightly different indices of refraction, and are independent on the assumed wavelength interval.

\section{\cherpho yield from secondary tracks}

\subsection{The maximum energy threshold \label{sec:emax}}

An important parameter for this study is the energy $E_{max} $ above which processes are considered \emph{catastrophic}.
Low-energy processes below $E_{max} $ occur rather frequently and can be 
considered quasi-continuous for the typical scales and spacing of 
sensors in large neutrino telescopes.
As a first step, we investigate  the occurrence of these 
 small energy losses as a function of $E_{max}$.

\begin{figure}[htp]
    \subfloat[][]{
        \includegraphics[width=.49\textwidth]{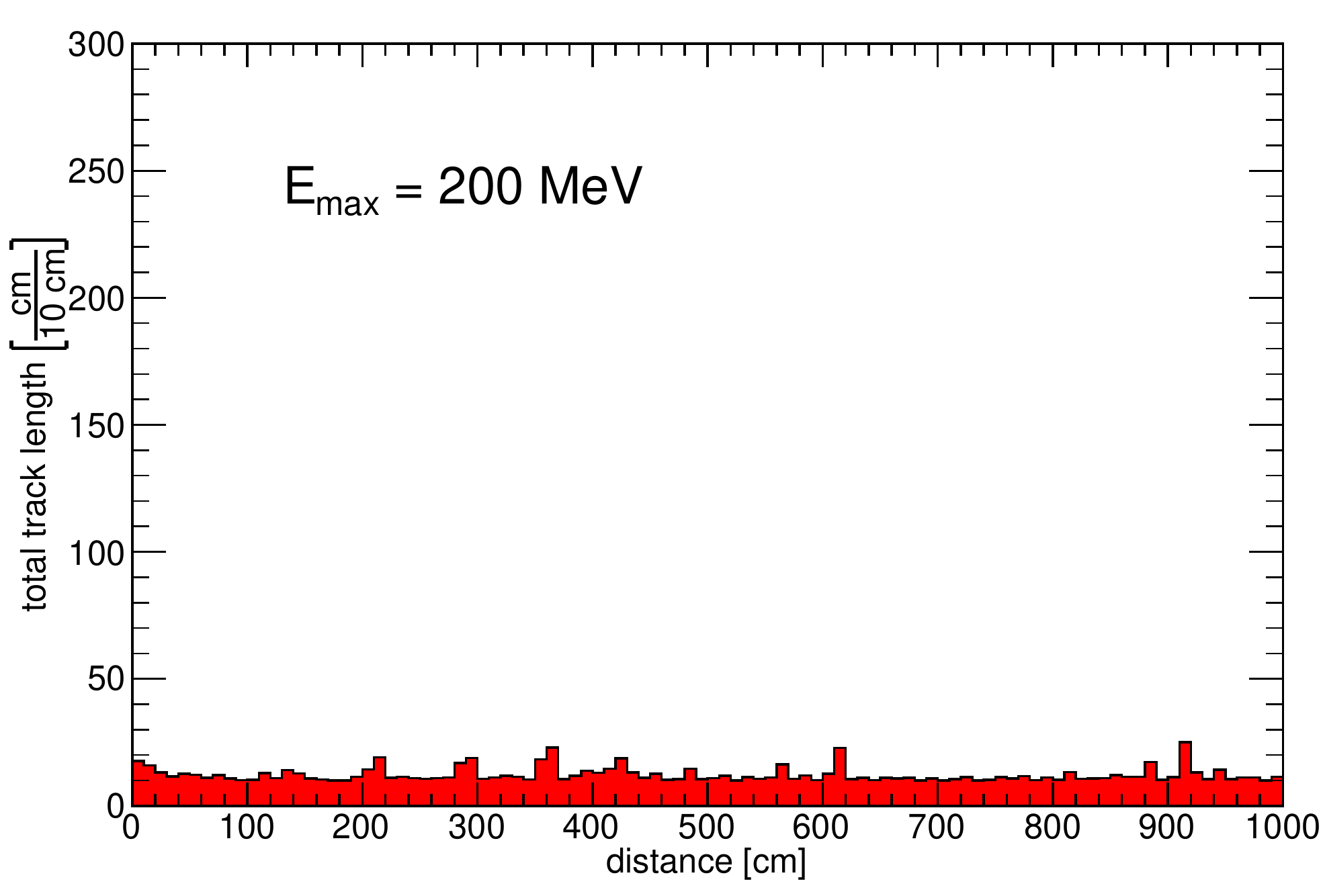}
    } \hfill
    \subfloat[][]{
        \includegraphics[width=.49\textwidth]{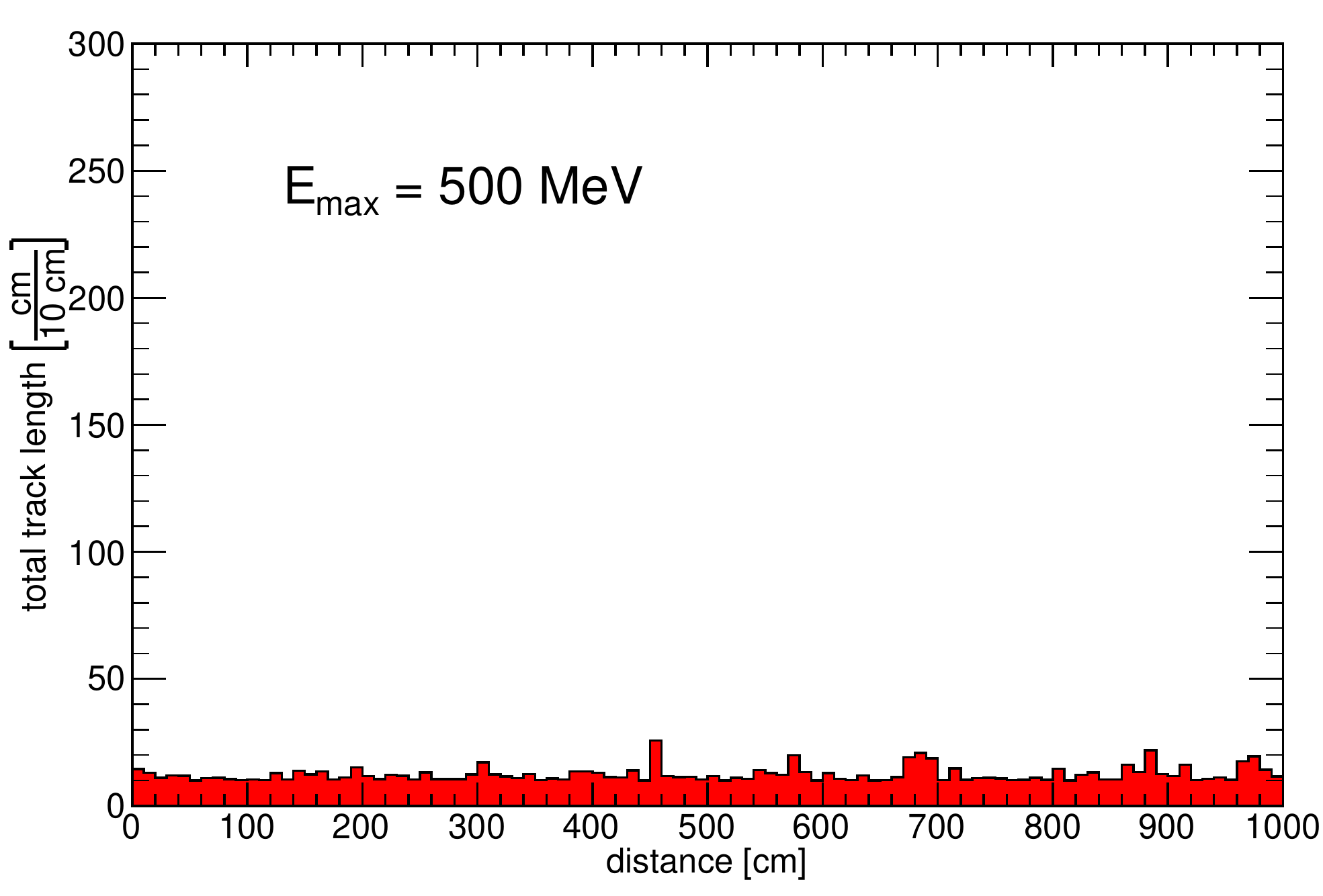}
    }\newline
    \subfloat[][]{
        \includegraphics[width=.49\textwidth]{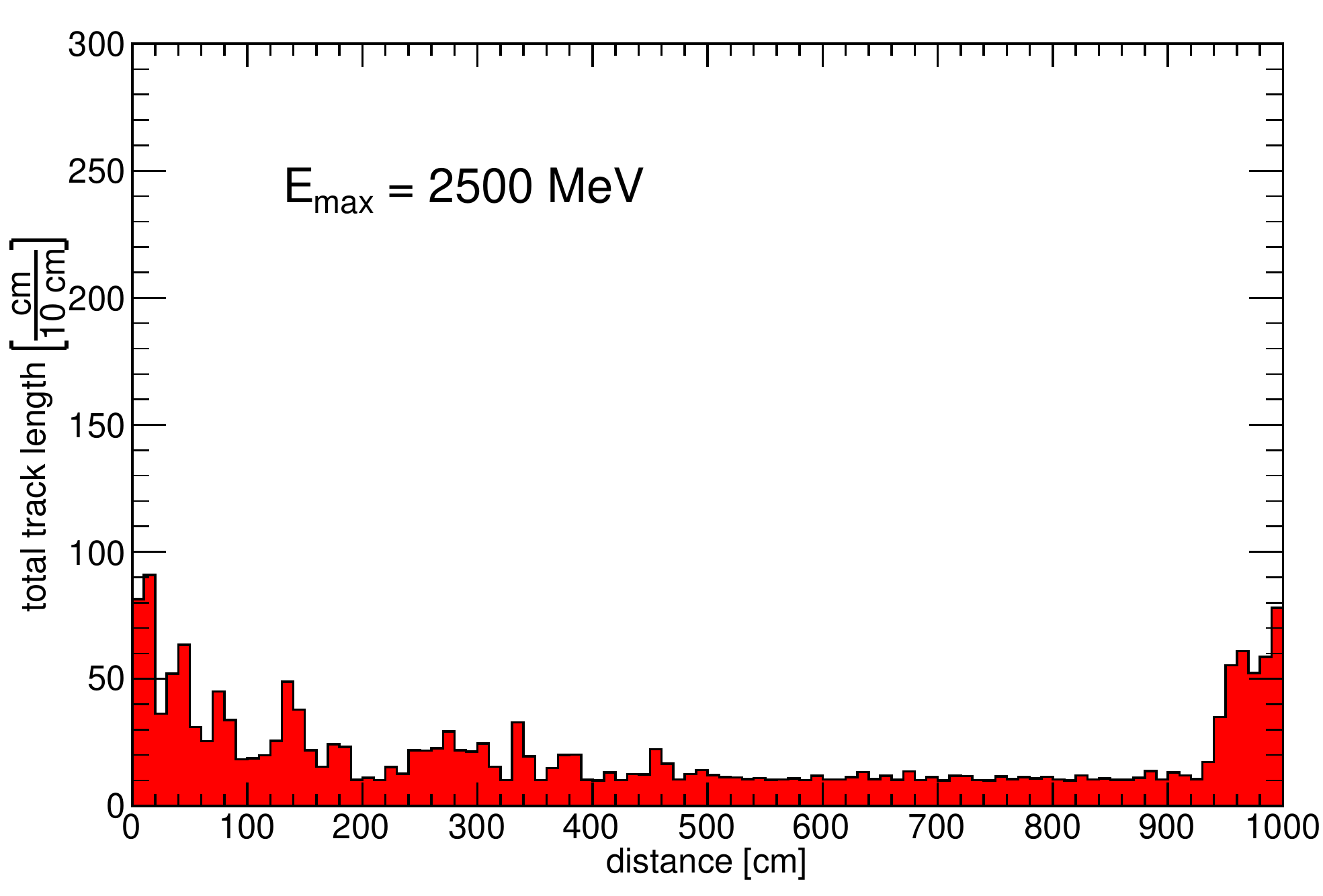}
    } \hfill
    \subfloat[][]{
        \includegraphics[width=.49\textwidth]{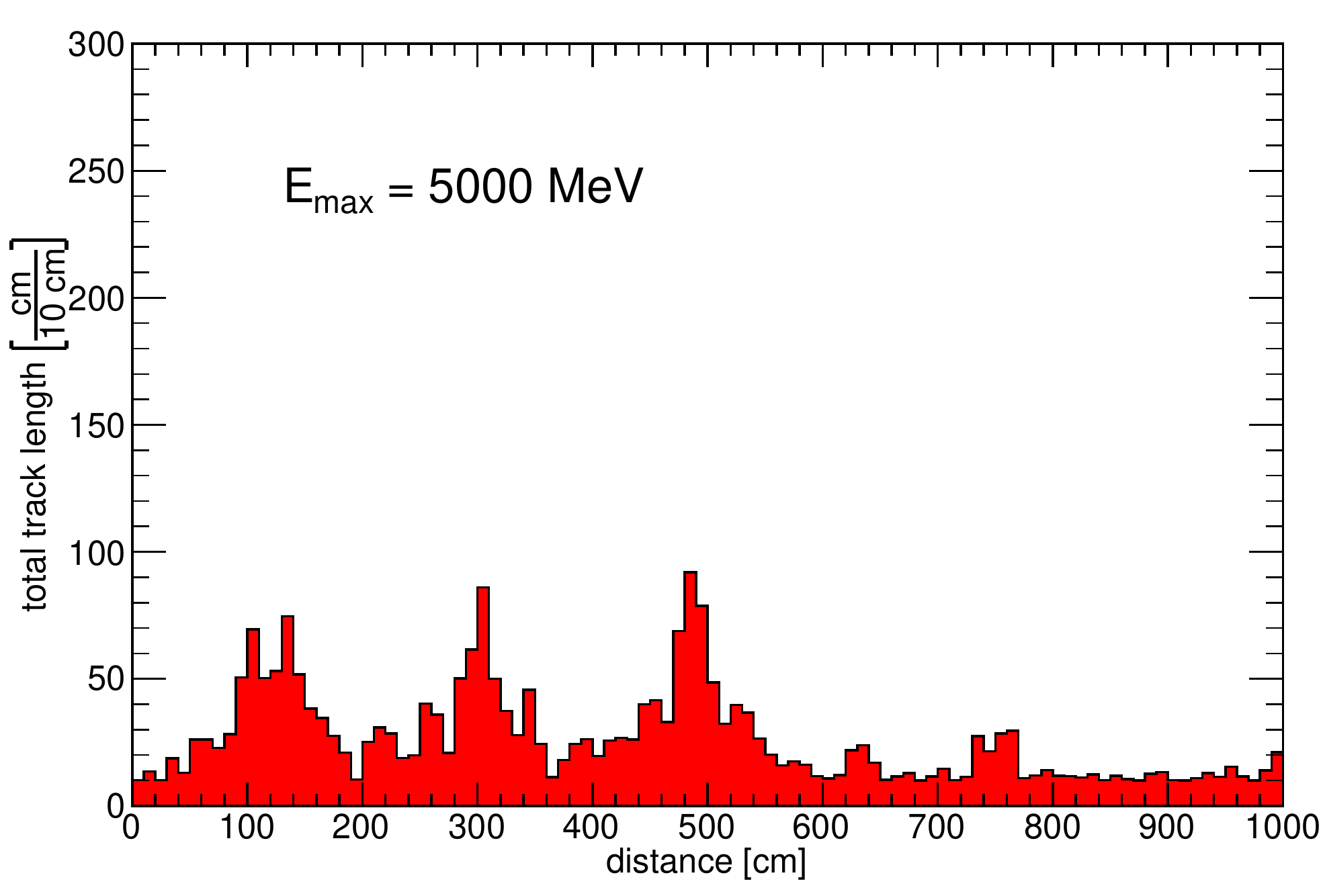}
    }

	\caption{Examples for the occurrence of charged particles tracks
along a $5$\,TeV  muon track, which is propagated $10$\,m 
through water. Shown is the track length summed over all charged tracks 
along the direction of the muon with a bin size of $10$\,cm. 
In addition to the $10$\,cm muon track in each bin,  all 
 secondary tracks have been counted if they originate from an energy loss
process below $E_{max}$ and if their energy is 
above the \cherthr. From left to right four 
individually simulated muons with rising $E_{max} $ ($ 0.2 $\,GeV, $ 0.5 $\, GeV,  $ 2.5 $\,GeV,  $ 5 $\,GeV) are shown.
 \label{fig:intro:emax}}
\end{figure}

The effect of  $E_{max} $ is illustrated in figure \ref{fig:intro:emax}.
It  shows for four randomly chosen muon tracks 
the occurrence of  secondary particle tracks  
along the path for  those particles, which originate from energy losses below
$E_{max}$. It is obvious that the stochastic nature
of catastrophic losses increases with rising $E_{max} $.
For  $E_{max} = 0.2$\,GeV rarely bins with large individual 
losses, e.g. with more than
a factor $2$ of the muon light yield occur. These losses seem to be 
largely uncorrelated and distributed randomly and almost continuously
 along the track.
However, for  $E_{max} = 2.5$\,GeV  and   $E_{max} = 5$\,GeV 
large local fluctuations are observed and 
 the muon light yield can be exceeded by a factor $10$ or 
more, for several subsequent bins.

\begin{figure}[htp]

    \subfloat[][]{
        \includegraphics[width=.49\textwidth]{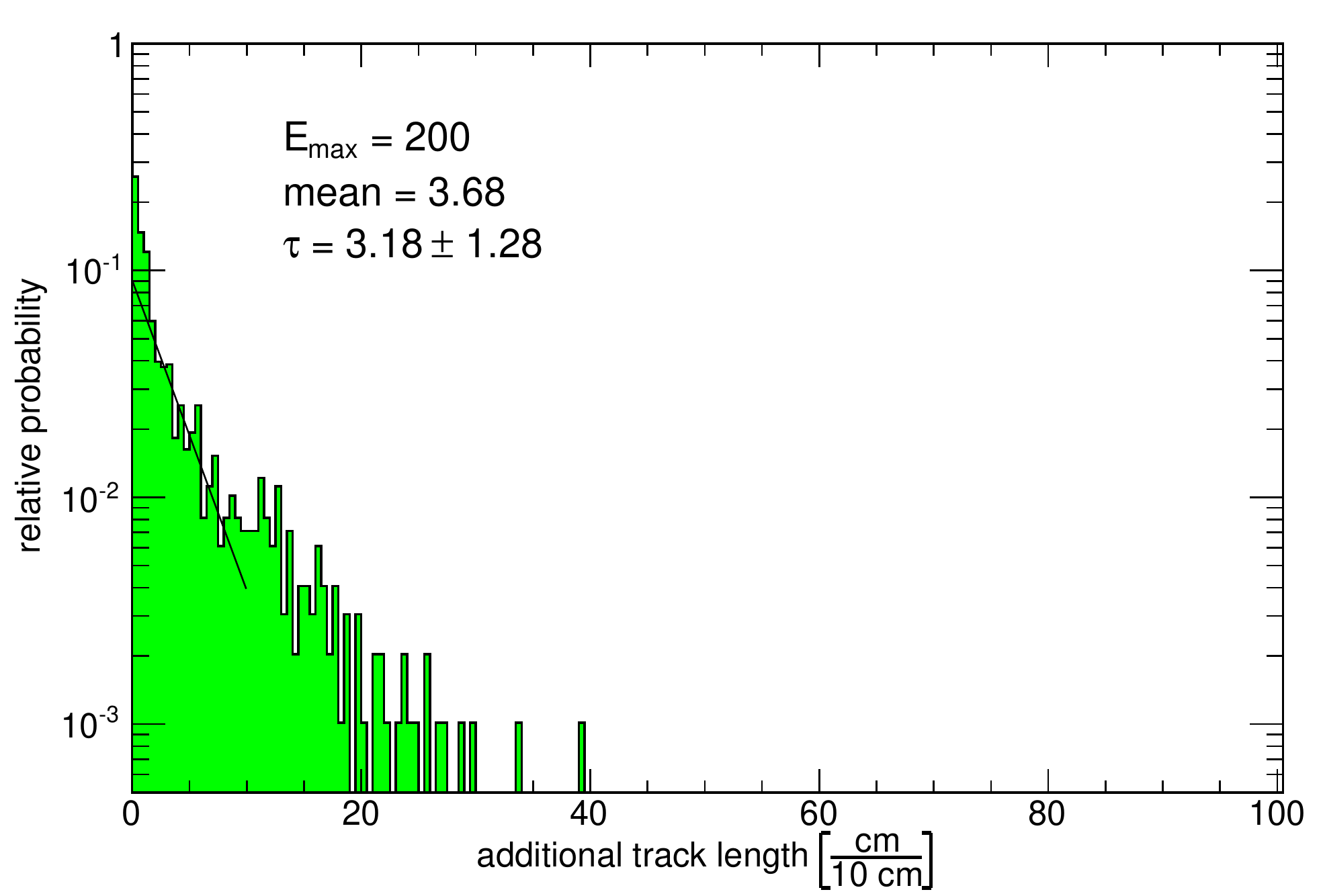}
    } \hfill
    \subfloat[][]{
        \includegraphics[width=.49\textwidth]{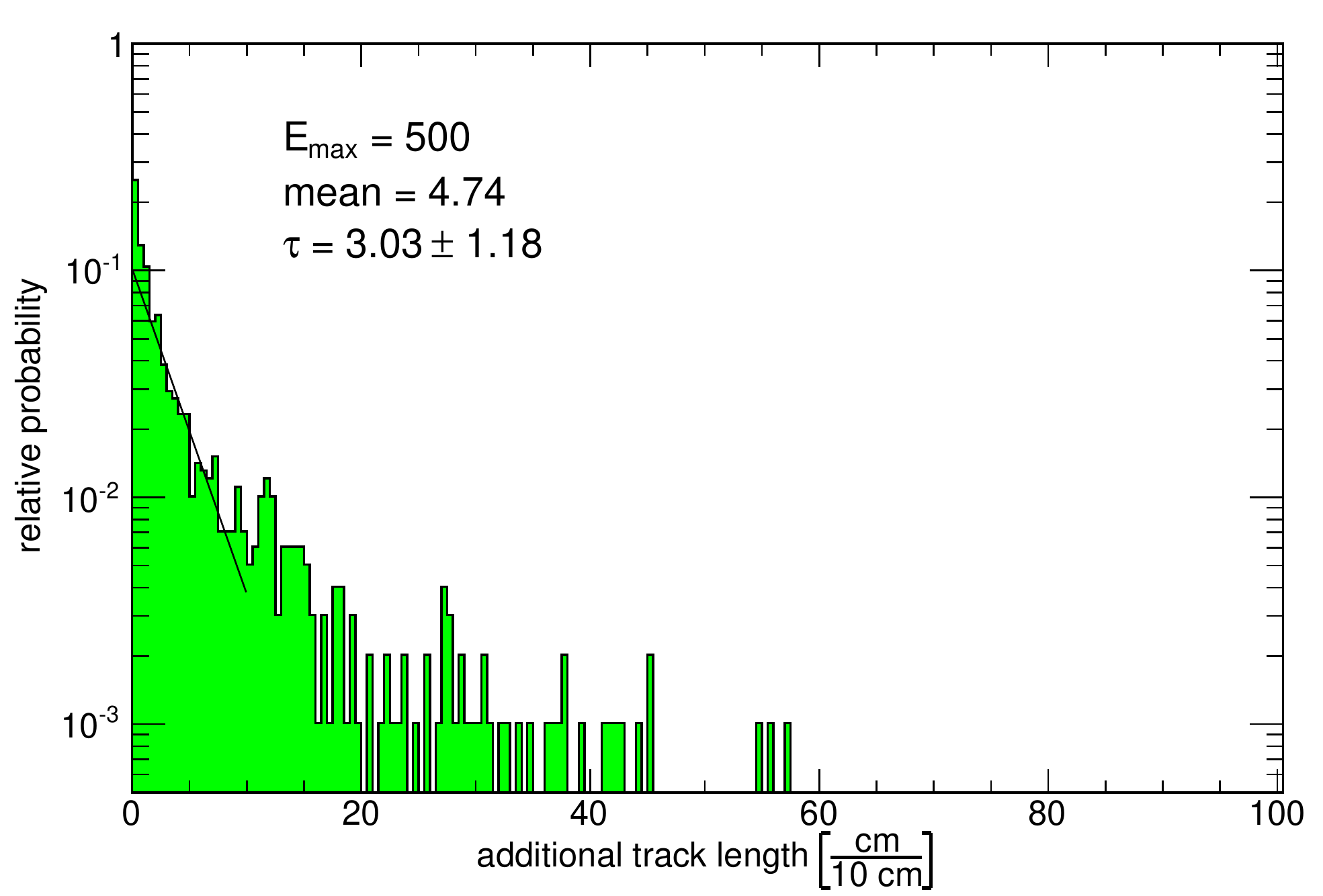}
    }\\
    \subfloat[][]{
        \includegraphics[width=.49\textwidth]{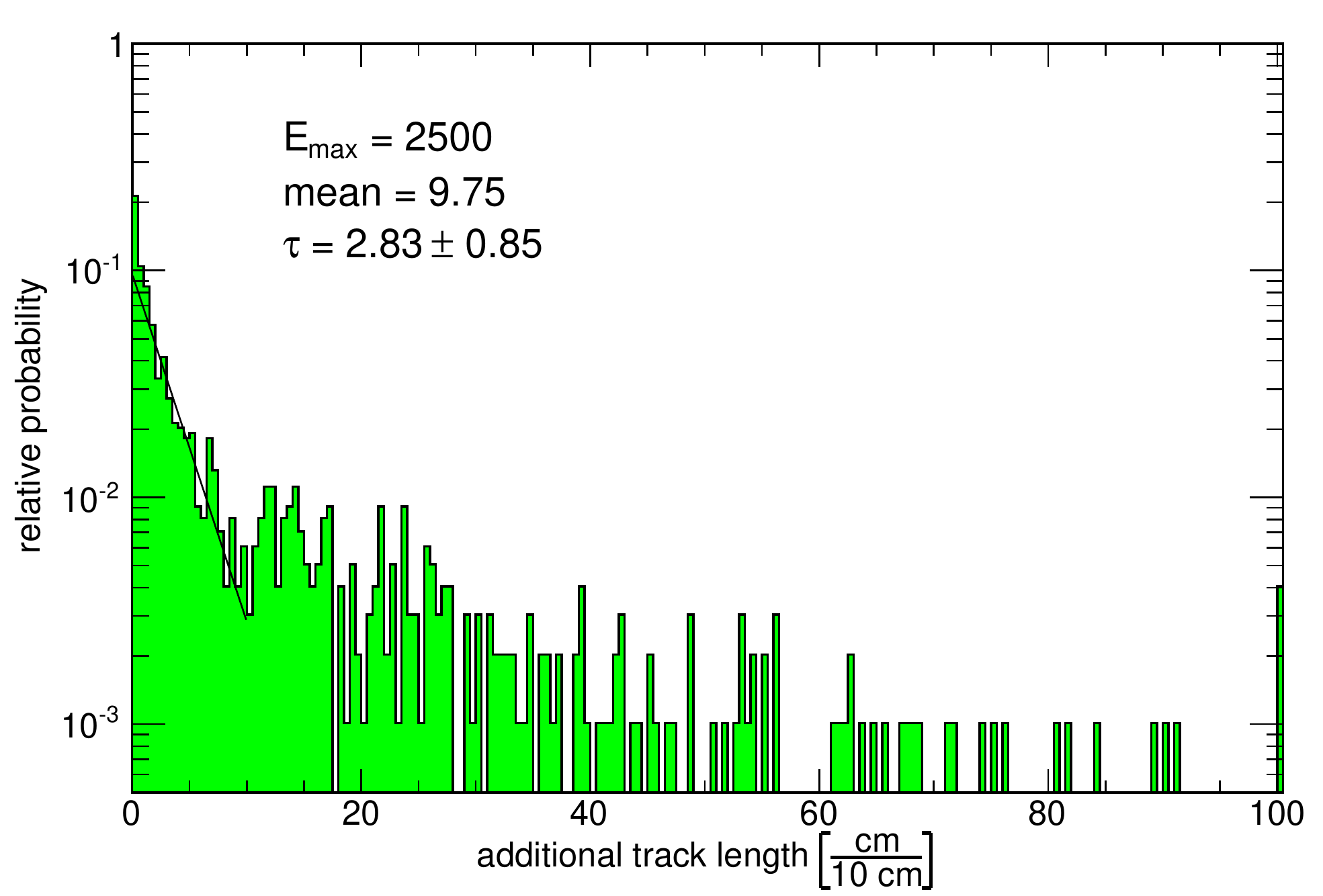}
    } \hfill
    \subfloat[][]{
        \includegraphics[width=.49\textwidth]{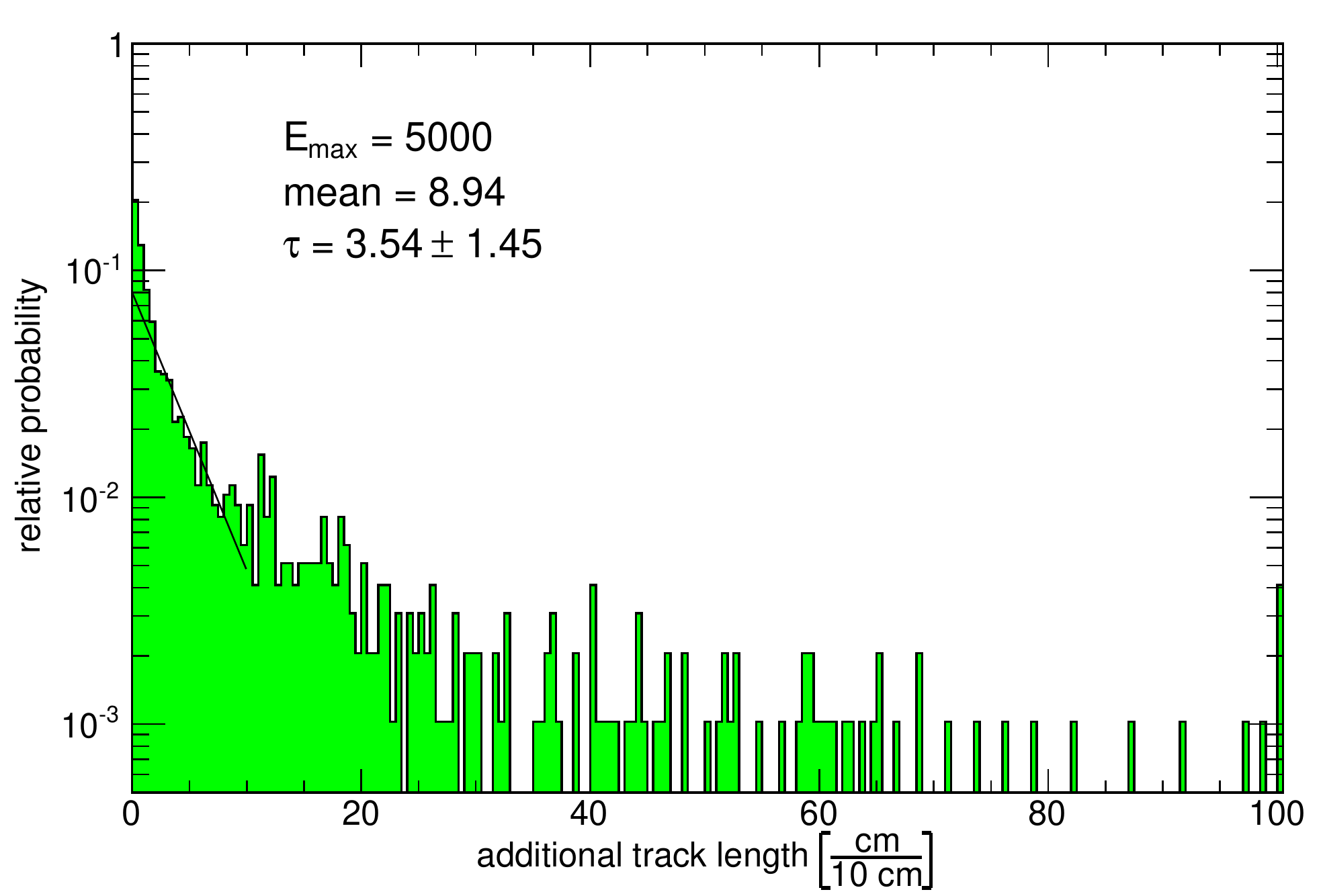}
    }

	\caption{Probability distribution of the occurrence of 
secondary track lengths for $10\unit{cm}$ bins.  \label{fig:intro:histoemax} 
The values $l_{add} $ exclude  the muon track length.
Each $10\unit{cm}$ bin 
in figure \ref{fig:intro:emax} has been histogramed for a total of 
$10$ simulated muons. The figures show the
 distributions for different  $E_{max} = 0.2\,\mathrm{GeV},  0.5\,\mathrm{GeV},  2.5\,\mathrm{GeV},  5.0\,\mathrm{GeV}$. 
The rightmost bin is the overflow bin. 
An exponential function $f(l_{add} ) = \mathrm{const} \cdot \exp{(- l_{add} /\tau)}$ is fitted to the data from $0\unit{cm}$ to $10\unit{cm}$.}
\end{figure}

The distributions of the size of these fluctuations are shown in 
figure \ref{fig:intro:histoemax}.
 For small  $E_{max} \le 0.5\unit{GeV}$, the deviation of the tails from an exponential distribution become smaller.
This  indicates
 that the occurrence of  additional track lengths from low-energy  processes 
can be approximated  by a Poissonian  process
for step sizes of the order of $10$\,cm. 
The typical value $\tau \approx 3$\,cm corresponds to an additional yield of $\sim 30$\% more \cherphos relative to the bare muon track.

For larger step-length  $l\gg 10$\,cm, e.g.  of the 
order of meters, the exponentially distributed additional track length can be approximated continously by a Gaussian distribution.
  According to the central limit theorem, the mean and variance  are then
\eqb
\langle l_{add} \rangle = \frac{l_\mu}{10\mathrm{cm}} \cdot \tau 
\quad \mbox{and} \quad
\sigma_{ l_{add}} =  \sqrt{\frac{l_\mu}{10\mathrm{cm}}} \cdot \tau 
\eqe

For  $E_{max} \ge 0.5\unit{GeV}$  the distributions of track lengths show
an increasingly non-exponential tail to large $l_{add} $ and
large local fluctuations
appear. The fitted $\tau$ remains reasonably constant for different cut-off energies,
but the difference between the mean value of ${l}_{add}$ and the fitted $\tau$ increases with the cut-off energy.
Therefore, the occurrence of secondary tracks above $E_{max} \gtrsim 0.5$\,GeV
cannot  be considered  Poissonian and quasi-continuously. 
In the following the value   $E_{max} =0.5$\,GeV is used as default, 
unless noted otherwise.

\subsection{Total secondary track  yield \label{sec:tracklength}}

The total amount of additional \cherphos depends (see eq.\ref{eq:tamm}) 
on the Lorentz factor $\beta$ of each secondary 
track.

 A large number of muons is propagated as described in section \ref{sec:mcframe} for $10 $\,m through ice and the track length $l_{add}$, 
additional to the muon, is 
recorded and investigated as a function of $\beta $.

\begin{figure}[htp]

    \subfloat[][]{
        \includegraphics[width=.49\textwidth]{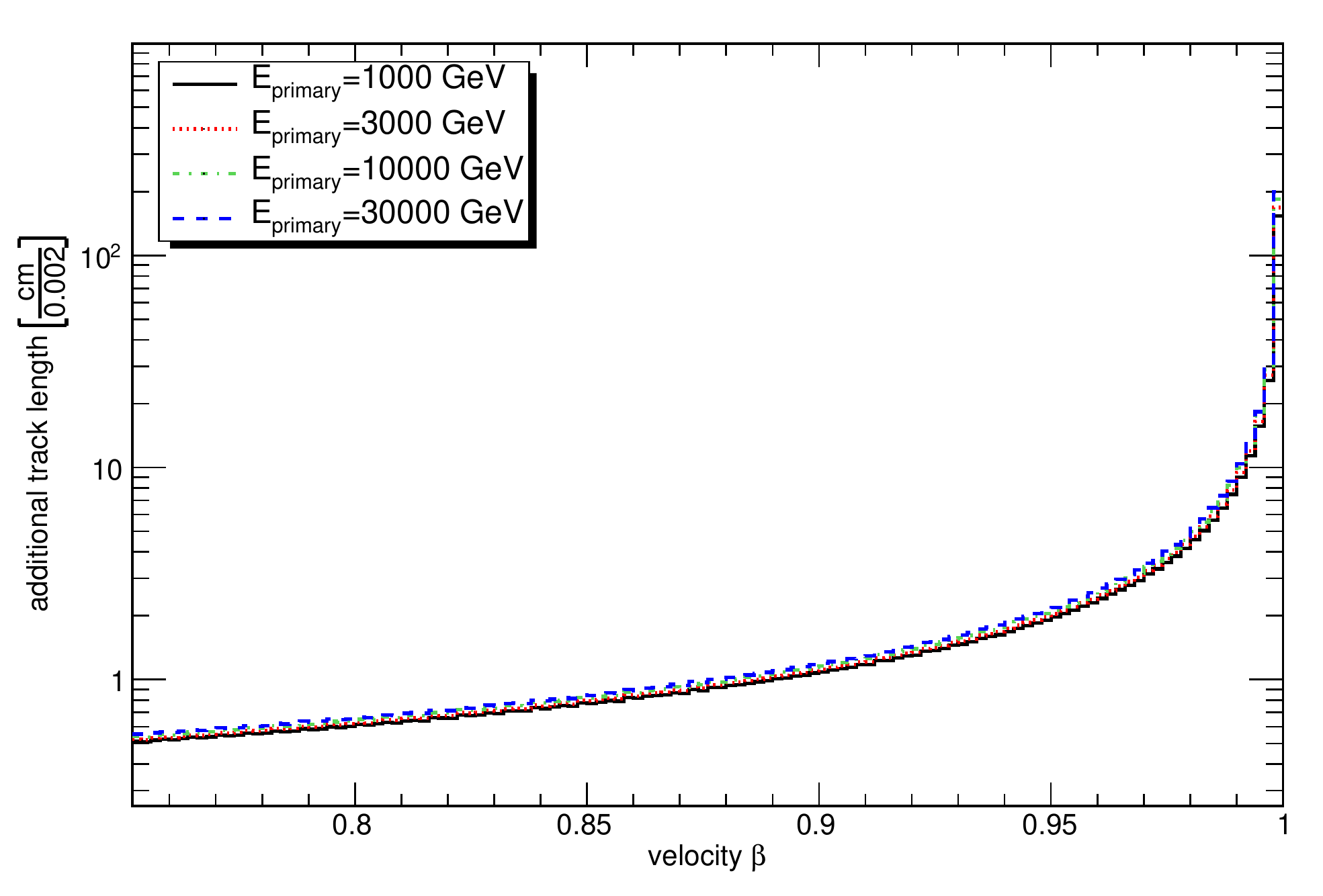}
    } \hfill
    \subfloat[][]{
        \includegraphics[width=.49\textwidth]{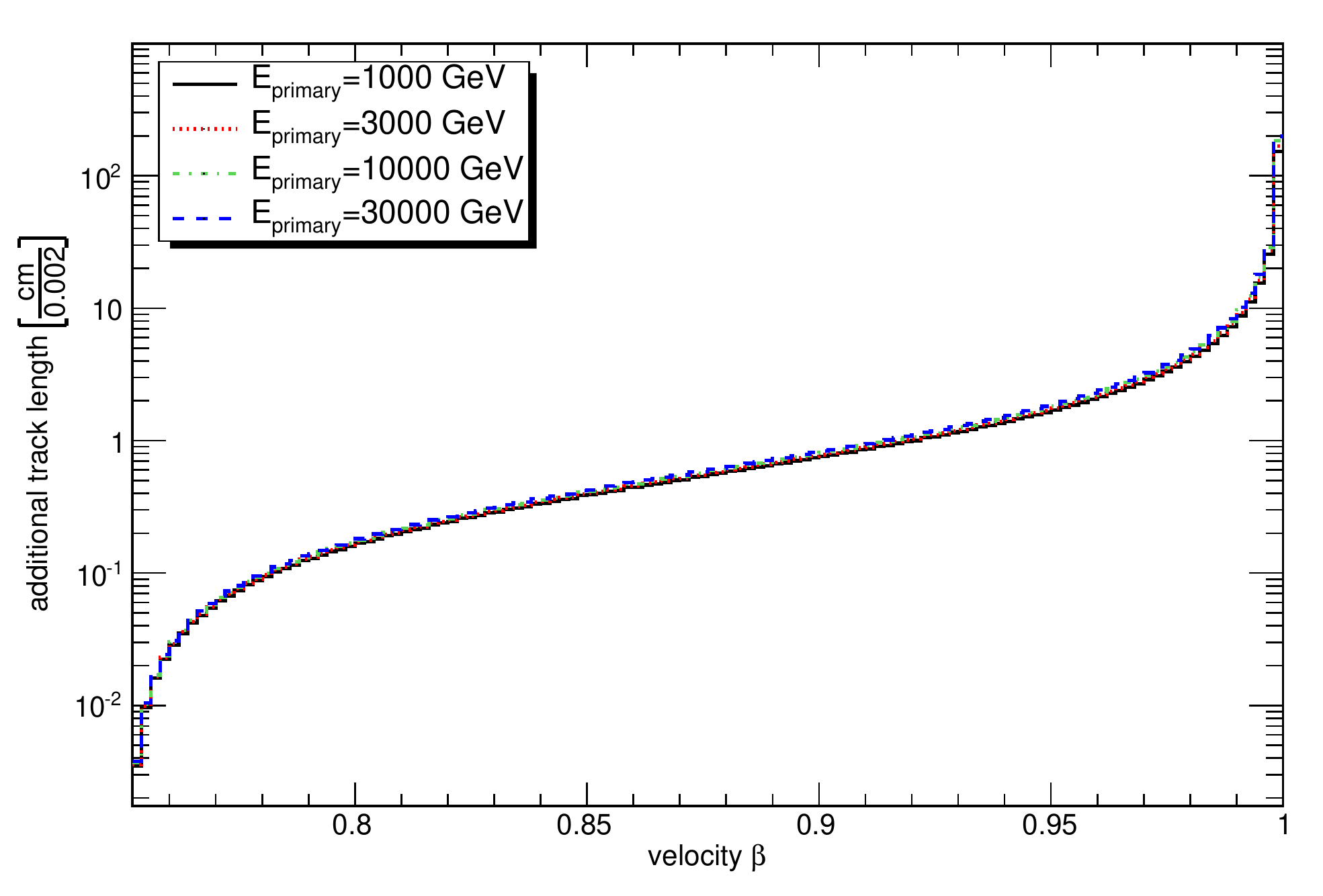}
    }\\
    \subfloat[][]{
        \includegraphics[width=.49\textwidth]{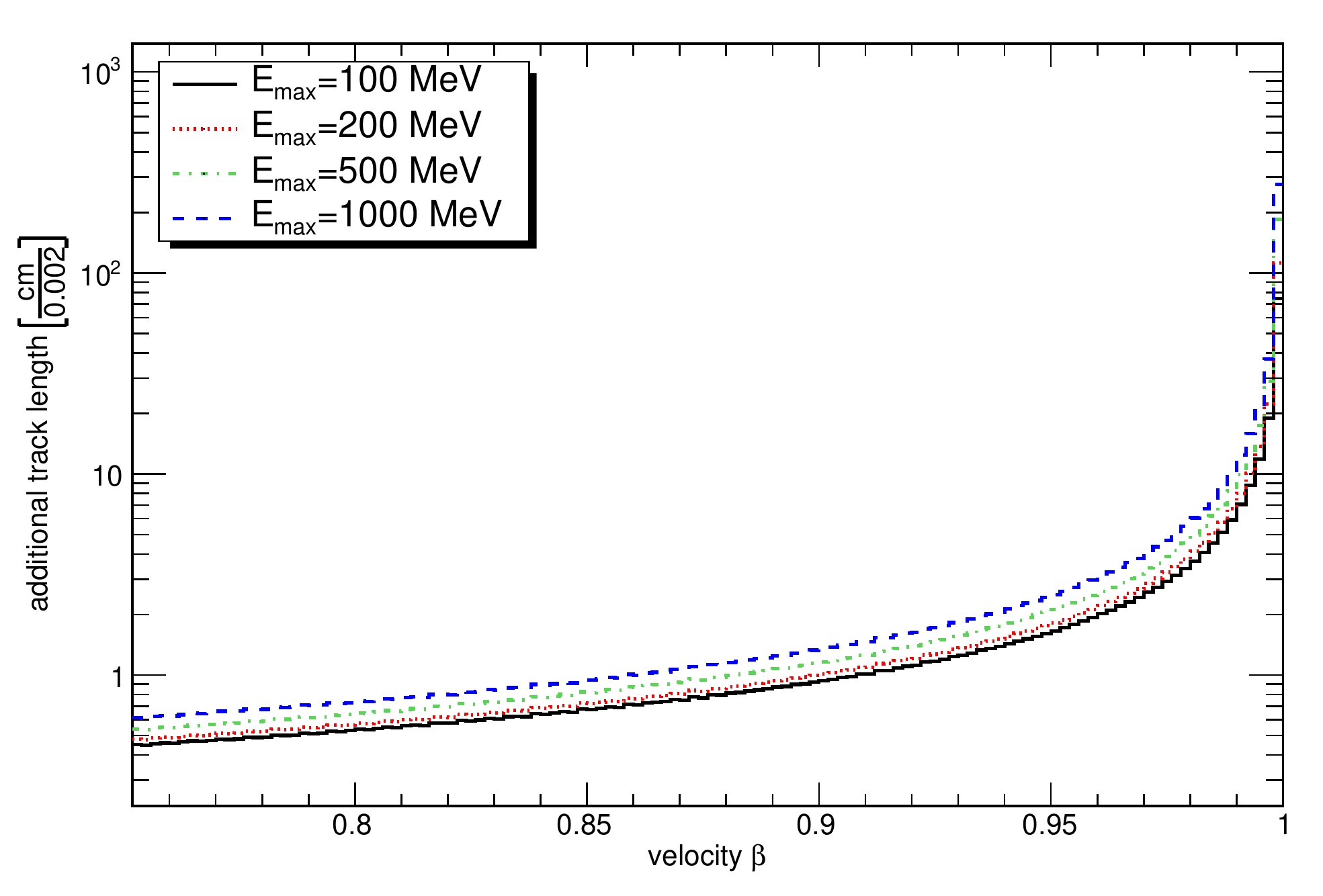}
    }\hfill
    \subfloat[][]{
        \includegraphics[width=.49\textwidth]{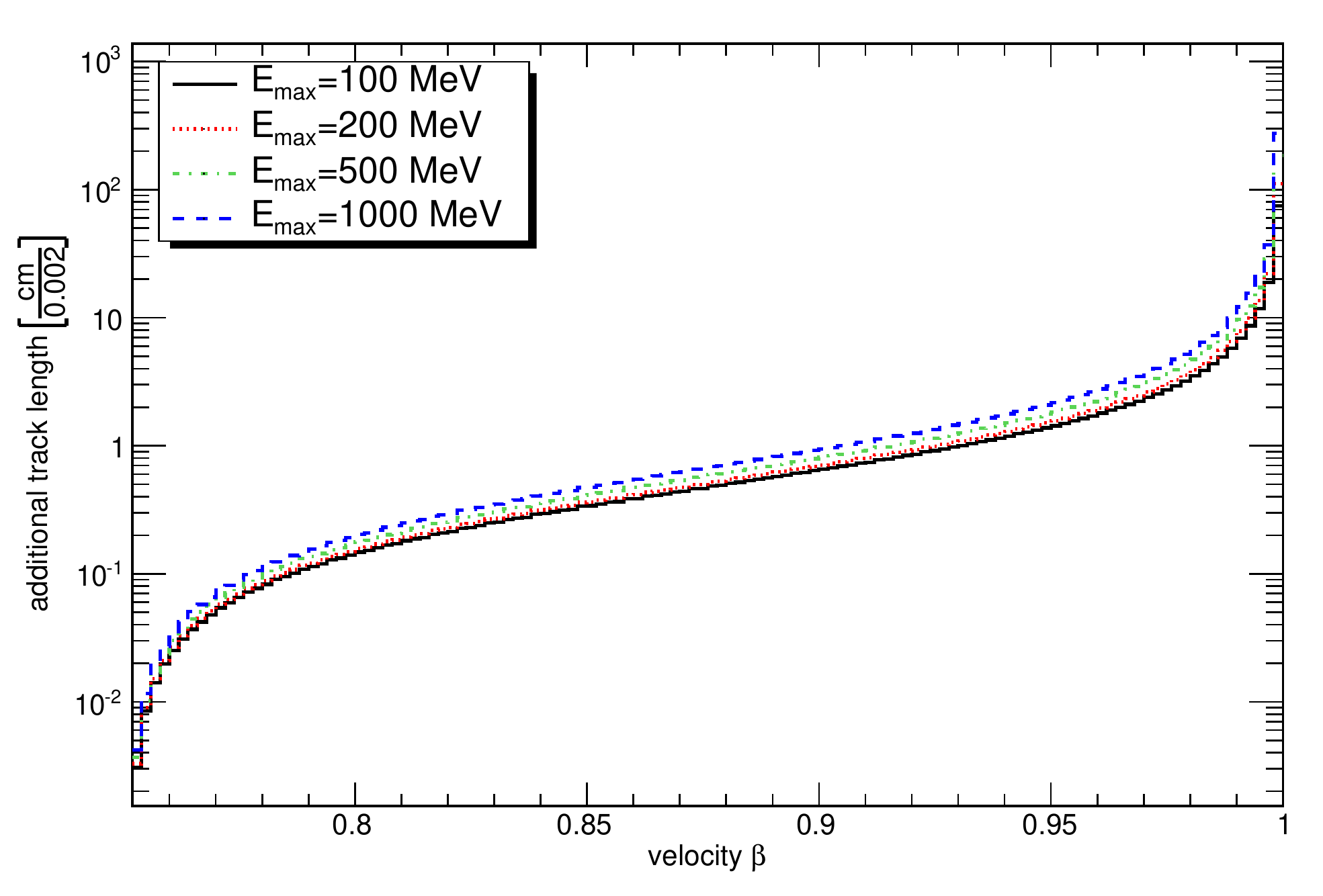}
    }
	\caption{Velocity distribution of secondary track lengths.
Shown is the additional track length $l_{add} $ per $10$\,m muon track
versus the Lorentz factor  $\beta$ for bins of $0.002$ in 
$\beta $. 
The top figures a) and b) show the distributions for 
different muon energies with $E_{max}=0.5\unit{GeV} $ and the bottom figures c) and d) the distributions 
for different $E_{max} $ for a $10\unit{TeV}$ muon.
In the left figures each track segment enters with its physical length, whereas 
 in the figures on the right side
 each length  is  weighted with the \tammf
(eq. \ref{eq:tamm:factor}) and the distribution corresponds to $\hat{l}_{add} $, 
which is proportional to the effective \cherpho yield.
 \label{fig:track:beta:tamm}}
\end{figure}

Figure \ref{fig:track:beta:tamm} shows the distributions $\Delta l_{add}/\Delta\beta $
for different $E_\mu $ and $E_{max} $. 
Most of the secondary track-segments have a large velocity $\beta \simeq 1$.
Interestingly, the shape of the distributions are remarkably invariant for different  $E_\mu $ and $E_{max} $
and only the total $l_{add} $ changes.  
The application of the \tammf  (eq.\ref{eq:tamm:factor}) shows that the 
majority of the additional \cherphos is strongly dominated 
by secondary track-segments with $\beta \ge 0.99 $.

\begin{figure}[htp]
    \centering
	\includegraphics*[width=.69\textwidth]{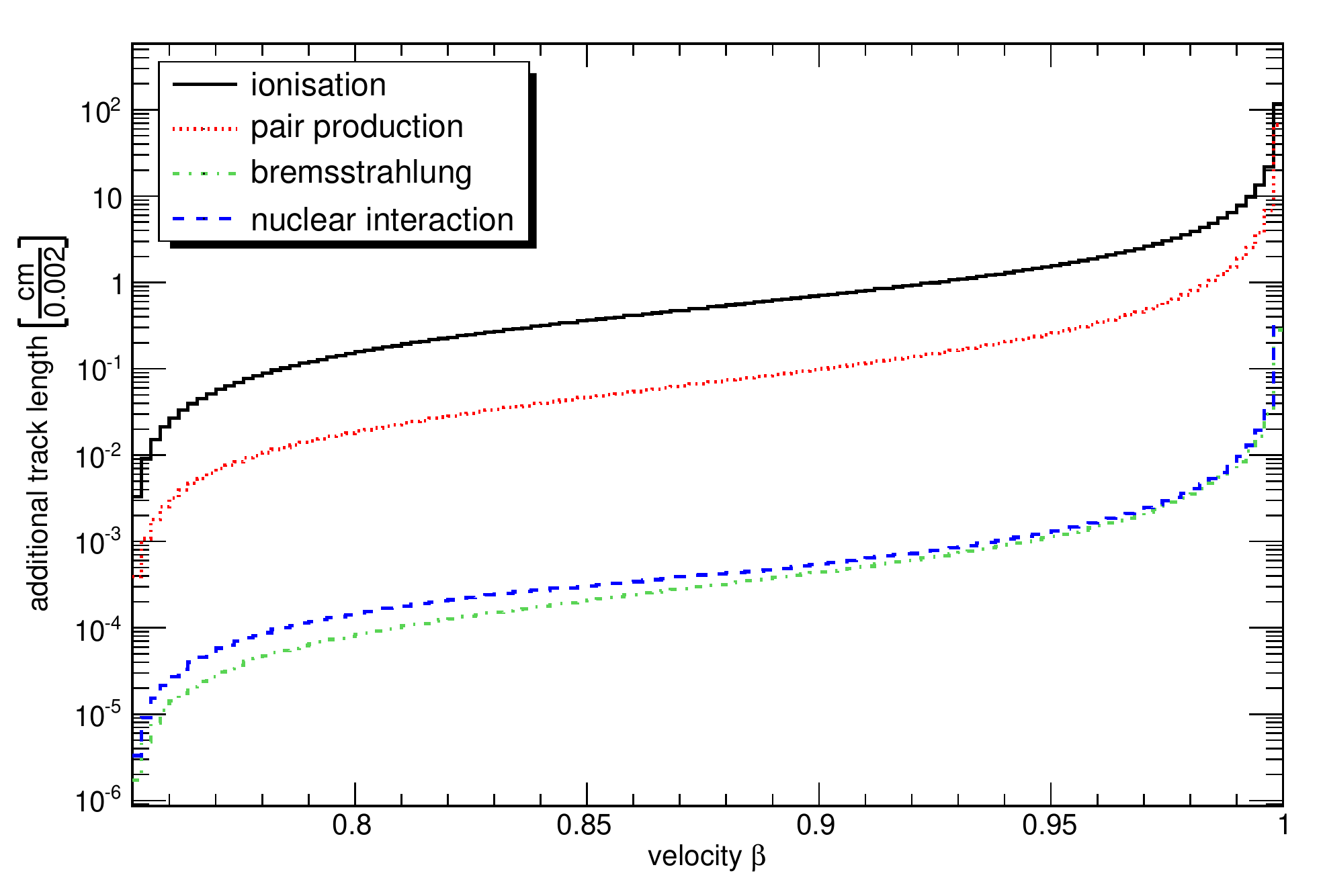}
	\caption{Velocity distribution of secondary track lengths. \label{fig:track:beta:process}
Shown is the  additional track length $\hat{l}_{add} $ per $10$\,m muon track
versus the Lorentz factor $\beta$ for bins of $0.002$ in $\beta $
 for $E_\mu = 10$\,TeV and $E_{max}=0.5$\,GeV. The 
contributions of  different  secondary energy loss processes are shown. The \tammf (eq. \ref{eq:tamm:factor}) has been applied.
}
\end{figure}

In order to understand the weak dependence on the primary energy,
figure \ref{fig:track:beta:process} shows the distributions $\Delta l_{add}/\Delta\beta $
for the different energy loss processes. It becomes obvious that  
the processes below $E_{max} $ are dominated 
by ionization, 
followed by pair-production, which contributes about $5\%-25\%$. 
The nuclear-interaction process contributes to about 
 $0.1$\% and  bremsstrahlung even less. At high energies these processes 
usually contribute significantly 
to the muon energy loss \cite{MMC}. However, they usually result
in large energy transfers, which rarely fall below $E_{max} $.
Therefore,  bremsstrahlung, being the hardest energy loss process, 
has the smallest contribution to the  energy losses below $E_{max}$.

\begin{figure}[htp]
\includegraphics*[width=.49\textwidth]{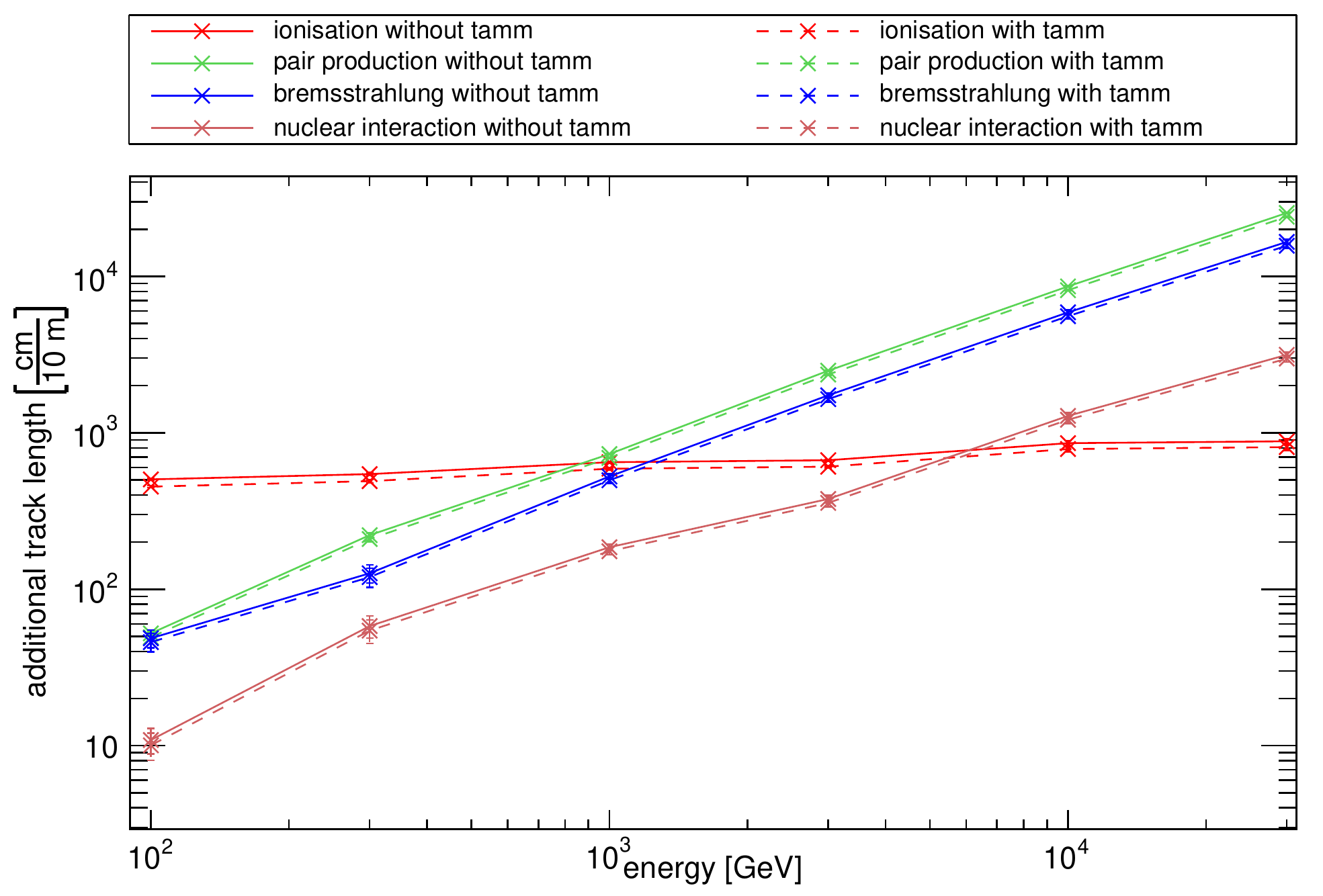}
\hfill
\includegraphics*[width=.49\textwidth]{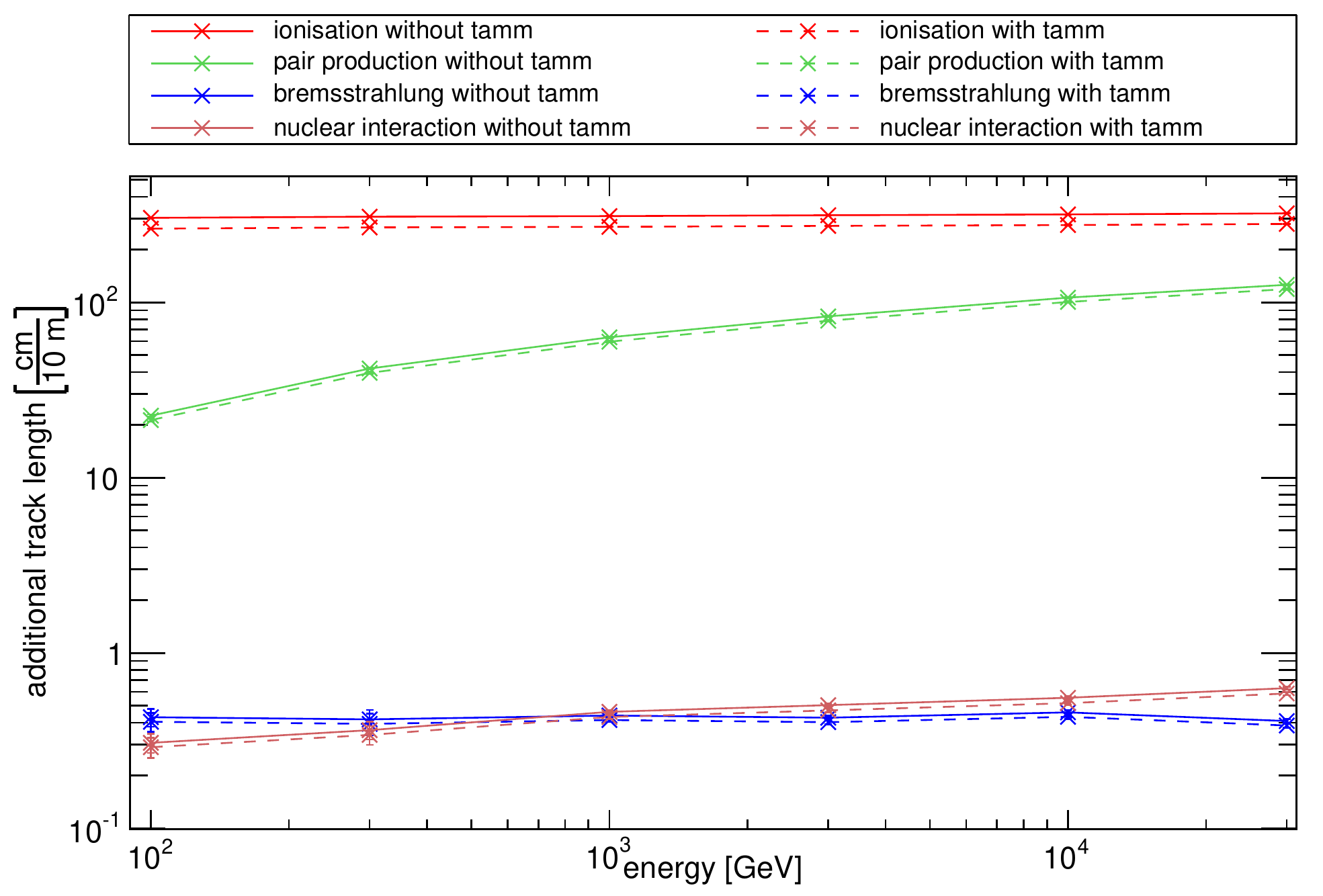}
\caption{Additional track length without (left) and with (right) a maximum energy cut $E_{max} =0.5 \unit{GeV}$ as a function of the muon energy. 
The solid lines show the result  without and the dashed lines with application of the \tammf.}
 \label{fig:track:energloss:noemax}
\end{figure}

This can be verified when investigating the  energy dependence of the 
amount of secondary tracks
for the different processes. This  is  shown in figure \ref{fig:track:energloss:noemax}.
In the left figure the  maximum energy cut has not been applied and the 
radiative processes show the expected linear increase with energy 
while ionization 
only weakly depends on energy. 
The application of 
$E_{max} $ in the right figure results in a change of the relative strength and 
ionization becomes the most important process with only a weak 
($\sim 3$\%) increase with energy.
Bremsstrahlung and photo-nuclear interactions are marginally relevant 
and hence the weak energy dependence can be identified as a result of an 
increasing contribution of the pair-production process.

The weak energy dependence is thus a consequence of the strong suppression 
of radiative processes.

\subsection{Parameterization of the secondary \cherpho yield}

\begin{figure}[htp]
\includegraphics*[width=.49\textwidth]{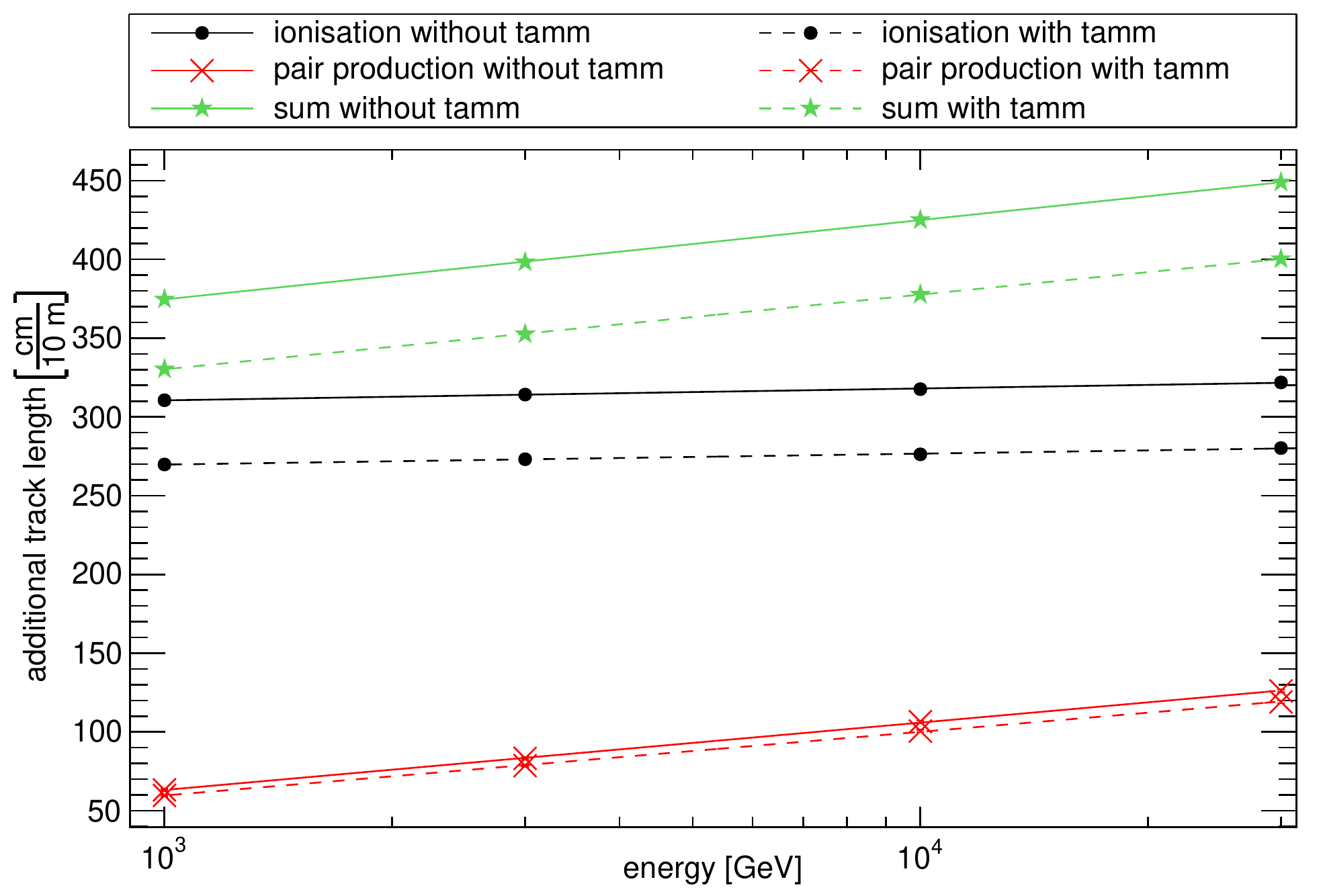}
\hfill
\includegraphics*[width=.49\textwidth]{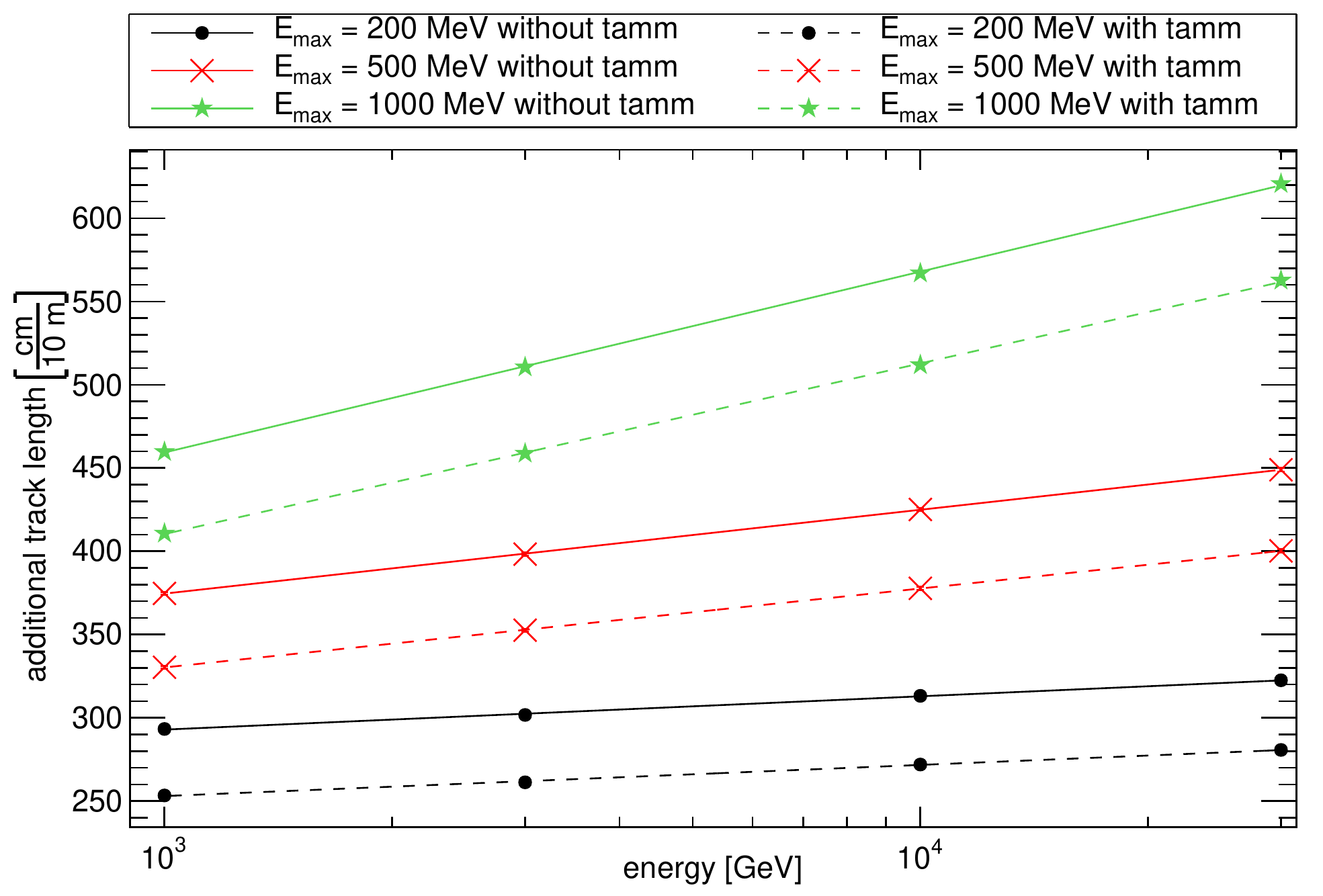}
\caption{Parametrization of the additional track length as function of $E_\mu $. Shown left are the parameterization of the total
and separately parameterizations of the two strongest processes ionisation, and pair production  for $ E_{max} = 0.5$\,GeV The right figure shows the
 parameterization of the total for different values of $E_{max} $.
Each parameterization is done for $l$ without and $\hat l$ with the \tammf, 
}
 \label{fig:track:energloss:param}
\end{figure}

In the following we parameterize the (effective) additional 
track length $l_{add}$ and $\hat l_{add}$ as a function of 
the muon energy for different $E_{max} $ with the expression
\eqb \label{eq:yield:fit}
\frac{l_{add}}{l_\mu} = \lambda_0 +  \kappa \cdot \ln \lbra{E_\mu \over \unit{GeV}}\rbra
\quad \mbox{and} \quad
\frac{\hat l_{add}}{l_\mu} = \hat \lambda_0 +  \hat \kappa \cdot \ln\lbra {E_\mu \over\unit{GeV}} \rbra
\eqe
The result is shown in figure \ref{fig:track:energloss:param}.
The parameterization result for different processes and $E_{max} $ are summarized in table \ref{tab:tracklength} in \ref{app:param:ang}.

As a typical value the amount of \cher photons is increased by about $38\% $
for $E_\mu = 10  $\,TeV, when the \tammf is applied.

As expected the effect of pair production and the slope of the
 energy dependency increases with increasing $E_{max} $. 
Changing $E_{max} $ by a factor $5$  from 
$ 0.2 $\,GeV to $1$\,GeV
leads to a change of the  additional track length by typically a factor 2.

The effect of the \tammf reduces the number of  additional photons by
about $10-12 \%$. The reduction is largely independent of $E_\mu $ and $E_{max} $.

The dependency on  the index of refraction is  small.
Though the total number of generated photons by the muon and all secondaries
increases according to equation \ref{eq:tamm} the fractional 
amount of secondary photons relative to the muon changes only because 
of a slightly changed Cherenkov threshold. However, only very few \cherphos are created by secondary particles close to the 
\cherthr anyway.

For the comparison  with the  result in \cite{CHWPHD}, 
(see eq.\ref{eq:cwresult}), one has to consider that the older calculation used different
 values $E_{max} =1  $\,GeV for photons
and $E_{max} = 0.5 $\,GeV for electrons and positrons.
However, in table \ref{tab:tracklength} it can be seen that
that parameterization is consistently between our parameterizations for 
 $E_{max} = 0.5 $\,GeV and   $E_{max} =1  $\,GeV.


\subsection{Water versus ice}

The same simulations have been repeated using water ($\rho = 1 \unit{g}\unit{cm}^{-3} $) and seawater ($\rho = 1.039\unit{g}\unit{cm}^{-3}$) instead of ice 
($\rho = 0.91 \unit{g}\unit{cm}^{-3} $)\footnote{Note, that the density of South Pole ice at the depth of the IceCube detector is $\rho=0.9216\unit{g}\unit{cm}^{-3}$ \cite{DIMA}.}.
For these simulations we have used pure water and ocean water, which contains
small traces of other elements (for details see table \ref{table:geant:materials},  \ref{app:geant:conf}).

\begin{figure}[htp]
\centering
\includegraphics*[width=0.49\textwidth]{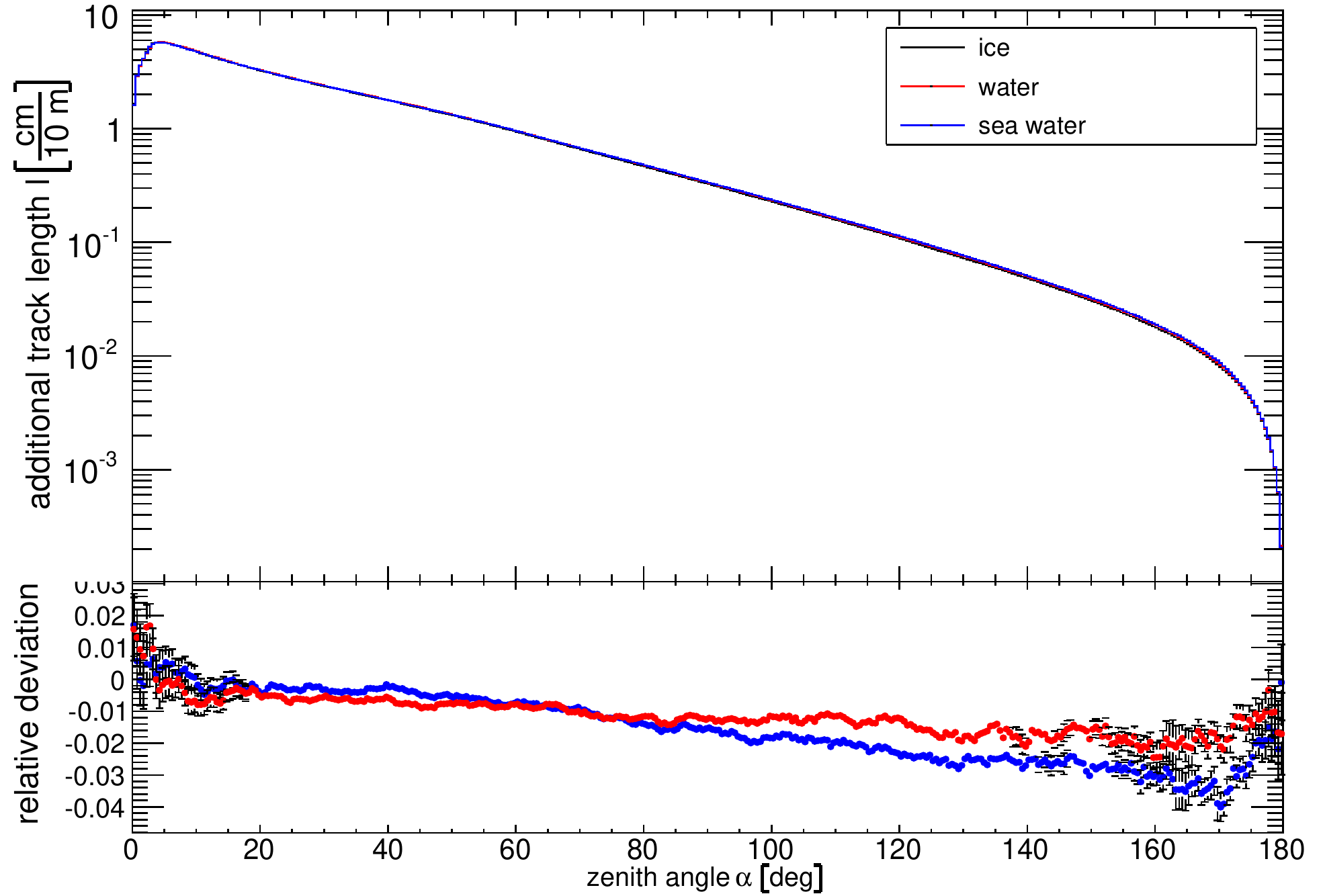}
\hfill
\includegraphics*[width=0.49\textwidth]{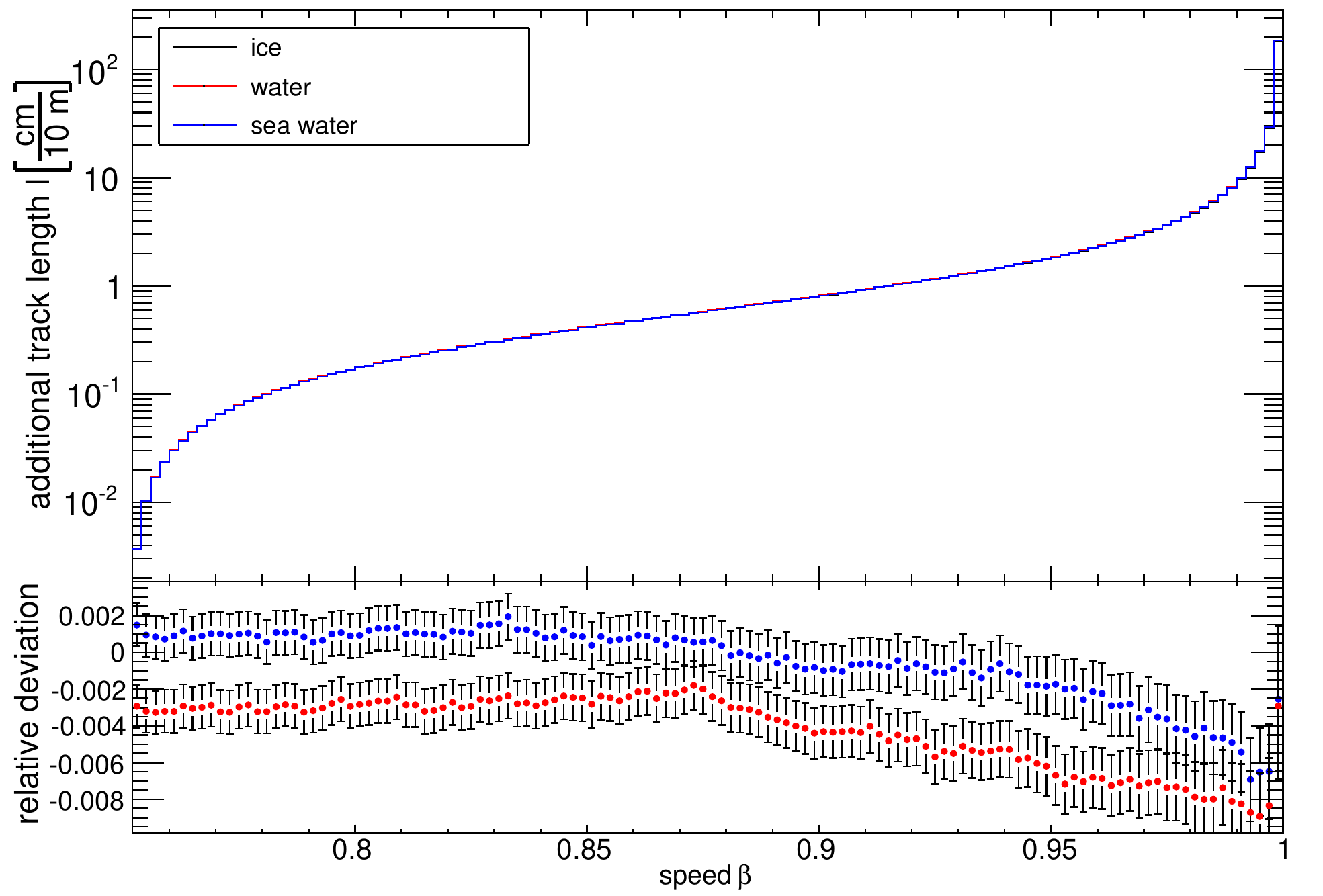}
\caption{Distribution of angular directions $\alpha $ (left) 
and Lorentz factors $\beta $ (right)
of secondary particles in pure water, sea water and ice. The simulated 
parameters are   $E_{\mu}=10\unit{TeV}$ and $E_{\mathrm{max}}=500\unit{MeV}$ in water and ice. Shown at the bottom are also the relative deviations
$(i-w)/(i+w)$  of the two types of water ($w$) relative to ice ($i$)
(Red: relative deviation ice/pure water; Blue: ratio ice/sea water)
 \label{fig:track:water:ice}}
\end{figure}

As shown in figure \ref{fig:track:water:ice}, the number of photons does not change strongly.
Naively, one could expect that the number of photons increases proportionally to the density change,
that this assumption is wrong can be explained by two canceling effects:
Though the number of secondary tracks per unit track of the muon increases with the density ratio, the length of these tracks decreases by the same ratio, 
because of larger energy losses and fewer emitted \cher photons.
Hence, the resulting number of photons in water is similar to the number
of photons in ice within $\lesssim 1\% $, comparable to the uncertainties
of the chemical composition.
We conclude, that the here presented parameterizations, though simulated for ice,
 can be applied to water without a density correction.

\section{Angular distribution \label{sec:angular}}

The angular distribution of secondary 
\cherphos with respect to the muon axis 
depends on two parameters:  the zenith angles $\alpha $ of the  
secondary tracks (see figure \ref{fig:intro:principle}) 
and on the Lorentz factors $\beta $ of these
tracks (see equation \ref{eq:tamm:factor}).

\begin{figure}[htp]
\centering
	\includegraphics*[width=.69\textwidth]{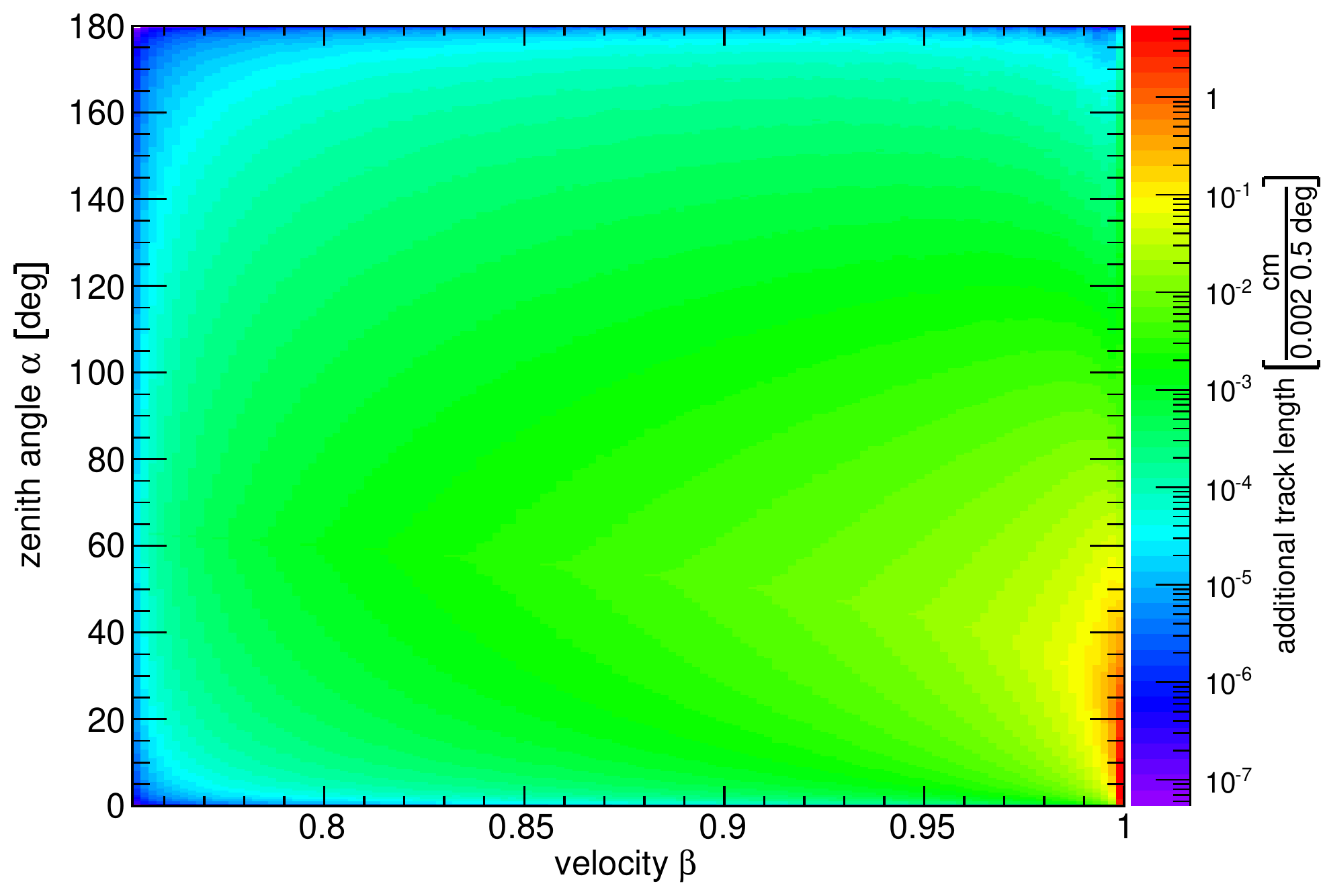}
	\caption{Density distribution of the relative track length  versus the 
zenith angle $\alpha $ and Lorentz factor $\beta $ for a $10$\,TeV muon.
The vertical color codes corresponds to the histogrammed lengths $\hat l$ for a $10 \unit{m} $ muon with the \tammf applied.
 \label{fig:alpha:beta}}
\end{figure}

An example of the density distribution of secondary track length relative to these parameters $ \frac{\Delta^2 \hat{l}_{\mathrm{add}}}{\Delta\alpha \, \Delta\beta} $  is shown in 
figure \ref{fig:alpha:beta}. 
Most tracks are emitted into forward direction with large values of 
$\beta $.

\begin{figure}[htp]
	\includegraphics*[width=.49\textwidth]{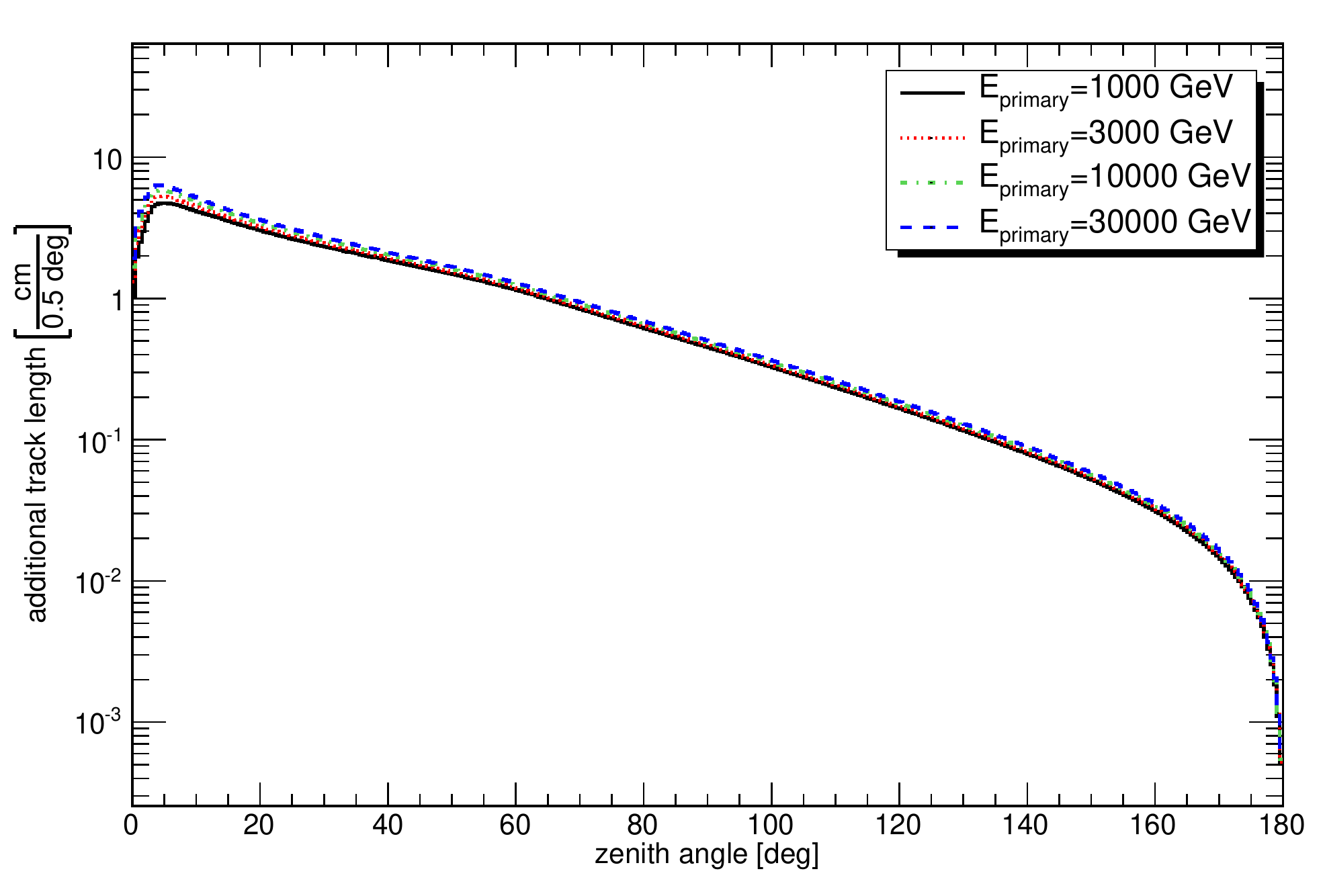}
\hfill
	\includegraphics*[width=.49\textwidth]{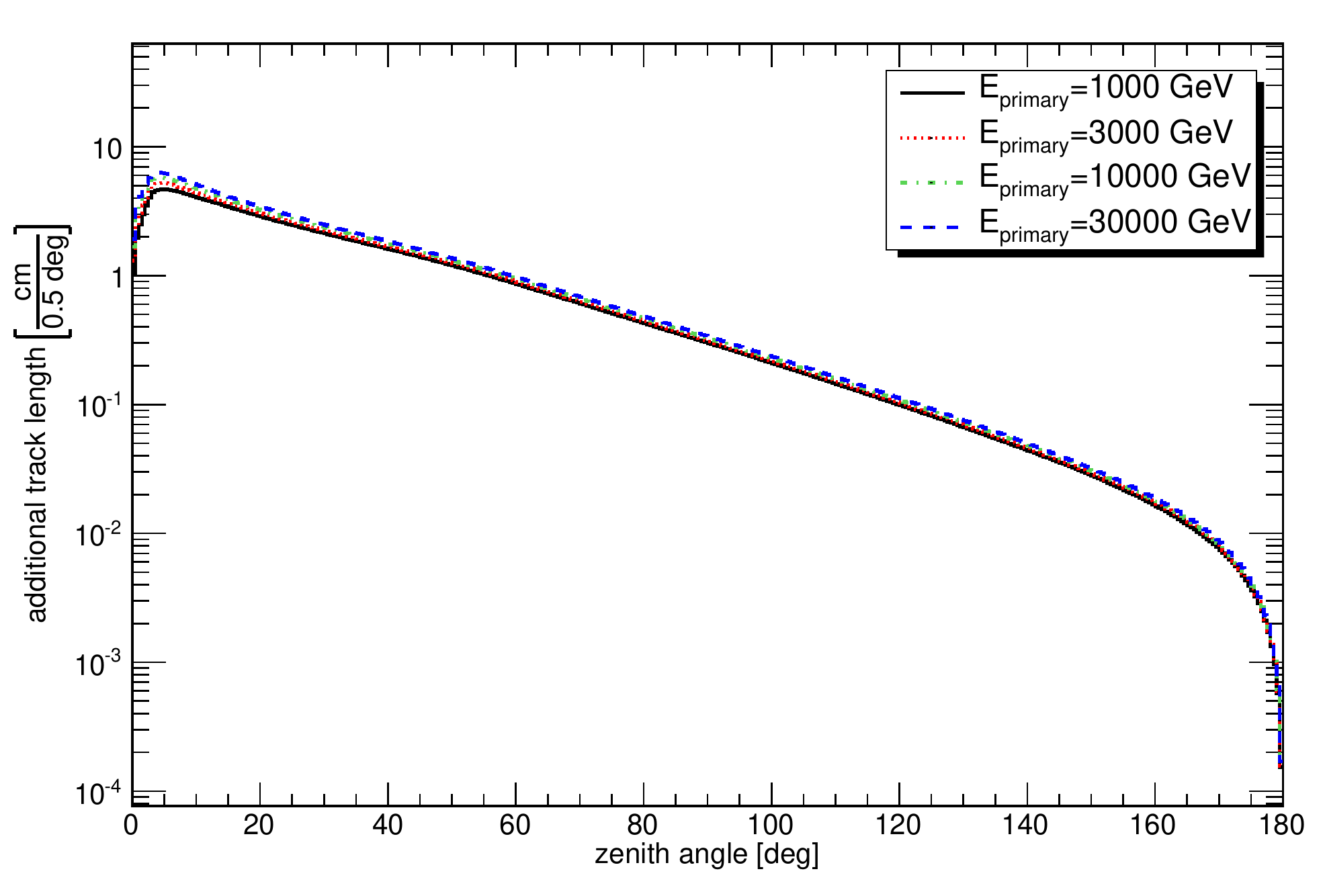}
	\includegraphics*[width=.49\textwidth]{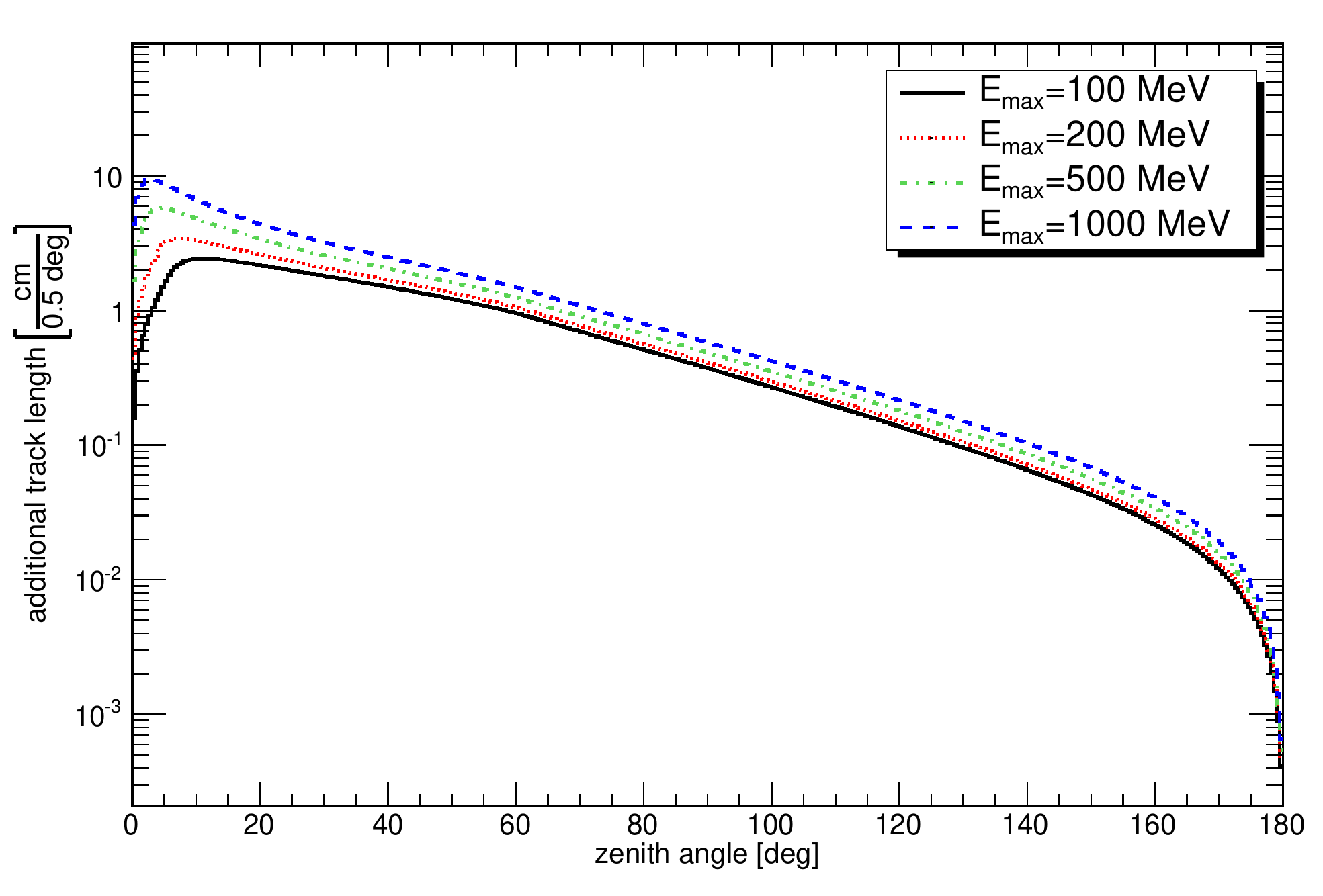}
\hfill
	\includegraphics*[width=.49\textwidth]{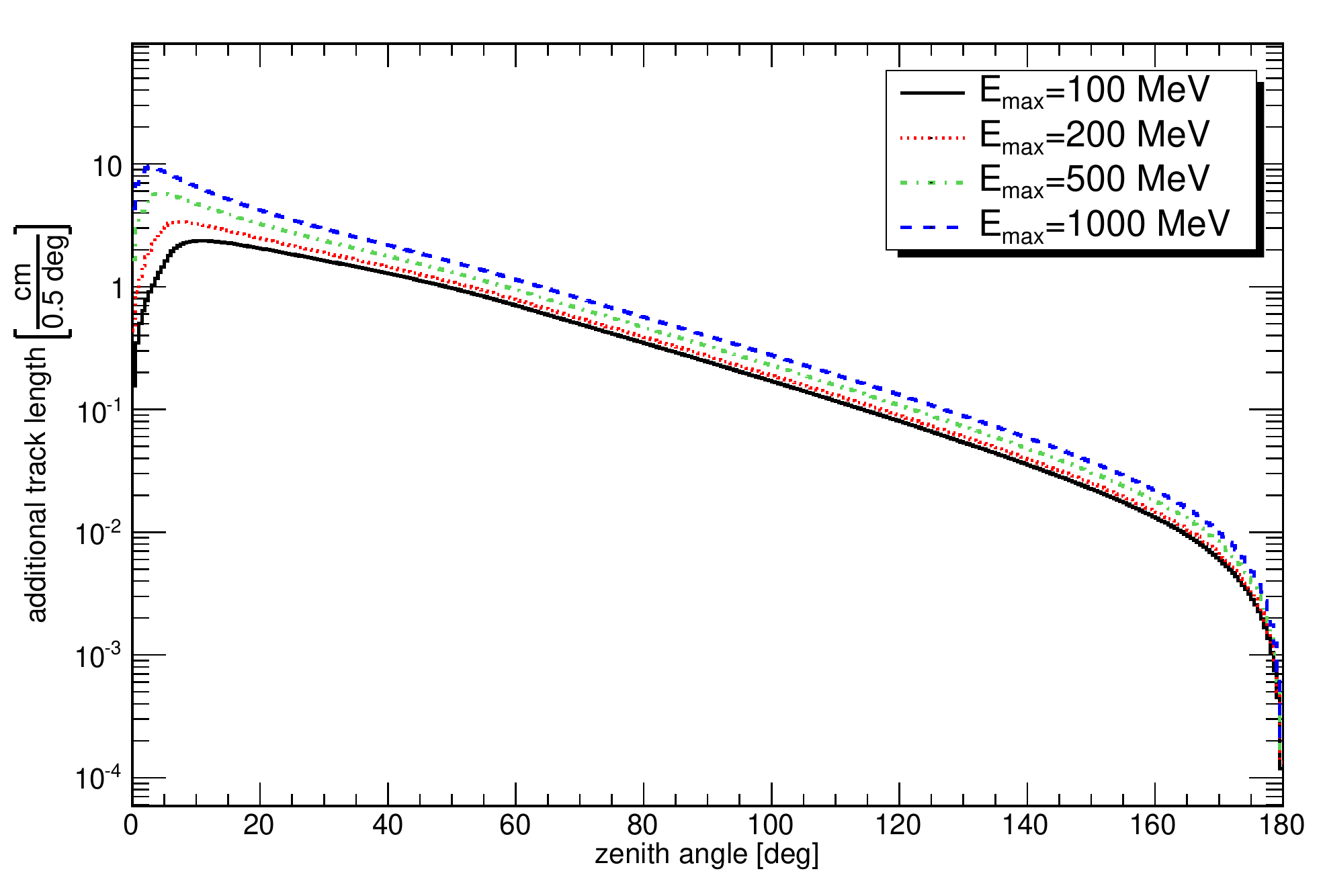}
	\caption{Histograms of the additional track length per $10\unit{m}$
 muon track as a function of the zenith angle $\alpha $  of the track.
The top figures show the results for $E_{max} = 500 \unit{MeV} $ and different $E_{\mu} $ and the bottom figures the results for $E_\mu = 10 \unit{TeV} $ and different $ E_{max} $. The left figures show the physical track length and the 
\tammf has been applied additionally to the tracks in the right figures.
 \label{fig:alpha:dist}}
\end{figure}

In figure \ref{fig:alpha:dist}  the  dependence of the angular distribution
$ \frac{\Delta \hat{l}_{\mathrm{add}}}{\Delta \alpha } $ on the parameter $E_{max} $ and  the muon energy
$E_\mu $ is shown. The corresponding figures for the parameter $ \beta $ 
have been shown in figure \ref{fig:track:beta:tamm}.
It can be seen that the angular distribution depends only weakly 
on the muon energy.  Also the \tammf is only a small correction to the overall shape. The parameter $E_{max} $,  has a larger effect. Not only the 
total track length increases with larger $E_{max} $, but 
the additional tracks are pointing preferably into forward direction.
This corresponds to the larger boost of higher-energy secondaries.

When assuming azimuthal symmetry, 
the angular distribution of \cherphos can be calculated according to the procedure described in  \ref{sec:trafo}, if the Lorentz factor and zenith angle $\alpha $ for each track segment is known.

An  advantage of the transformation method is that \cherphos 
do not need to be generated and propagated
 during the \geantfours, which greatly reduces the computing time.
Furthermore, during the \geantfours no specific wavelength interval 
needs to be assumed and the transformation can be applied later with 
different values of $n$ on the same data set.
Disadvantage of the method is the assumption of azimuthal symmetry for 
the occurrence  of track elements. This is a good assumption, e.g. for 
high statistics of small secondary tracks, which are produced with 
random orientation along the muon track. 
However, for a single low-energy electron of a few MeV multiple scattering
 will lead to a specific path, which is not symmetric in azimuth.

\begin{figure}[htp]
\centering
\includegraphics*[width=.49\textwidth]{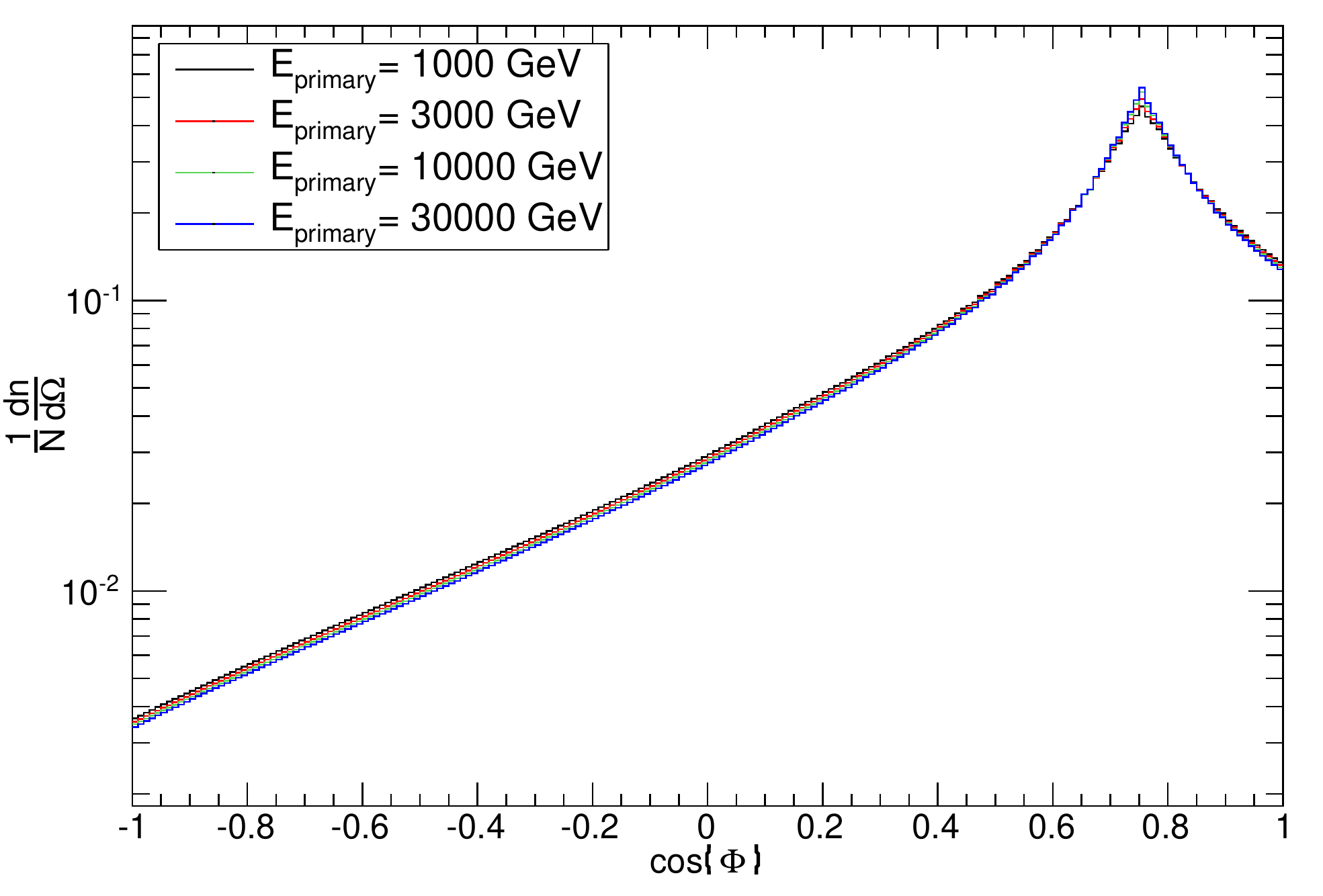}\hfill
\includegraphics*[width=.49\textwidth]{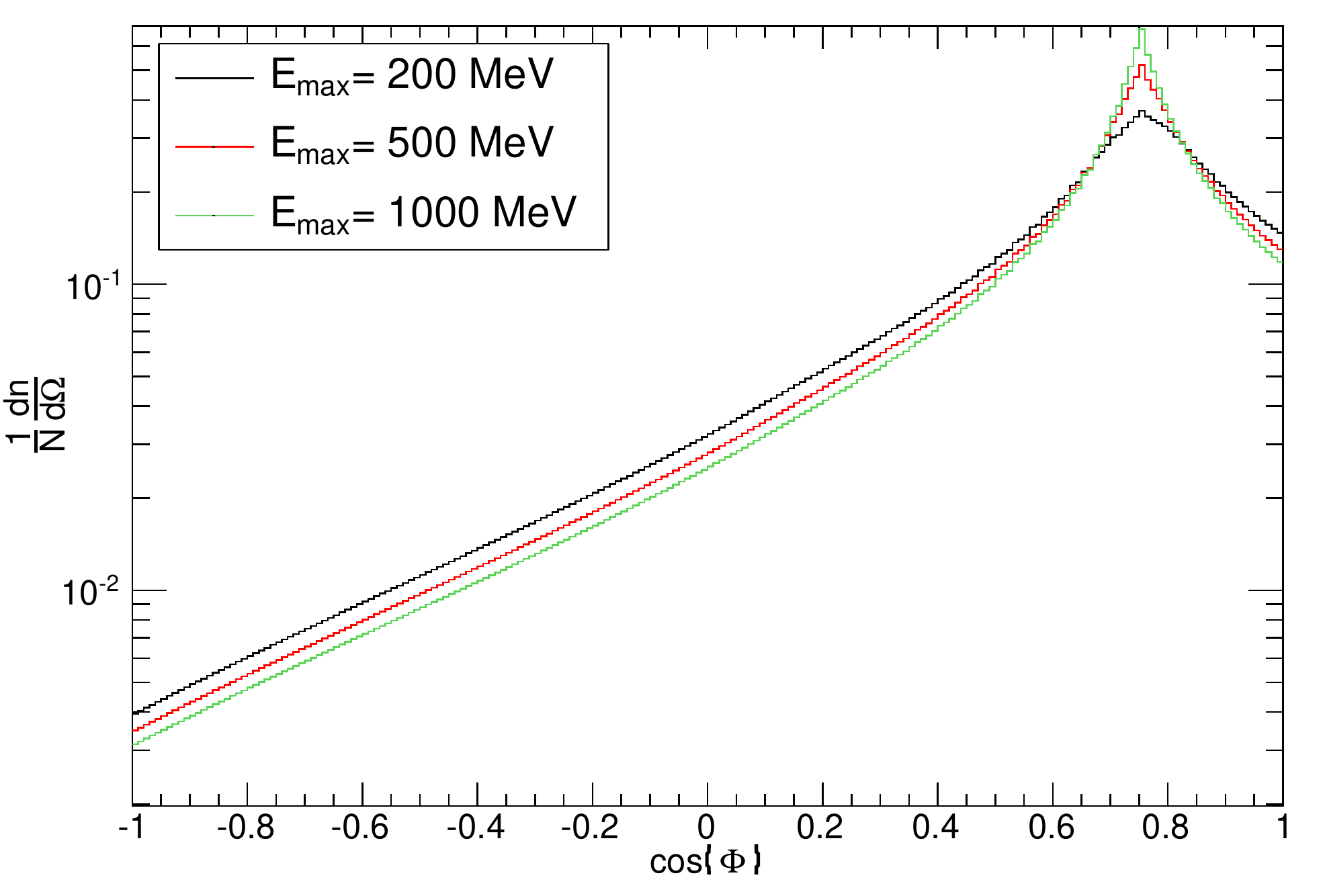}
\caption{The angular distribution of secondary \cherphos for different $E_\mu$ with $E_{max}=500\unit{MeV}$ (left) and for different  $E_{max}$ with $E_{\mu}=10\unit{TeV}$ (right)
 \label{fig:result:lightangle}. Shown are the normalized angular distribution per photon and steradian.}
\end{figure}

The calculated angular distributions are shown in figure \ref{fig:result:lightangle}. The distribution depends only weakly on the  muon energy where
higher-energy muons produce a slightly more pronounced \cher peak.
The dependence on $E_{max} $  is stronger. Correspondingly
to the enhancement of tracks into forward direction 
(see figure \ref{fig:alpha:dist})
also the Cherenkov peak becomes more pronounced.
 For larger $E_{max} $ the relative contribution of pair production increases whereas the contribution from ionization remains constant.

\begin{figure}[htp]
\centering
\includegraphics*[width=.49\textwidth]{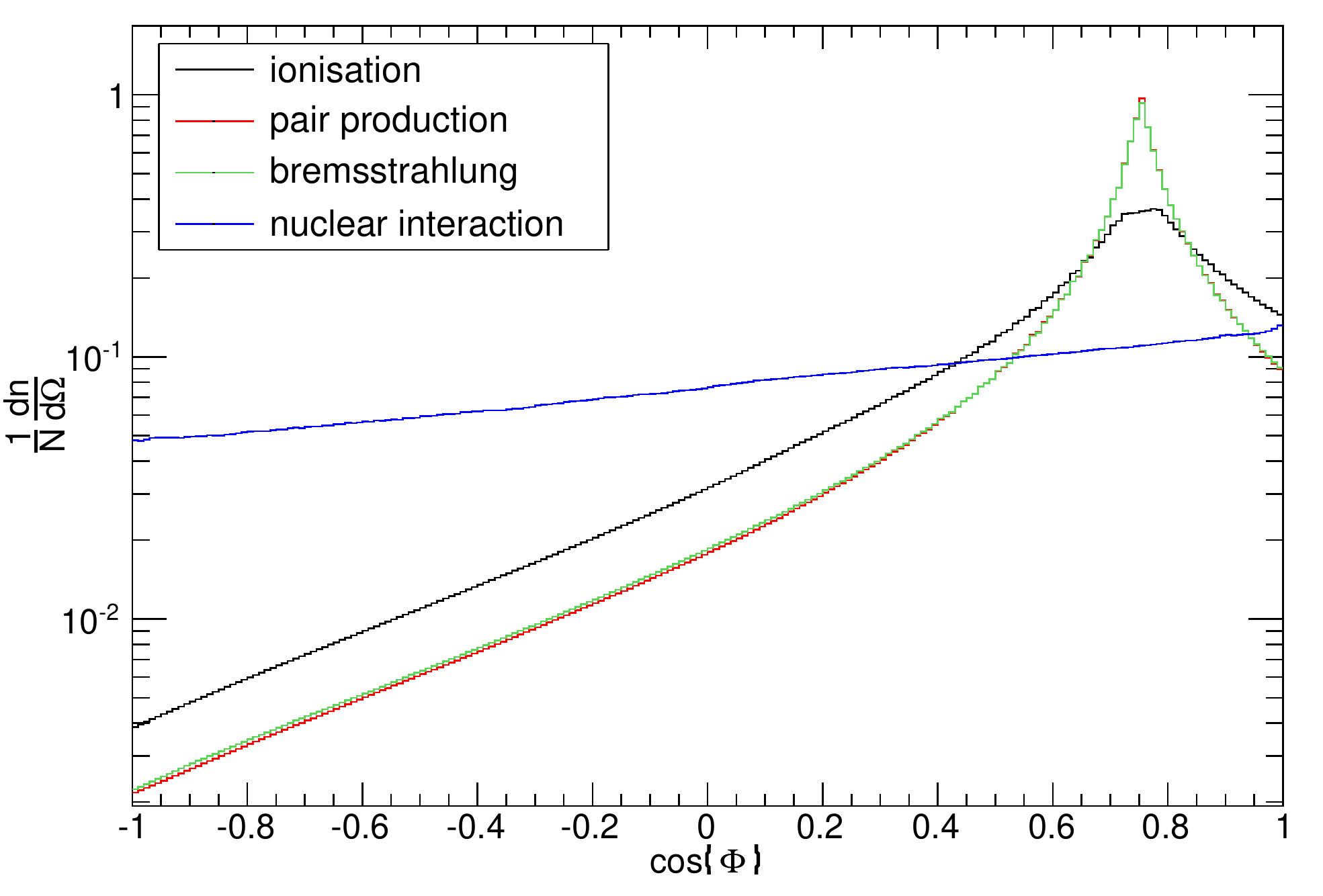} \hfill
\includegraphics*[width=.49\textwidth]{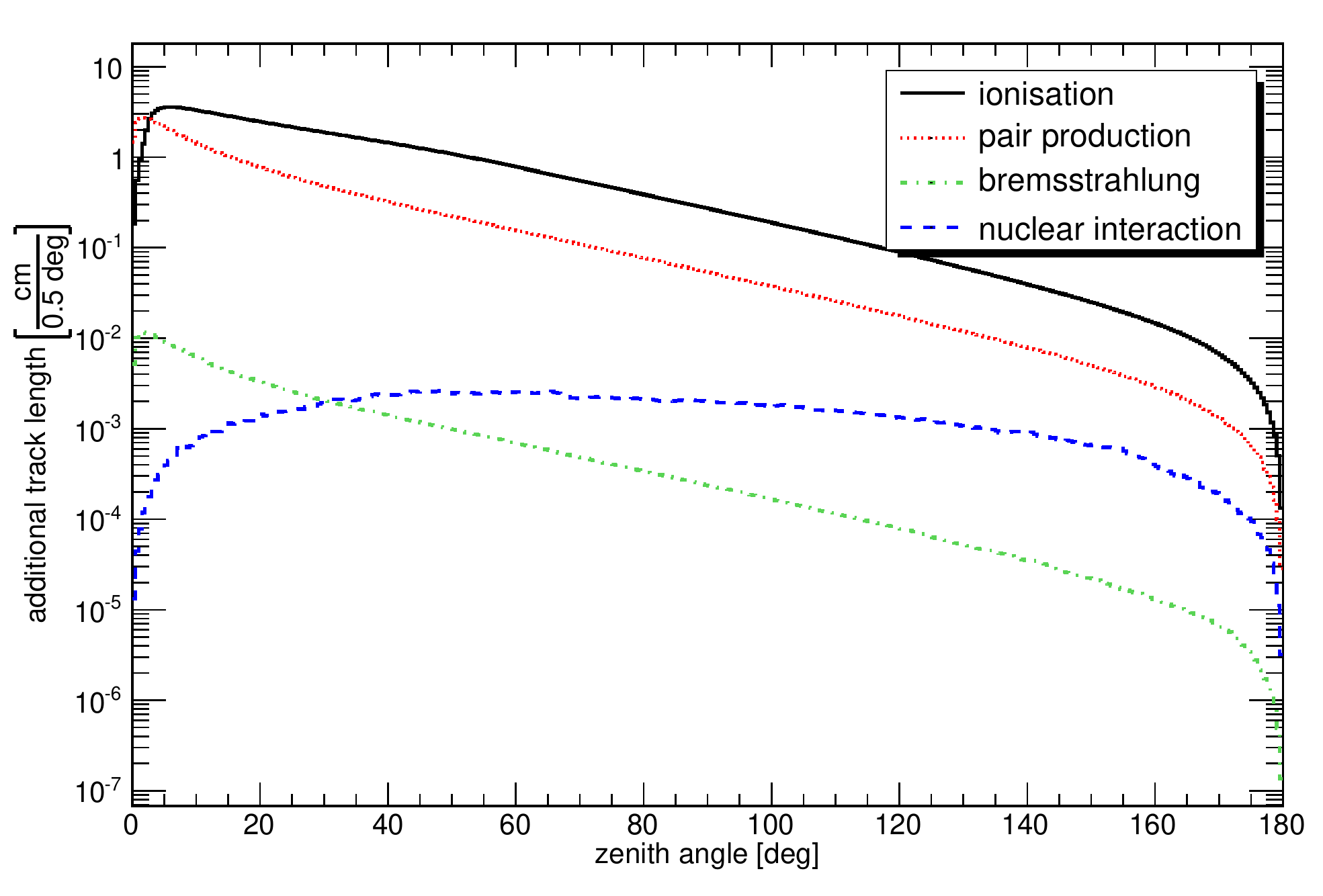}
\caption{The angular distribution of  \cherphos (left) 
and the corresponding
 angular distribution of secondary track length (right)
 for the different processes and $E_{max} = 500$\,MeV  and $E_{\mu} = 10$\,TeV.
 \label{fig:result:trackprocessangle}
For the \cherphos the  normalized angular distributions per photon and steradian
are shown. For the track length the \tammf has been applied.
}
\end{figure}

When looking at the individual energy loss processes large 
differences are observed.
The  distributions of \cherphos and the corresponding distributions of 
track length are shown in figure 
\ref{fig:result:trackprocessangle}.
Bremsstrahlung and pair production exhibit a similarly
sharp distribution.
The ionization process has a less pronounced (``chopped off'') 
\cher peak. This is due to the fact, that kinematics 
impose a minimum angle $\alpha$ as a function of $E_{max}$,
which is of the order of a few degrees 
(see equation \ref{equation:alpha_beta}). 
The Cherenkov light from nuclear interactions is largely featureless with 
approximately a factor 2 more 
photons into forward than into backward direction.
Correspondingly to the maximum energy threshold $E_{max} =0.5 \unit{GeV}$, 
the  considered  momentum transfers are  not large compared to the transversal 
fermi motion of interacting nuclear partons. Hence large scattering 
angles appear and the  forward direction is less pronounced compared to   
electro-magnetic interactions.

\begin{figure}[htp]
\includegraphics*[width=.49\textwidth]{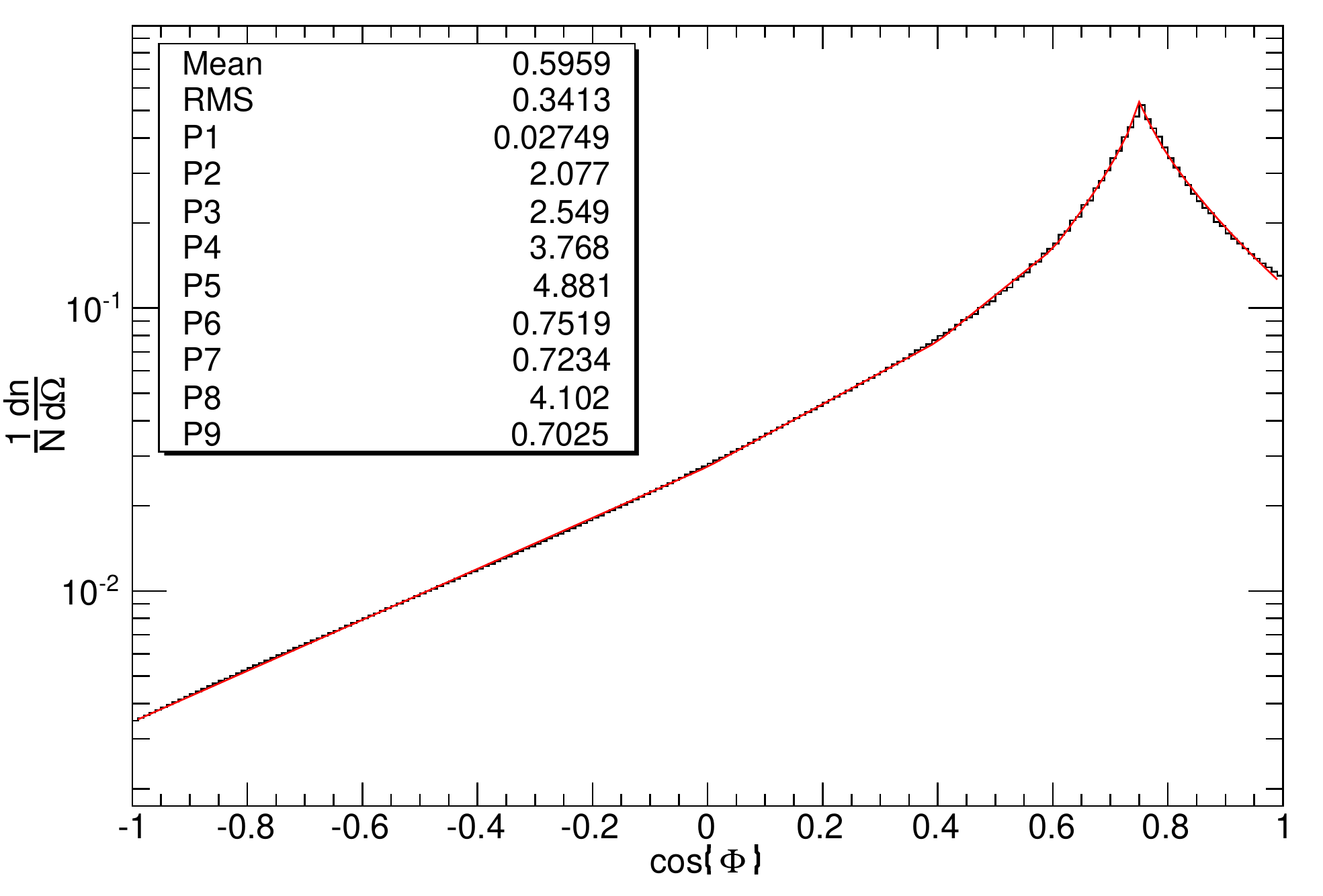}
\includegraphics*[width=.49\textwidth]{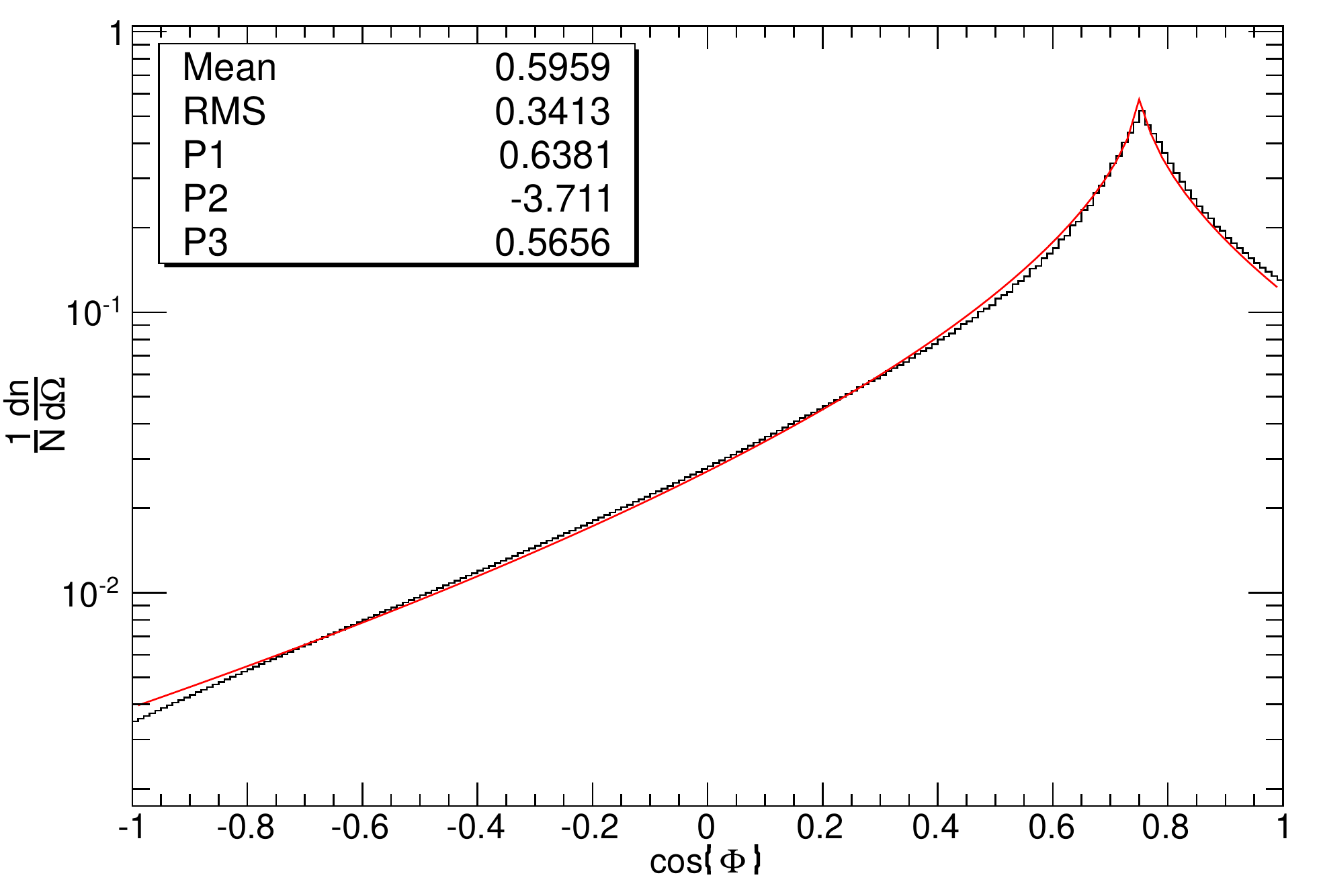}
	\caption{Example of the  parameterization of the angular distribution 
for $E_\mu = 10 $\,TeV and $E_{max} = 500 $\,MeV. Left is 
the asymmetric parameterization function and right the 
symmetric parameterization.
 \label{fig:result:para}}
\end{figure}

The angular distribution has been parameterized with two functions, in the following called asymmetric and simple parameterization.
The asymmetric parameterization function is
\begin{equation}
    \frac{1}{N}\frac{\mathrm{d}N}{\mathrm{d}\Omega}(x) = a
    \begin{cases}
        \mathrm{e}^{bx} & \text{if } -1\leq x < 0, \\
        \mathrm{e}^{cx} & \text{if } 0\leq x < 0.4, \\
        \mathrm{e}^{0.4(c-d)} \mathrm{e}^{dx} & \text{if } 0.4\leq x < 0.6 \\
        \mathrm{e}^{0.4(c-d)} \mathrm{e}^{d0.6} \mathrm{e}^{e(f-0.6)^{g}} \mathrm{e}^{-e(f-x)^{g}} & \text{if } 0.6\leq x < f, \\
        \mathrm{e}^{0.4(c-d)} \mathrm{e}^{d0.6} \mathrm{e}^{e(f-0.6)^{g}} \mathrm{e}^{-h(x-f)^{i}} & \text{if } f\leq x \leq 1,
    \end{cases}
\end{equation}
with $x=\cos\phi$ and $f=\cos(\theta_{c,0})=1/n$, where $n$ is the index of refraction. This function fits the parts left and right of the \cher peak individually. It is a piecewise continuous function consisting of an exponential function from -1.0 to 0.0, from 0.0 to 0.4 and from 0.4 to 0.6. Around the \cher peak a function $\exp ax^{b}$ is fitted on both sides. 

The
simple parameterization function is
\begin{equation}
    \frac{1}{N}\frac{\mathrm{d}N}{\mathrm{d}\Omega}(x) = a \mathrm{e}^{b\left|\cos(\theta_{c,0})-x\right|^{c}},
\end{equation}
with $x=\cos\phi$ and $\cos(\theta_{c,0})=1/n$, where $n$ is the index of refraction. This parameterization is fit with only three parameters and describes the data reasonably well but slightly worse than the asymmetric parameterization. Note, that the parameter $a$ 
could be eliminated by normalization of the resulting distributions.
Examples of the fits are shown in  figure \ref{fig:result:para}.
The resulting fit parameters for different $E_\mu $ and $E_{max} $ for
 the asymmetric and the simple fit are listed  in  \ref{app:param:ang}.

\begin{figure}[htp]
\includegraphics*[width=.49\textwidth]{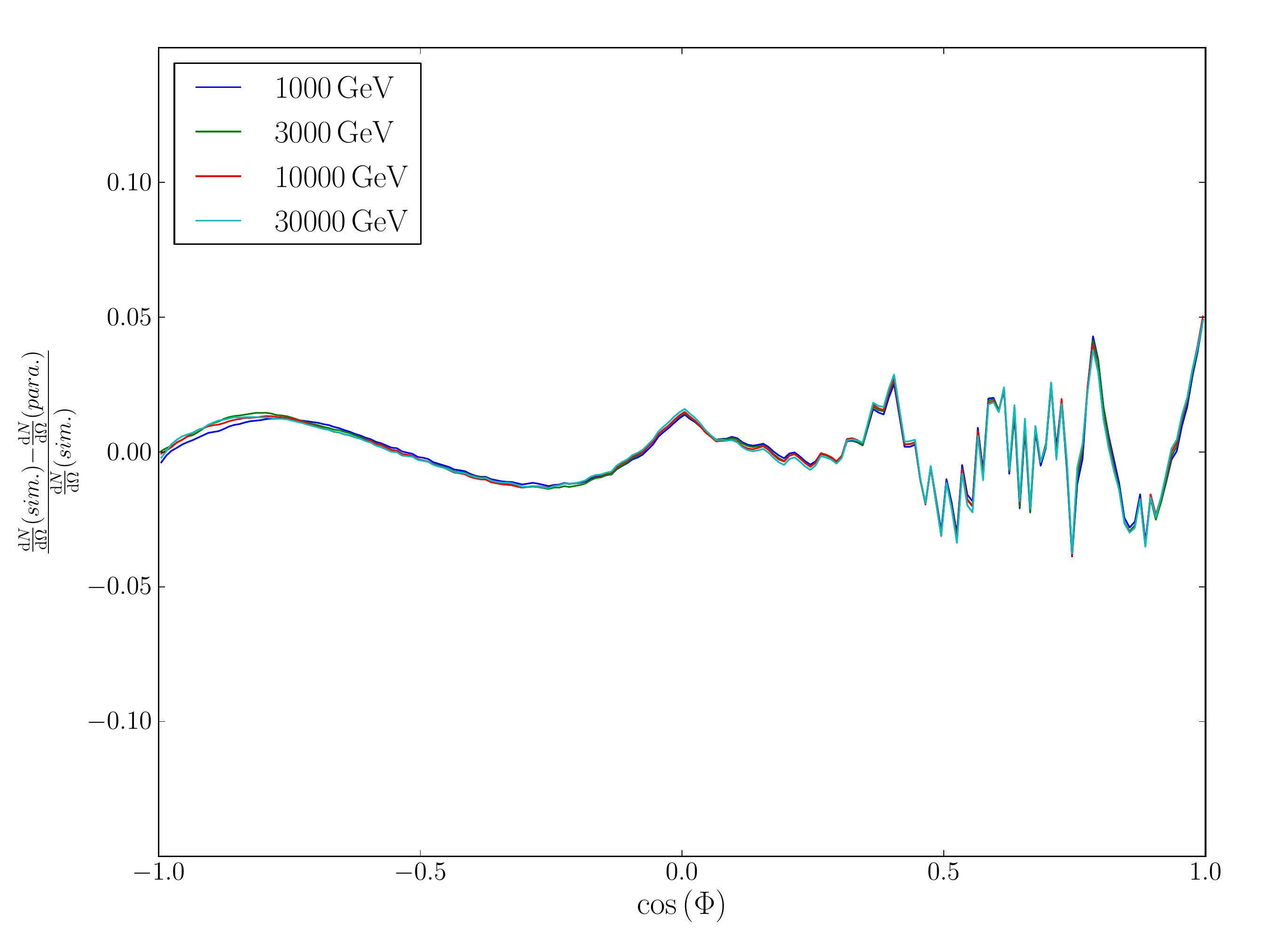}
\hfill
    \includegraphics*[width=.49\textwidth]{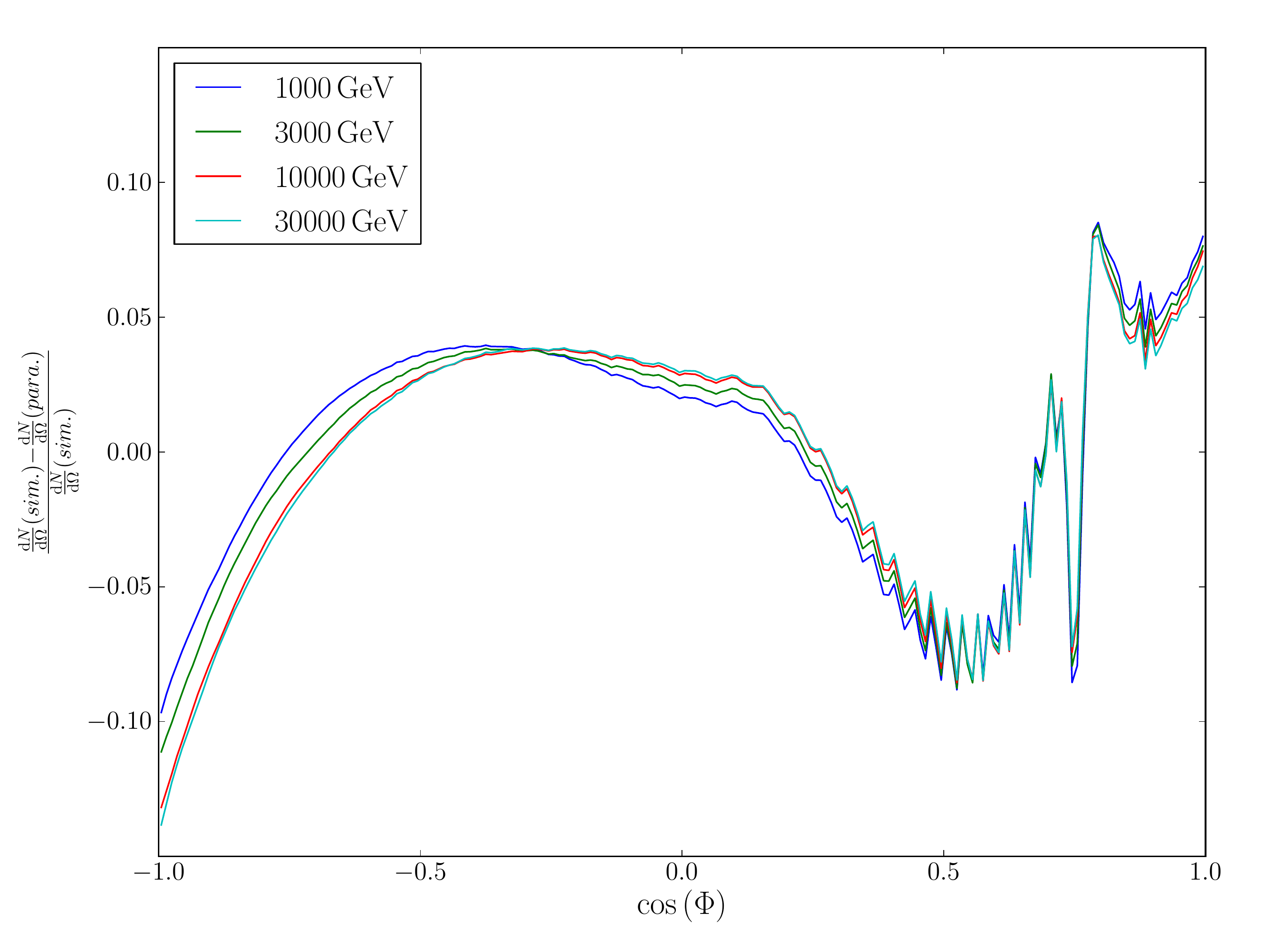}
    \caption{Relative deviation of the asymmetric parameterization (left)
and the simple parameterization (right) from the data. The deviations are shown
for parameterizations of different $E_\mu$ and $E_{max} = 500 $\,MeV 
(see tables \ref{tab:result:paraasym} and \ref{tab:result:parasimple}).
 \label{fig:result:parabenchmark}}
\end{figure}

Figure  \ref{fig:result:parabenchmark} shows the relative residuals of the parameterizations for different $E_\mu $. The accuracy of the asymmetric parameterization is for all angles within a few percent. The 
simple parameterization is less accurate with relative deviations up to $10$\,\%.
The accuracy is found to be very similar for different energies.

\begin{figure}[htp]
    \includegraphics*[width=.49\textwidth]{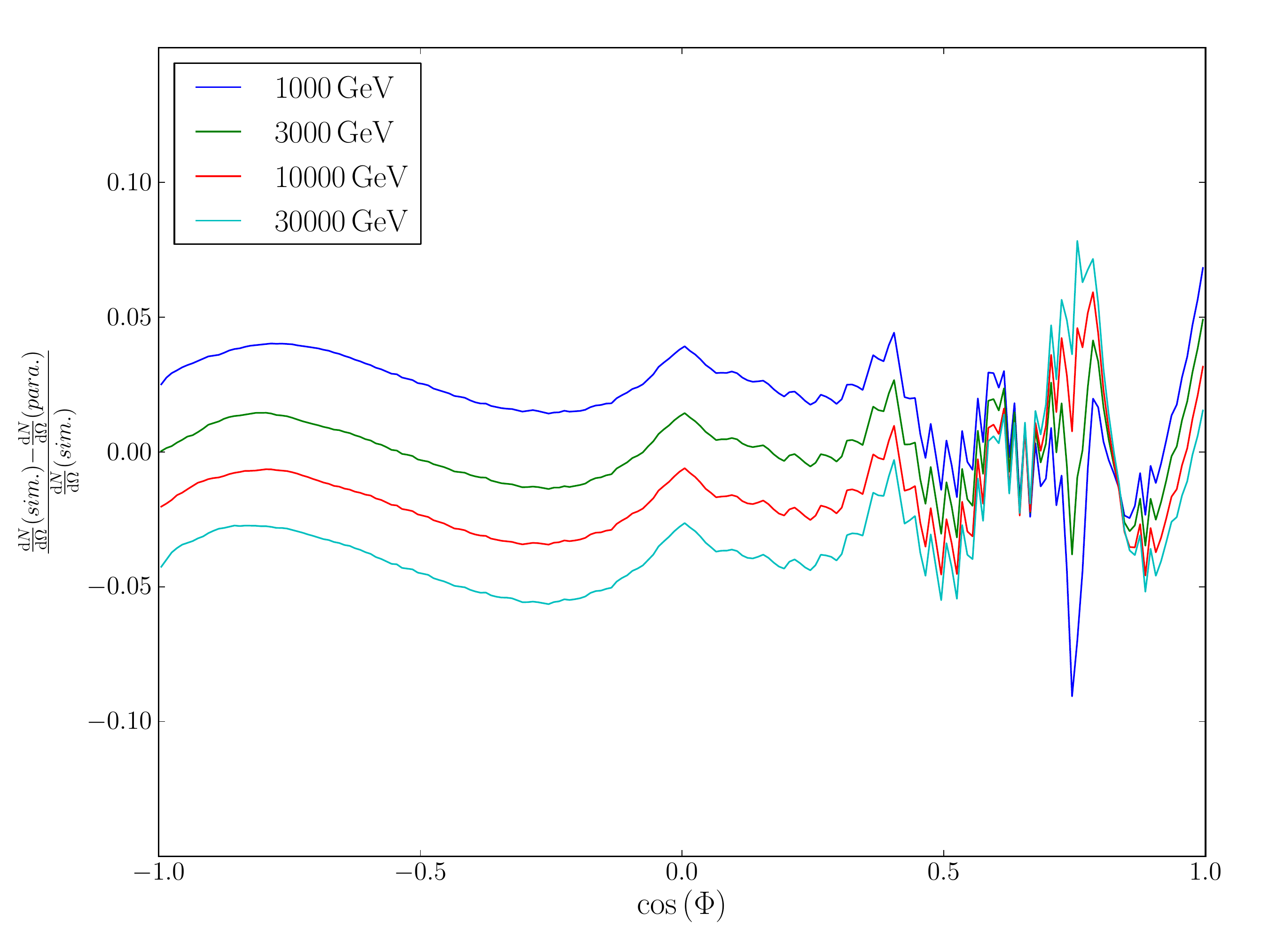} \hfill
    \includegraphics*[width=.49\textwidth]{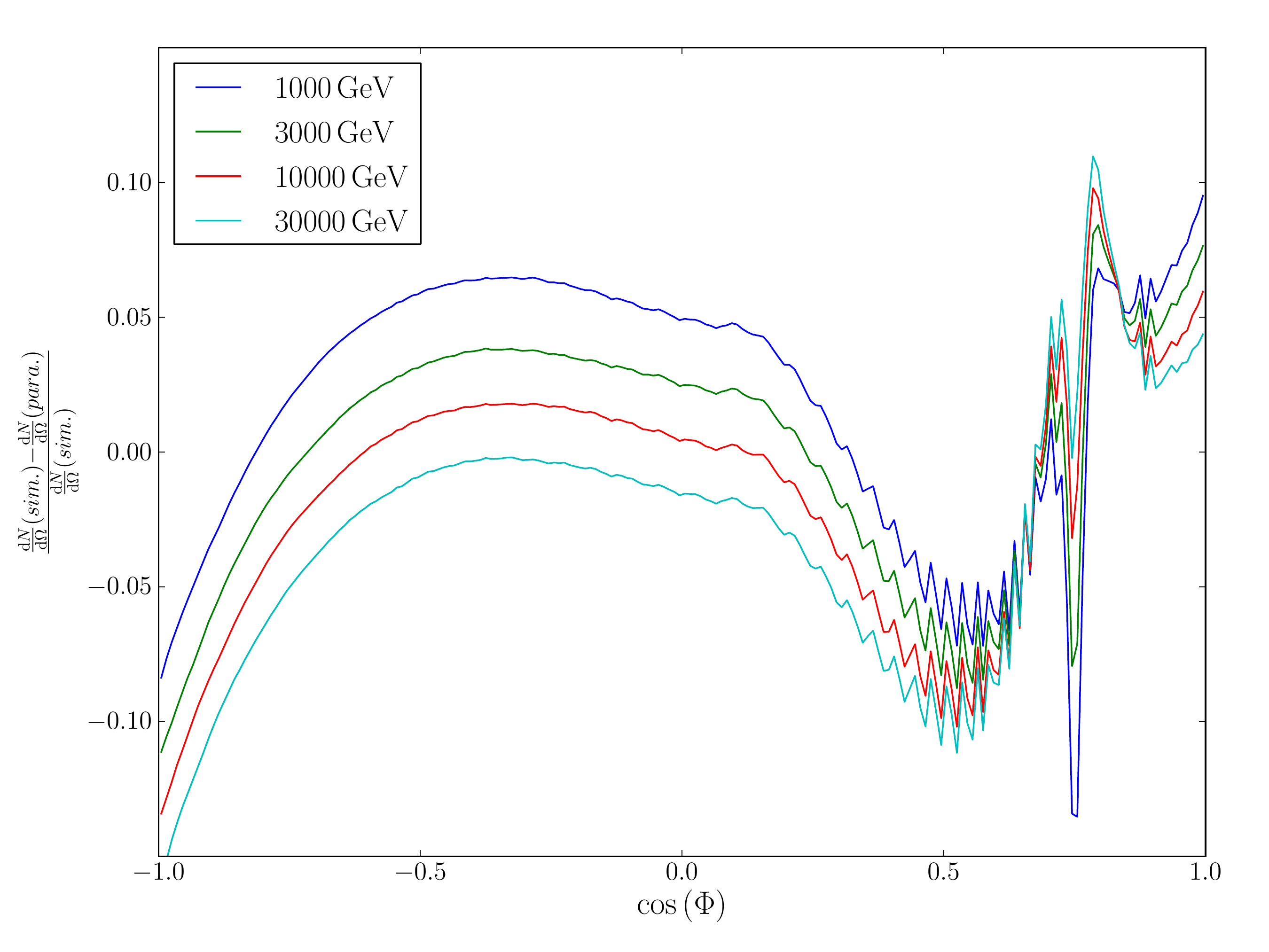}
	\caption{Relative deviation of the parameterization for $E_\mu = 3$\,TeV
from the data for different $E_\mu $ and $E_{max} =500$\,MeV. Left is 
 the asymmetric parameterization (left) and right the simple 
parameterization (see tables \ref{tab:result:paraasym} and \ref{tab:result:parasimple}).
 \label{fig:result:datafixedpara}}
\end{figure}

Figure \ref{fig:result:datafixedpara} shows the difference of 
different energies with 
a parameterization of fixed energy. It becomes obvious that without
the proper treatment of the energy dependence,
 the good accuracy of the asymmetric parameterization would be 
substantially degraded. On the other hand  for the simple fit, the energy dependence is smaller than the overall uncertainty and the parameterization could be
used approximately for a fixed energy.

\begin{figure}[htp]
\includegraphics*[width=.49\textwidth]{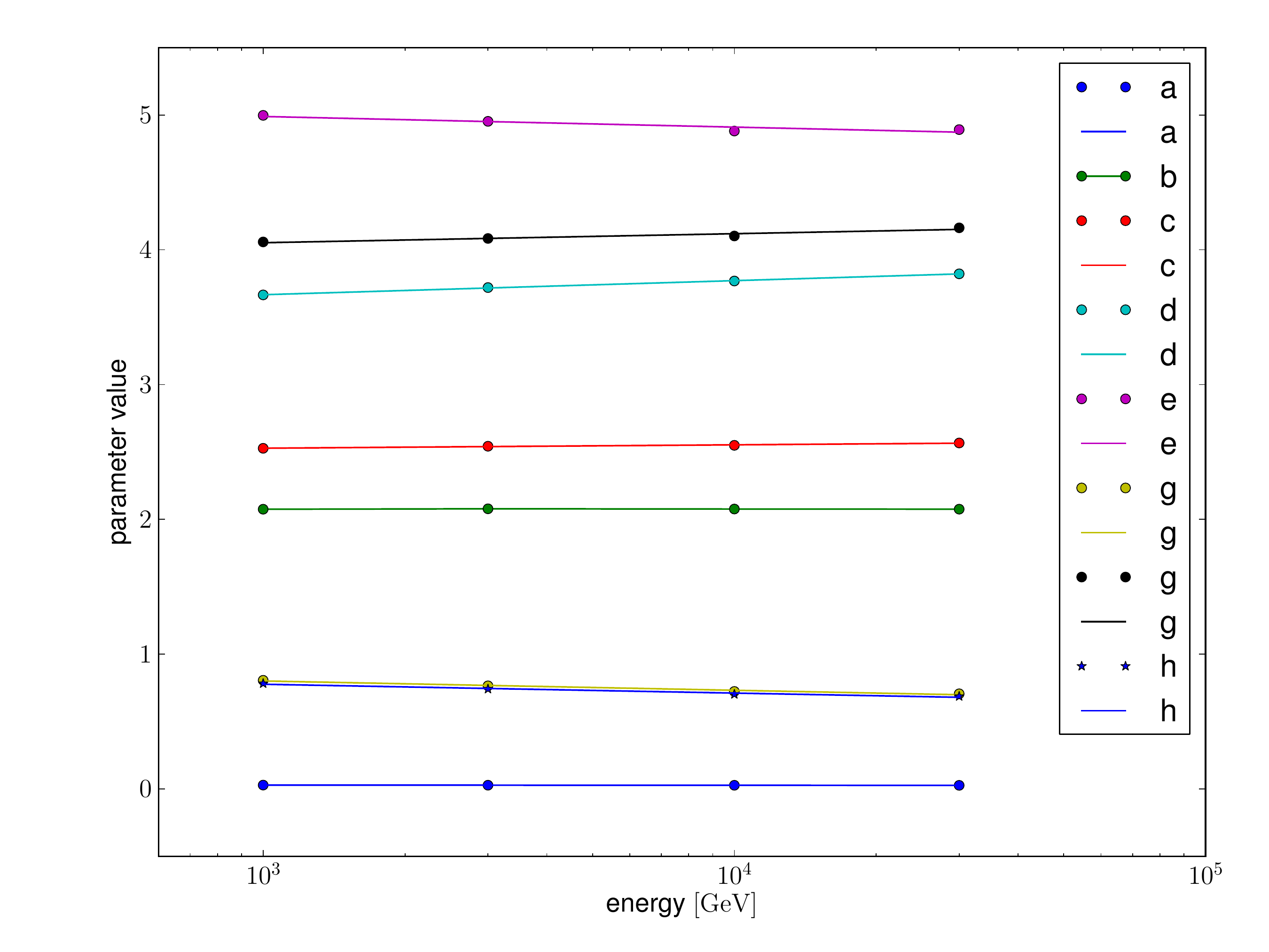}
\hfill
    \includegraphics*[width=.49\textwidth]{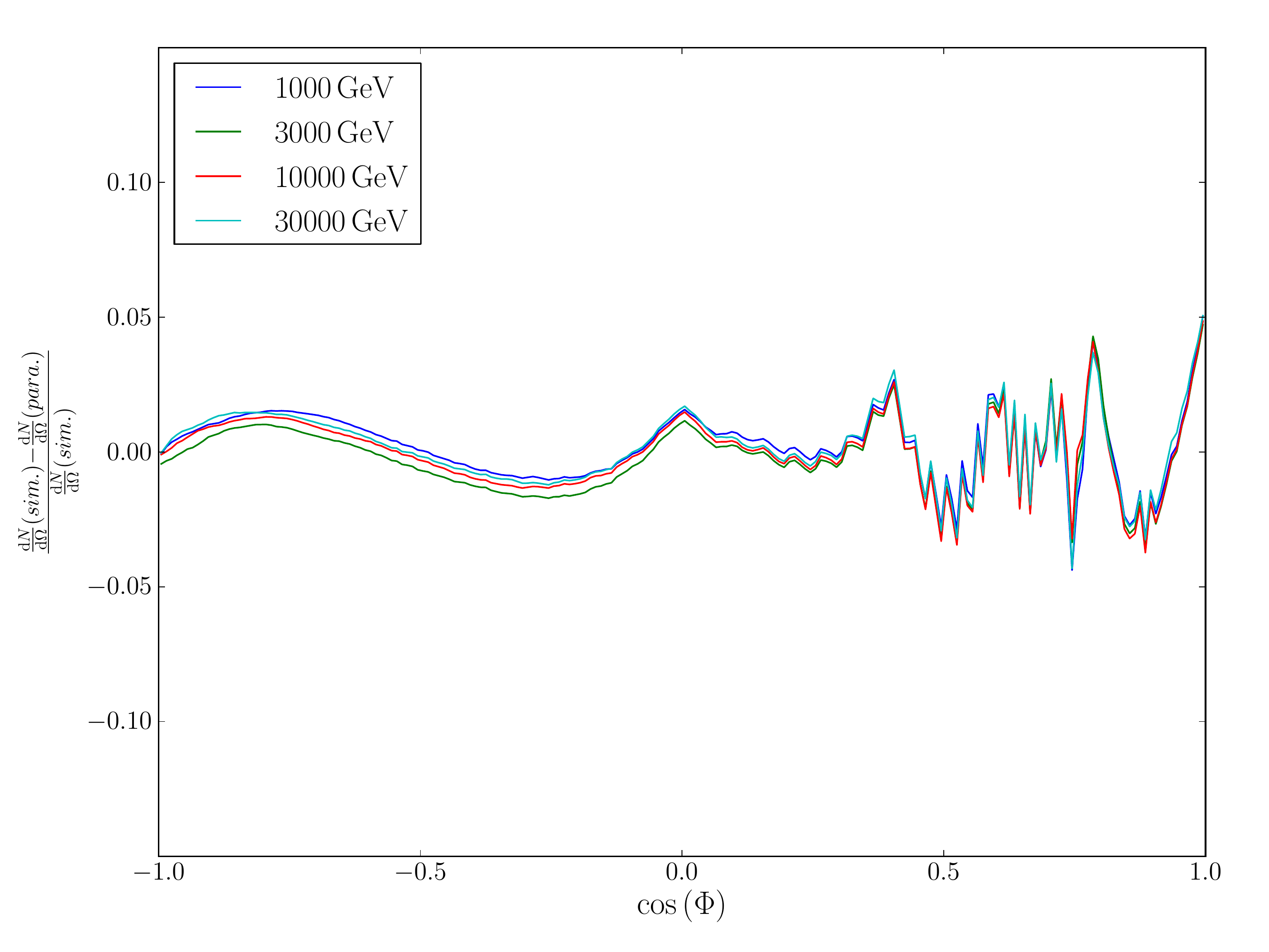}
	\caption{Parameters of the asymmetric fit versus energy (left). 
Markers represent the parameters and the solid lines are best fit lines
 \label{fig:result:paraen}
The right figure shows the relative deviations of the parameterization when the
energy dependence is used from the linear fits
\label{fig:result:dataenpara}}
\end{figure}

 The energy dependence of each  parameter $p_i$ 
 of the asymmetric parameterization has been fit
with a linear ansatz
\eqb
p_i = \lambda_0 + \lambda_1 \cdot \ln \lbra \frac{E_\mu}{\unit{GeV}} \rbra. 
\eqe
Figure \ref{fig:result:paraen} (left) 
shows the fit energy dependence of the parameters of the asymmetric parameterization. The numerical fit parameters are given in table \ref{tab:result:paraen} in 
 \ref{app:param:ang}.
Figure \ref{fig:result:dataenpara} (right) shows the relative deviation of the
data from the  parameterization with this linear ansatz.
The parameterization error related to the energy dependence is substantially 
smaller than the overall uncertainty.

\section{Summary and Conclusions}

In this paper we have focused on a very specific subject: 
the parameterization of \cherl from low-energy secondary particles, 
which accompany high-energy muons due to energy loss in water and ice. 

While high-energy catastrophic energy losses are usually 
treated stochastically by muon propagation codes,
 a similar  treatment of the 
low-energy tail for these processes would require a large amount of CPU.
 This is unnecessary because  these processes are almost continuous. For this
 we present a parameterization 
of the total \cherl light yield and its angular distribution.

The total amount of \cherl from these low-energy  processes is of the order of a few ten percent of the light from the bare muon and small compared to catastrophic  processes. However, the proper consideration of this light is required
for determining a correct energy scale e.g. for high-energy neutrino telescopes,
 in particular  to achieve  systematic uncertainties of the order of a few percent. 
As an example for the IceCube neutrino telescope, the parametrization 
\cite{CHWPHD} is used. That parametrization is based on an out-dated 
\geantdrei simulation
and does not include the correct \tammf for the photon yield. Furthermore, 
no  parameterization for the angular distribution 
of \cherphos is provided.
Here, we present a substantially improved parametrization of the total photon yield and the angular distribution, based on \geantfour. 

\medskip

In the first step of these  simulations it
is verified that the  light yield from low-energy processes 
increases logarithmically with the muon energy for a fixed high-energy 
cut-off  $E_{max} $. It is  found to be weakly dependent on  $E_{max} $.
A reasonable value for the transition between continuous and stochastic
treatment is $E_{max} \simeq 0.5 \unit{GeV} $, similar to the findings in \cite{CHWPHD}.

We find that the proper treatment of the \tammf leads to a small reducton of roughly  $10$\,\% of the additional photon yield. We find this reduction largely independent on $E_\mu $ and $E_{max}$.

A more detailed investigation reveals, that the  \cherl yield is dominated by  ionization processes, followed by pair production while  
bremsstrahlung and nuclear interactions are only marginally 
contributing.

For the calculation of the angular \cherl distribution, we present an improved version of a transformation method introduced in \cite{CHWPHD}. This
method  allows to calculate the photon distribution from a distribution
of track directions and their velocities. 
The resulting angular distribution of photons depends only weakly on the energy but more strongly on $E_{max} $. We are presenting a simple 3-parameter fit
and a more complex 8 parameter fit to these distributions and the energy dependence of these parameters.

\medskip

For the interpretation of the accuracy of these results it has to be 
considered that the additional light yield corresponds to only  a few 
ten percent of the light yield of a bare muon track. 
Hence, the errors of this calculation will have a small effect on the calculation of the total light yield for a muon track.
For high muon energies, $E_\mu\gtrsim 1 \unit{TeV} $, catastrophic 
processes dominate the light yield and outshine the muon as well as the 
small secondary tracks, which are considered  here.

The accuracy of the here presented calculations are limited by
 the accuracy of muon propagation code in \geantfour.
For high-energies, $E_\mu > 1 \unit{TeV} $, the cross sections 
in this code are poorly constrained by experimental data 
and rely on the proper calculation of electromagnetic processes.
The typical accuracy of these calculations is a few \% in the energy loss, 
see e.g. \cite{MMC}. For our  calculations we have identified   ionization 
 as the dominating process and this process 
has comparably smaller errors than other radiative processes. 
Hence, we  estimate the total 
systematic uncertainty of the light yield of the order of $1 \unit{\%} $ and
similar the differential error for the energy dependence, because
the light yield increases only logarithmically with  $E_\mu $.

For the angular distribution, we find that the relative 
accuracy of the  parametrization is about $10\unit{\%} $ for the simple
 and about $2-3 \unit{\%} $ for the asymmetric parameterization.

Corresponding to the $10\unit{m}$ length
of the muon track its energy is not constant but that uncertainty is small.
E.g. at $100 \unit{GeV} $ the mean energy loss is about $1 \unit{GeV} $
corresponding to $1\unit{\%} $ uncertainty, which could be corrected for.

For the application of the here presented results 
it has to be considered that an additional uncertainty  appears
because the low-energy 
secondary processes are assumed
to appear quasi-continuously and the direction of secondary tracks is 
azimuthally  symmetric. 
This is addressed in section \ref{sec:emax} where it is 
found that the chosen value $E_{max} $  is  a compromise with respect to
computing efficiency. A residual 
non-Poissonian component has to be considered.
However, the here used step-length of $10\unit{cm}$ is rather small 
compared to the length of a muon track in large neutrino telescopes of 
several hundred $\unit{m} $. 
Obviously, a systematic error related to the non-continuous light emission
and non-azimuthal symmetry will be largely suppressed 
for reasonably-long muon tracks. 
Furthermore, during propagation the \cherphos  are largely 
affected by scattering and change their direction.

In the here presented calculations an index of refraction of $n=1.33$
is used which corresponds to a specific \chera.
If required, a different index could be applied to the here presented 
parameterizations, because  the light yield is calculated 
in units of equivalent  track length and the resulting uncertainties
 due to the velocity distribution of secondary tracks close to the \cherthr
 are small.

No specific wavelength-interval, e.g. related to the  sensitivity of e.g. 
photo-sensors or the material transparency, 
has been assumed for the here presented calculations.

\begin{appendix}

\section{Transformation method for the calculation of the angular distribution 
of \cherpho from an angular distribution of track segments\label{sec:trafo}}

Goal of the transformation-method is to calculate the angular distribution
of the resulting \cherphos $dN \over d\phi $ with respect to the 
direction of the injected muon track.
 For the muon-track, which is aligned with the z-axis
 one expects a pure cone and all photons are emitted under the 
\chera $${dN\over d \phi } = N \cdot \delta(\theta_c) $$
where $\theta_c $  is defined by equation \ref{eq:cerangle} and $\delta(x)$ is the delta function.
However, for inclined track-segments, e.g. for secondary electrons,
one expects that the \cherc cone is inclined with respect to the $z$-axis
and the photons are emitted  in a broader  angular region with respect to the $z$-axis. If  the track-segment has a smaller velocity $\beta < 1$,
the number of photons is reduced according to equation \ref{eq:tamm}
and the \chera $\theta_c $ shrinks.

During the Geant-simulations we have stored the distribution of all 
track segments 
$l_i$ with respect to  the rapidity $\beta $ and the polar angle $\alpha $  to the $z$-axis. The $z$-axis is 
defined by the direction of the injected muon. This procedure therefore 
results in a 2-dimensional histogram (e.g. see figure \ref{fig:alpha:beta}): 
$$d^2l \over d\alpha d \beta$$

\begin{figure}[htp]
	\includegraphics*[width=.99\textwidth]{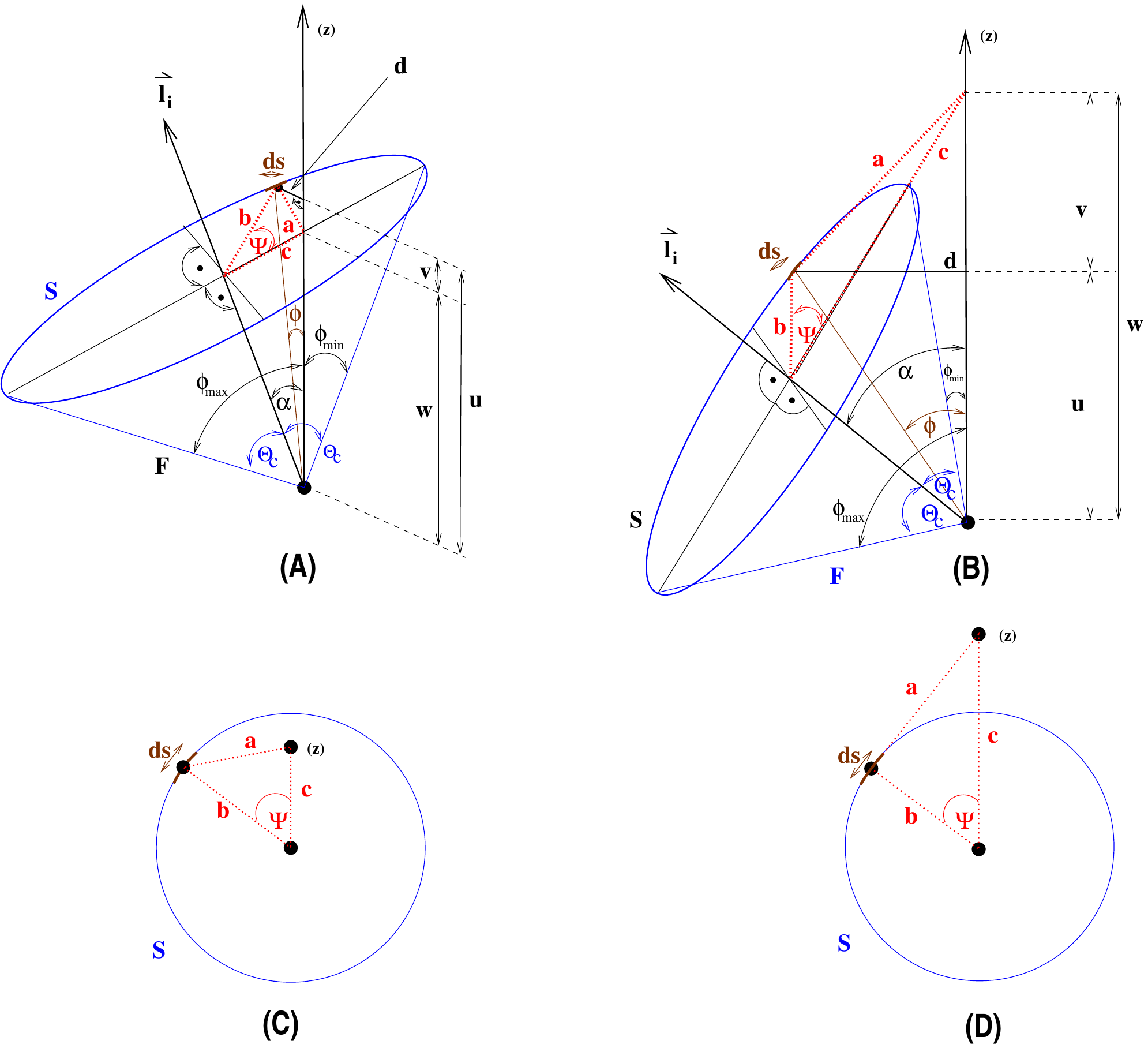}

	\caption{Illustration of the transformation geometry. 
\label{fig:app:trafo} The left figures (A), (C) 
shows a track $l_i$ with a small zenith $\alpha< \theta_c$, 
while for the right figures (B), (D) show a larger zenith 
$\alpha > \theta_c$. The top
 figures (A), (B) show a perspective side-view of the \cherc, 
while the bottom figures (C), (D) show a projected 
view aligned with the direction of $l_i$ in the plane of the cone.}
\end{figure}

The \cherphos will be distributed randomly along the rim of the 
\cherc with a total circumference $S$ (see figure \ref{fig:app:trafo}). The relative 
fraction of the cone's 
circumference $\Delta s \over S $, which points into the direction 
$\phi\pm \Delta\phi $ is proportional to the fraction of photons emitted into that
 direction.
Idea of the transformation algorithm is to calculate for all 
track elements and directions the fraction of photons, which are 
emitted into that direction. The method had been initially developed
in \cite{CHWPHD}, however, there the calculations were done assuming 
$\beta =1$ for all tracks. Here,  we extend the 
method to variable values of $\beta$ in order to properly 
account for the changing  \chera and photon yield for each track-segment
(eq.\ref{eq:tamm}).

The geometry is illustrated in figure \ref{fig:app:trafo}. 
Note, that for the following calculation we use radians instead of degrees.

The rim 
of the  cone, which is inclined by the angle $\alpha $ will have zenith angles between 
\eqb \phi_{min} \le \phi \le  \phi_{max} \eqe
with
\eqb \label{eq:phiminmax1}
\phi_{min} = | \alpha - \theta_c | \qquad \mbox{and} \qquad \phi_{min} = \alpha + \theta_c 
\eqe
except for $\phi > \pi - \theta_c $ where
\eqb\label{eq:phiminmax2}
 \phi_{max} = 2 \cdot \pi - \alpha - \theta_c 
\eqe

As we are only interested in the relative fraction of the rim,
we normalize the neck of the rim $F \equiv 1$.
Then, the length of the cone is $\cos(\theta_c) $,
the radius of the circular rim is $b=\sin(\theta_c) $ and therefore
\eqb
S = 2 \pi \cdot b = 2 \pi \sin(\theta_c )
\eqe 
The track element $ds$, which points towards the direction
$ \phi$ can be represented by an azimuth angle $\psi $
in the plane of the cone's rim.
\eqb
ds  = \sin (\theta_c ) \cdot d \psi
\eqe
The fraction of the rim is then
\eqb
\frac{ ds( \alpha, \beta )}{S} = \frac{2 \cdot {ds \over d\psi}  \cdot d \psi  }{ 
2 \pi \sin (\theta_c ) }  = {1 \over \pi } \cdot 
 d \psi
\eqe
The factor $2 $ arises if we only allow for angles $0 \le \psi \le \pi $.
Because of  the symmetry of the cone the  
 elements $ds'$ for  $\psi' = 2\pi - \psi $ point
into the same direction.

In the following  express the angle $\psi $ as a function of 
 the known  angles $\theta_c(\beta) $, $\alpha (l_i) $ and the target 
$\phi $.
The cosine law (see  figure \ref{fig:app:trafo} (C),(D)) results in
\eqb \label{eq:cosine}
\cos (\psi ) = {b^2+c^2-a^2 \over 2 b c }
\eqe
From the figures \ref{fig:app:trafo} (C),(D) one can obtain the identities
\eqb
b= \sin (\theta_c ) \qquad \tan (\alpha) = {c \over \cos (\theta_c) }
\qquad a^2 = d^2 + v^2 
\eqe
and further
\eqb
 \sin (\phi ) = {d \over F} = d  ~,  \qquad v =|u-w| ~, \qquad  \cos (\phi)
= {u \over F} = u
~, \qquad \cos \alpha = {\cos \theta_c \over w }
\eqe
resolving for $a$, $b$, $c$ and inserting into equation \ref{eq:cosine}
gives
\eqb
\cos (\psi ) = \frac{\sin^2 (\theta_c ) + \tan^2 (\alpha ) \cdot \cos^2 (\theta_c ) -  \sin^2 (\phi) - \lbra \cos (\phi )  - {\cos (\theta_c) \over \cos (\alpha ) } \rbra^2   }{2 \cdot \sin (\theta_c ) \cdot \tan (\alpha ) \cdot \cos (\theta_c ) },
\eqe
which can be simplified to
\eqb \label{eq:finalpsi}
\cos (\psi ) = {\cot (\theta_c ) \over \sin (\alpha) } \cdot
\lbra  {\cos (\phi) \over \cos (\theta_c)}  - \cos (\alpha )\rbra
\eqe
With this formula the $\Delta \Psi $-range, which corresponds 
to the bin boundaries of the source histogram
$d^2l \over d\alpha d \beta$
and target histogram $dN \over d\phi $ can be calculated.

For practical reasons it is convenient not to 
 calculate the number of photons 
explicitly but rather to use a relative  light  yield with respect to a 
 relativistic track.
The relative photon yield for a track $\beta <1 $ relative to $N_0$
(for $\beta =1 $)
is
\eqb
\frac{N}{N_0} = \frac{sin^2(\theta_c)}{sin^2(\theta_{c,0})}
\eqe
In the following we will use 
\eqb
\hat{l}(\phi) \equiv l \cdot  \frac{sin^2(\theta_c)}{sin^2(\theta_{c,0})}  \cdot {d\psi \over \pi},
\eqe
which is the  fractional part  of the track length $l$, which illuminates
the direction $\phi $ rescaled to the photon-yield relative to a relativistic track.

This particular choice is advantageous, because it is
independent of a specific wavelength interval for the \cherphos.
The here presented results can  
be rescaled to arbitrary wavelength intervals.

\bigskip

The numerical algorithm proceeds as follows:
\enb
\item A loop over all bins in the 2-dimensional source-histogram $d^2l \over d\alpha d \beta $ gives the track length $l_{ij} $, which corresponds to
a certain bin  $\beta_i $ and $\alpha_j $ 

\item The \chera $\theta_c (\beta_i, n ) $ is calculated
according to equation \ref{eq:cerangle}.

\item The target range $\phi_{min} (\theta_c, \alpha_j ) $ and
 $\phi_{max} (\theta_c, \alpha_j ) $ are calculated according to equation \ref{eq:phiminmax1} and \ref{eq:phiminmax2}.

\item For all bins $\phi_k$ of the target histogram 
$d\hat{l} \over d\phi $, for which $\phi_{min} < \phi_k < \phi_{max} $
we can calculate the $\Psi$-range, which corresponds
to the bin boundaries according to equation \ref{eq:finalpsi}
\eqb
\Delta \psi \equiv \Psi (\theta_c, \alpha_j , min(\phi_{k+1},\phi_{max} )) 
- \Psi (\theta_c, \alpha_j , max(\phi_k,\phi_{min} )) 
\eqe
\item The target histogram bin ${d\hat{l} \over d\phi}( \phi_k) $ is incremented
with the value
\eqb
\Delta \hat{l}(\phi_k ) = l_{ij} \cdot \frac{\Delta \psi }{\pi} \cdot \frac{\sin^2 (\theta_c )}{sin^2(\theta_{c,0}) }
\eqe
The final  value $\hat{l}_k$ of  the target histogram then  
corresponds to the equivalent relativistic track length, emitting
the  total number of photons as observed under this zenith.

\item The procedure is numerically improved by the interpolation
of the  values $\beta $ and $\alpha $ within the bins of the input histogram.
\ene

\begin{figure}[htp]
	\includegraphics*[width=.99\textwidth, page=1]{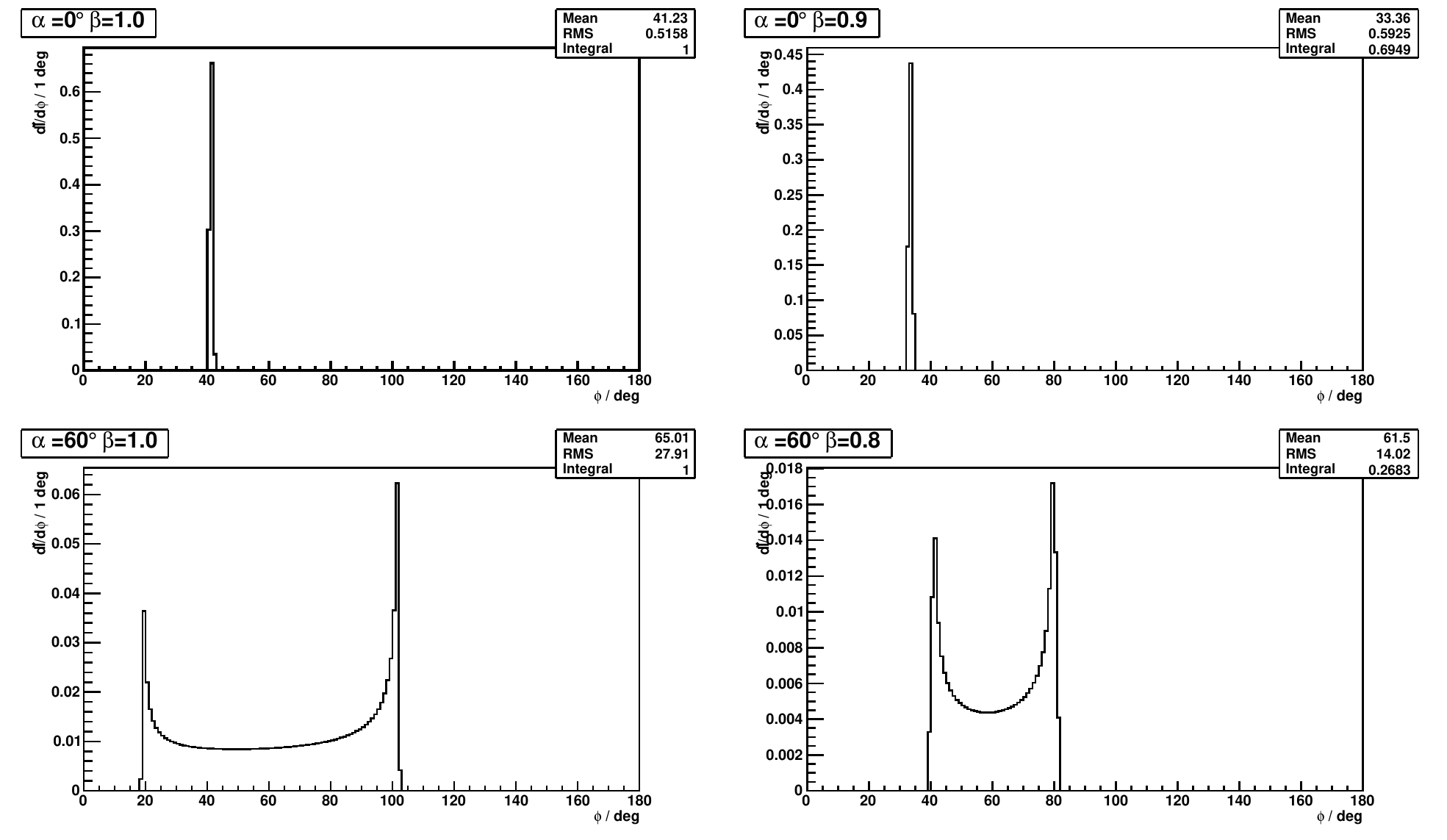}
	\caption{Examples of calculated effective light distribution. Shown is the resulting distribution: $d\hat{l} /d\phi $ for a single track of unit 
length with different directions $\alpha $ and velocities $\beta$ and a refraction index $n=1.33$.
\label{fig:app:exam} }
\end{figure}

Examples of the transformation are shown in figure
\ref{fig:app:exam}. For a track into direction (z) (top figures) 
the light cone is aligned
with the z-axis and all photons are emitted under a fixed angle. A relativistic track of unitlength 1 will produce a distribution\footnote{The finite 
width of the distribution results from the finite bin size of the 
source histogram and becomes slightly larger for smaller $\beta$.}, which is peaked at the \chera
and the integral corresponds to 1 unitlength (top left).
For smaller values of $\beta  $ the opening angle  of the cone shrinks
and  the total light yield is reduced (top right). 
 For
inclined tracks the photons are distributed over a large angular region (bottom left).
Also here  for smaller values of $\beta $ the opening angle  
of the cone shrinks
and  the total light yield is reduced (bottom right).

\section{Geant4 configuration parameters used for this study \label{app:geant:conf}}

In this chapter a summary of the defined media properties and physics processes is given.

\subsection{Materials}

\begin{table}[htp]
    \begin{tabular}{lcccc}
        \hline
        \hline
        Medium    & Density $\left[\frac{\mathrm{g}}{\mathrm{cm}^3}\right]$  & Element & Fraction of mass & Isotope abundance \\
        \hline
        Ice       & 0.910 & Hydrogen & 88.81\% & \\
                  &       & Oxygen   & 11.19\% & \\
        \hline
        Water     & 1.000 & Hydrogen & 88.81\% &\\
                  &       & $\mathrm{H}^{1}$  &         & 99.9\%\\
                  &       & $\mathrm{H}^{2}$  &         & 0.01\%\\
                  &       & Oxygen            & 11.19\% & \\
                  &       & $\mathrm{O}^{16}$ &         & 99.76\%\\
                  &       & $\mathrm{O}^{17}$ &         & 0.04\%\\
                  &       & $\mathrm{O}^{18}$ &         & 0.20\%\\
        \hline
        Sea water & 1.039 & Hydrogen  & 10.74\% & \\
                  &       & Oxygen    & 85.41\% & \\
                  &       & Chlorine  & 2.37\%  & \\
                  &       & Sodium    & 1.32\%  & \\
                  &       & Magnesium & 0.15\%  & \\
        \hline
        \hline
    \end{tabular}
    \caption{Composition of ice, water and sea water as used in the \geantfour-simulation. Water is defined as 
    \texttt{G4\_WATER} from the NIST database.}
    \label{table:geant:materials}
\end{table}

In \geantfour macroscopic properties of matter are described by \texttt{G4Material} and the atomic
properties  are described by \texttt{G4Element}. A material can consist of multiple
elements and therefore represent a chemical compound, mixture as well as pure materials.

For the performed simulations, three different materials were used: Ice, water and sea water.
The properties are summarized in table \ref{table:geant:materials}.

Unless noted otherwise, the value $n=1.33 $ is used for the index of refraction.

\subsection{Physicslist}
\begin{table}[htp]
    \begin{tabular}{llc}
        \hline
        \hline
        Particle  & Process & Model \\
        \hline
        $\gamma$  & G4PhotoElectricEffect & G4PEEffectFluoModel \\
                  & G4ComptonScattering & G4KleinNishinaModel \\
                  & G4GammaConversion  & \\
        \hline
        $e^-$     & G4CoulombScattering & \\
                  & G4eIonisation & \\
                  & G4eBremsstrahlung & \\
        \hline
        $e^+$     & G4CoulombScattering & \\
                  & G4eIonisation & \\
                  & G4eBremsstrahlung & \\
                  & G4eplusAnnihilation & \\
        \hline
        $\mu^+,\mu^-$ & G4CoulombScattering & \\
                      & G4MuIonisation & \\
                      & G4MuBremsstrahlung & \\
                      & G4MuPairProduction & \\
                      & G4MuNuclearInteraction & \\
        \hline
        $pi^+,pi^-,K^+,K^-,p^+$ & G4CoulombScattering & \\
                                & G4hIonisation & \\
                                & G4hBremsstrahlung & \\
                                & G4hPairProduction & \\
        \hline
        $\alpha,\mathrm{He}^{3+}$ & G4ionIonisation & \\
                                  & G4CoulombScattering & G4IonCoulombScatteringModel \\
                                  & G4NuclearStopping & \\
        \hline
        all unstable particles & G4Decay & \\
        \hline
        \hline
    \end{tabular}
    \caption{Physics processes of most important particles used in the simulation. 
    If no model is specified the default model is used. For hadrons and ions that are
    not listed Coulomb scattering and ionisation are defined.}
    \label{table:geant:physicslist}
\end{table}

All physics processes, which are used during the simulation must be registered in 
\texttt{G4VUserPhysicsList}. For these simulations a customized version of the list \texttt{G4EmStandardPhysics\_option3} is used. 
Multiple scattering has been replaced by single scattering and the process \texttt{G4MuNuclearInteraction} has been added. 
In table \ref{table:geant:physicslist} all registered processes are summarized.

Single scattering is used in favor of multiple scattering because the step size in multiple
 scattering is not limited. This leads to an  overestimated track length in certain directions.
This is observed in particular for $\delta$-electrons from the ionisation process.
The zenith angle $\alpha $ of the quasi free electron is kinematically closely related to the velocity $\beta $ and the muon energy $E_\mu $
\begin{equation}
    \cos\alpha=\frac{1-\sqrt{1-\beta^2}}{\beta}\frac{E_{\mu}+m_{e}}{\sqrt{E_{\mu}^{2}-m_{\mu}^2}}.
    \label{equation:alpha_beta}
\end{equation}
The directions of these lower-energy electrons is quickly smeared
due scattering. However, the process multiple scattering leads to an overestimated track length for 
the initial direction
and hence the distribution of \cherphos is biased.

The default maximum energy for the cross section tables and the calculation of  $\mathrm{d}E/\mathrm{d}x$ in
 \geantfour is $10\unit{TeV}$. However, muon processes can be extrapolated theoretically
  up to $10\unit{PeV}$ and
several  high-energy corrections are implemented in \geantfour, e.g. LPM suppression in gamma conversion and nuclear EM form-factors in single scattering models \cite{IVANCHENKO}.

\section{Parameterization results of the light yield and angular \cherl distribution \label{app:param:ang}}

\begin{table}[htp] \centering
\begin{tabular}{|l|c|c|cc|cc|}
\hline
Process & n & $E_{max} /$GeV & \ruo $ \hat \lambda_0 $ & $\hat \kappa$ &  $ \lambda_0 $ & $ \kappa$\\
\hline
total & \multirow{5}{*}{1.33} & \multirow{5}{*}{0.2}  &  0.1969 & 0.0081 &  0.2329 & 0.0087 \\
ion   &  &  &  0.2117 & 0.0024 &  0.2486 & 0.0026 \\
pair  &  &  & -0.0150 & 0.0058 & -0.0158 & 0.0061 \\
brems &  &  &  0.0002 & 0      &  0.0002 & 0      \\
nucl  &  &  &  0.0000 & 0      &  0.0000 & 0      \\
\hline
\hline
total & \multirow{5}{*}{1.33} & \multirow{5}{*}{0.5}  &  0.1880 & 0.0206 &  0.2235 & 0.0219 \\
ion   &  &  &  0.2489 & 0.0030 &  0.2879 & 0.0033 \\
pair  &  &  & -0.0626 & 0.0175 & -0.0651 & 0.0186 \\
brems &  &  &  0.0004 & 0      &  0.0004 & 0      \\
nucl  &  &  &  0.0005 & 0      &  0.0005 & 0      \\
\hline
\hline
total & \multirow{5}{*}{1.33} & \multirow{5}{*}{1.0}  &  0.1028 & 0.0445 &  0.1334 & 0.0472 \\
ion   &  &  &  0.2712 & 0.0043 &  0.3114 & 0.0046 \\
pair  &  &  & -0.1686 & 0.0400 & -0.1783 & 0.0423 \\
brems &  &  &  0.0008 & 0      &  0.0009 & 0      \\
nucl  &  &  &  0.0010 & 0      &  0.0011 & 0      \\
\hline
\hline
total & \multirow{5}{*}{1.30} & \multirow{5}{*}{0.5}  &  0.1842 & 0.0204 &  0.2196 & 0.0218 \\
ion   &  &  &  0.2247 & 0.0030 &  0.2837 & 0.0032 \\
pair  &  &  & -0.0612 & 0.0174 & -0.0648 & 0.0185 \\
brems &  &  &  0.0004 & 0      &  0.0004 & 0      \\
nucl  &  &  &  0.0005 & 0      &  0.0005 & 0      \\
\hline
\hline
total & \multirow{5}{*}{1.36} & \multirow{5}{*}{0.5}  &  0.1916 & 0.0207 &  0.2268 & 0.0220 \\
ion   &  &  &  0.2528 & 0.0030 &  0.2914 & 0.0087 \\
pair  &  &  & -0.0619 & 0.0176 & -0.0654 & 0.0186 \\
brems &  &  &  0.0004 & 0      &  0.0004 & 0      \\
nucl  &  &  &  0.0005 & 0      &  0.0005 & 0      \\
\hline
\hline
total \cite{CHWPHD} & 1.33 & 0.5-1.0 & - & - & 0.172 & 0.032 \\
\hline
\hline
\end{tabular}
\caption{Parameterization of the additional light yield
for different $n$ and $E_{max} $
\label{tab:tracklength}}
\end{table}

\begin{table}[htp]
    \begin{tabular}{llccccccccc}
        \hline
        \hline
        \parbox{1cm}{$E_{max}$\\ {MeV}} &
 \parbox{1cm}{$E_{primary} $ \\ {GeV} }  & a & b & c & d & e & g & h & i \\
        \hline
        200  & 1000   & 0.0325  & 2.090 & 2.486 & 3.412 & 5.910 & 1.134 & 4.639 & 1.169 \\
             & 3000   & 0.0320  & 2.090 & 2.498 & 3.438 & 5.846 & 1.104 & 4.535 & 1.124\\
             & 10000  & 0.0317  & 2.090 & 2.500 & 3.469 & 5.530 & 1.046 & 4.344 & 1.060 \\
             & 30000  & 0.0313  & 2.090 & 2.514 & 3.497 & 5.451 & 1.024 & 4.309 & 1.035\\
        \hline
        500  & 1000   & 0.0288  & 2.075 & 2.527 & 3.665 & 4.997 & 0.806 & 4.058 & 0.782\\
             & 3000   & 0.0281  & 2.079 & 2.543 & 3.720 & 4.953 & 0.765 & 4.084 & 0.742 \\
             & 10000  & 0.0275  & 2.077 & 2.550 & 3.768 & 4.881 & 0.723 & 4.102 & 0.703 \\
             & 30000  & 0.0269  & 2.076 & 2.567 & 3.822 & 4.892 & 0.706 & 4.163 & 0.687\\
        \hline
        1000 & 1000   & 0.0264  & 2.071 & 2.557 & 3.855 & 4.711 & 0.629 & 4.105 & 0.613\\
             & 3000   & 0.0254  & 2.074 & 2.580 & 3.927 & 4.767 & 0.589 & 4.224 & 0.575 \\
             & 10000  & 0.0246  & 2.073 & 2.588 & 4.007 & 4.820 & 0.561 & 4.343 & 0.550\\
             & 30000  & 0.0239  & 2.071 & 2.606 & 4.081 & 4.857 & 0.540 & 4.437 & 0.532 \\
        \hline
        \hline
    \end{tabular}
	\caption{Parameters of the asymmetric parameterization, with 
$n=1.33$ ($f = 1/n \approx 0.7519$)
    \label{tab:result:paraasym}}
\end{table}

\begin{table}[htp]
\centering
    \begin{tabular}{llccccc}
        \hline
        \hline
        \parbox{1cm}{$E_{max}$\\ {MeV}} &
 \parbox{1cm}{$E_{primary} $ \\ {GeV} }  & a & b & c \\
        \hline
        200  & 1000  & 0.4036 & -3.075 & 0.7299 \\
             & 3000  & 0.4147 & -3.118 & 0.7183 \\
             & 10000 & 0.4282 & -3.163 & 0.7038 \\
             & 30000 & 0.4367 & -3.195 & 0.6967 \\
        \hline
        500  & 1000  & 0.5597 & -3.531 & 0.6033 \\
             & 3000  & 0.5983 & -3.626 & 0.5840 \\
             & 30000 & 0.6660 & -3.776 & 0.5554 \\
             & 10000 & 0.6381 & -3.711 & 0.5656 \\
        \hline
        1000 & 1000  & 0.7403 & -3.899 & 0.5261 \\
             & 3000  & 0.8266 & -4.049 & 0.5024 \\
             & 10000 & 0.9084 & -4.175 & 0.4837 \\
             & 30000 & 0.9778 & -4.278 & 0.4709 \\
        \hline
        \hline
    \end{tabular}
	\caption{Parameters of the simple parameterization, with $n=1.33$ ($f = 1/n \approx 0.7519$)
    \label{tab:result:parasimple}}
\end{table}

\begin{table}[htp]
    \begin{tabular}{lccccc}
        \hline
        \hline
         $p_i $ & $\lambda_0$ & $\lambda_1 $ \\
        \hline
         a & 0.03257 & -0.00055 \\
         b & 2.07658 &  0 \\
         c & 2.45248 &  0.01092 \\
         d & 3.35320 &  0.04536 \\
         e & 5.22537 & -0.03425 \\
         g & 1.00817 & -0.02997 \\
         h & 3.85166 &  0.02905 \\
         i & 0.97332 & -0.02845 \\
        \hline
        \hline
    \end{tabular}
\hfill
    \begin{tabular}{lccccc}
        \hline
        \hline
         $p_i$ & $\lambda_0$ & $\lambda_1 $ \\
        \hline
        \hline
         a &  0.34485492 &  0.03144662 \\
         b & -3.04159806 & -0.07192665 \\
         c &  0.69937264 & -0.01420755 \\
        \hline
        \hline
    \end{tabular}

	\caption{Parameterization of energy dependence of fit parameters for $n=1.33$ and $E_{max}=500\unit{MeV}$ for the asymmetric parameterization (left) and the simple parameterization (right).
    \label{tab:result:paraen}}
\end{table}


\end{appendix}

\clearpage

\section*{Acknowledgement}

We  thank the IceCube group at the RWTH Aachen University for fruitful 
discussions. In particular Karim Laihem was of great help in discussing and setting up the simulation. The layout of the simulation geometry has been suggested to 
us by Marius Wallraff.
We thank Dima Chirkin for reading the manuscript and 
suggestions for the angular parameterizations.

%
%

\bibliographystyle{elsarticle-num}
\bibliography{references}



\end{document}